\documentclass[9pt,twoside]{extarticle}

\usepackage[margin=0.75in, top=0.75in, bottom=0.75in,
            columnsep=0.25in]{geometry}
\usepackage{multicol}

\usepackage[T1]{fontenc}
\usepackage{mathptmx}
\usepackage[scaled=0.86]{helvet}
\usepackage{microtype}

\usepackage{amsmath,amssymb,bm}
\usepackage{mathtools}

\usepackage{float}

\usepackage{graphicx}
\usepackage{booktabs}
\usepackage{dcolumn}
\usepackage[rightcaption]{sidecap}
\usepackage[labelfont=bf,
            font=small,
            labelsep=period]{caption}
\usepackage{enumitem}

\usepackage{xcolor}
\usepackage{url}
\newcommand{\doi}[1]{\href{https://doi.org/#1}{\detokenize{#1}}}
\usepackage{hyperref}
\usepackage{hycolor}
\hypersetup{colorlinks=true,citecolor=[rgb]{0,0.18,0.68},
urlcolor=[rgb]{0,0.18,0.68}, linkcolor=[rgb]{0,0.18,0.68}, final = true}
\renewcommand{\url}[1]{\href{#1}{\detokenize{#1}}}
\usepackage{relsize}

\usepackage[super,compress,sort]{natbib}
\setcitestyle{numbers,super,open={},close={}}

\usepackage{titlesec}
\titleformat{\section}{\small\bfseries\sffamily\uppercase}{\thesection.}{0.5em}{}
\titleformat{\subsection}{\small\bfseries\sffamily}{\thesubsection.}{0.5em}{}
\titleformat{\subsubsection}{\small\itshape}{\thesubsubsection.}{0.5em}{}

\usepackage{fancyhdr}
\usepackage{pdfpages}

\usepackage{abstract}

\usepackage{siunitx}
\titleformat{\section}
    {\fontsize{12}{10}\bfseries\sffamily\uppercase}
    {\thesection.}{0.5em}{}

  \def\be{\begin{equation*}}
  \def\ee{\end{equation*}}
  \def\ba{\begin{eqnarray}}
  \def\ea{\end{eqnarray}}
  
  \def\fref#1{Fig.~\ref{#1}}

  \def\bt{\textrm} 
  
  \def\nsb#1{\noindent\textbf{\bt{#1~}}}
  \def\nsi#1{\noindent\textit{\bt{#1~--}}}
  \definecolor{or}{RGB}{234,142,53}
  \definecolor{gr}{RGB}{150,150,150}
  \definecolor{bl}{RGB}{54,152,187}

  \newcommand{\ie}{\textit{i.e.}}
  \newcommand{\eg}{\textit{e.g.}}

  \definecolor{YKB}{rgb}{0.00,0.18,0.65}

\begin{document}

\title{\LARGE\bfseries\sffamily Evolutionary modelling reveals
  melodic and harmonic constraints on global scale structure.}

\author{
    John M. McBride\textsuperscript{1,*} \and
    Steven Brown\textsuperscript{2} \and
    Elizabeth Phillips\textsuperscript{2} \and
    Patrick E. Savage\textsuperscript{3,4} \and
    Tsvi Tlusty\textsuperscript{1}
}

\date{
    \small
    \textsuperscript{1}Department of Physics, Ulsan National
    Institute of Science and Technology, South Korea\\[0.5em]
    \textsuperscript{2}McMaster University, Canada\\
    \textsuperscript{3}University of Auckland, New Zealand\\
    \textsuperscript{4}Keio University, Japan\\
    \textsuperscript{*}Correspondence:
    \href{mailto:jmmcbride@protonmail.com}{jmmcbride@protonmail.com}\\
}

\twocolumn[
  \begin{@twocolumnfalse}
    \maketitle
  \end{@twocolumnfalse}
]
\includegraphics[width=0.33\textwidth]{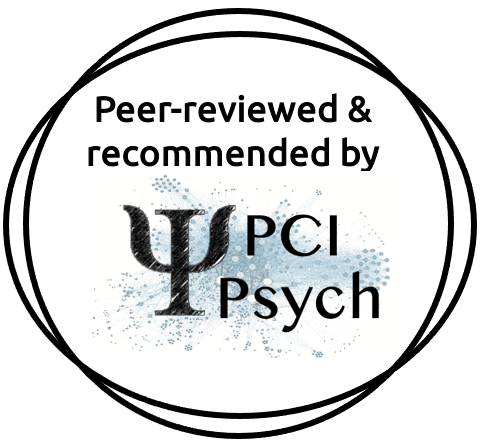}~

\begin{abstract}
  Since antiquity, musical scales have been explained by harmony rather than melody.
  This view relies on the mathematically designed scales of a few traditions,
  and was never directly tested. Testing it requires cross-cultural data and a
  method that judges theories by what they get wrong as well as right.
  We provide both, modelling scale evolution across 1,314 scales from 96 
  countries. A Melody model explains the near-universal preference for step-sizes
  of 1-3 semitones, and matches independent data from melodies, singing,
  and psychoacoustics. Harmony does far less: it explains the music-theoretic scales,
  but in those measured from performance it adds only a weak bias towards fourths,
  fifths, and octaves. Harmony's importance has been overstated, likely due to the
  historical focus on music-theoretic rather than measured scales.
  Melody is the primary driver of global scale structure; harmonic constraints are
  less impactful and mainly reflect musicological theory over musical performance.
  \\
  \\ This article was reviewed and recommended by Peer Community in Psychology.~\cite{kimRecommendation2026}

\end{abstract}
\vspace{0.5cm}
\vspace{0.5cm}

\section*{Introduction}

  Scales are sets of frequency ratios, or intervals, that form the basic
  building blocks of melodies\cite{ellisMusical1885,mcbrideWhat2026}. They are
  one of the most universal and defining features of human music
  \cite{savageStatistical2015,mehrUniversality2019,harwoodUniversals1976,brownUniversals2013,net00,savageComparative2022}.
  Some scales are known to be thousands of years old, but in general scales
  change over time through cultural evolution\cite{creanzaCultural2017,youngbloodCultural2023a}.
  This change does not appear to be random. Scales across the world tend to be more
  similar than what would be predicted by chance, and certain scales (\eg, the
  minor pentatonic) appear repeatedly in far-flung corners of the
  world\cite{khePentatonic1977,rechbergerScales2018,mcbrideConvergent2023b,brownMusical2025a}.
  This suggests that there is some selection process or conscious innovation
  common to different groups of people that leads to the use of similar scales
  (\textbf{SI Section S1})\cite{mcbrideConvergent2023b,mcdermottIndifference2016b,
  mcphersonPerceptual2020a,jacobyUniversal2019a,aucouturierHypothesis2008a,
  gillBiological2009,berezovskyStructure2019,mcbrideCrosscultural2020,buecheleCrystals2025,marjiehTimbral2024c,pelofiAsymmetry2021a}.

  The dominant theory of scales can be traced back to the Pythagorean
  theory that certain harmonic intervals (\eg, octaves and fifths) are inherently
  consonant\cite{nicomachusManual1994}. While 21\textsuperscript{st} century
  computational modelling and psychoacoustic experiments have challenged the
  relative importance of different aspects of harmony, they have re-affirmed
  the traditional view that the origin of musical scales lies in harmony
  \cite{bowlingVocal2023a,aucouturierHypothesis2008a,
  gillBiological2009,berezovskyStructure2019,marjiehTimbral2024c}.
  Modern theories of the innate harmonic properties of scale intervals are
  supported by two psychoacoustic
  phenomena\cite{greenAcoustics2002,harrisonSimultaneous2020b}: \textit{Tonal
  fusion}\cite{stumpfTonpsychologie1890} occurs when two complex tones have overlapping
  partials (also called overtones, or harmonics),
  making them difficult to distinguish. \textit{Auditory
  roughness}\cite{helmholtzSensations1885,plompTonal1965,mcbrideMusical2025} is perceived
  when interference between the partials of complex tones lead to
  audible amplitude-modulation patterns. These phenomena can be described, respectively,
  by harmonicity and interference models\cite{harrisonSimultaneous2020b}. Despite robust empirical support
  for these phenomena\cite{mcphersonPerceptual2020a,marjiehTimbral2024c,
  bidelmanNeural2009,mcdermottIndividual2010a,cousineauBasis2012,
  armitageCulture2023,plompTonal1965,setharesTuning2005,dewittTonal1987,
  demanyPerception2021},
  we do not know how they affect scale evolution.

  Three hypotheses link scale evolution to tonal fusion and/or auditory
  roughness without clearly distinguishing between the two
  phenomena. The first we have mentioned: that humans have an innate
  preference for harmonic intervals\cite{marjiehTimbral2024c,harrisonSimultaneous2020b,
  terhardtConcept1984,bowlingNature2017,distefanoConsonance2022b,bowlingVocal2023a},
  for which some evolutionary explanations have been proposed
  \cite{bowlingVocal2023a,arnalRough2025}. However, the cross-cultural
  evidence for this hypothesis is mixed\cite{mcdermottIndifference2016b,
  lahdelmaSweetness2021,milneEvidence2023a}, being confounded by culture
  \cite{armitageCulture2023,mclachlanConsonance2013,
  parncuttPsychocultural2018,lahdelmaCultural2020,lahdelmaDatadriven2023a}.
  Another theory is that group harmony facilitates social bonding through
  perceived synchrony\cite{savageMusic2021}. Alternatively, a
  hypothesis based on instruments is that these
  phenomena enable reliable tuning of intervals (\eg, octaves and fifths) as a
  precursor to tuning technologies (\eg, monochords, pitch
  pipes)\cite{mcbrideConvergent2023b,rahnWas2022,hallPerception1984}. Note that
  the first two hypotheses only apply to polyphonic music, while the third
  hypothesis also applies to (instrumental) monophonic music. Despite the
  mechanistic differences between these hypotheses, they make the same
  predictions: they predict a bias towards the use of harmonic intervals in scales.

  \begin{figure*}[th!]
  \centering
  \includegraphics[width=\textwidth]{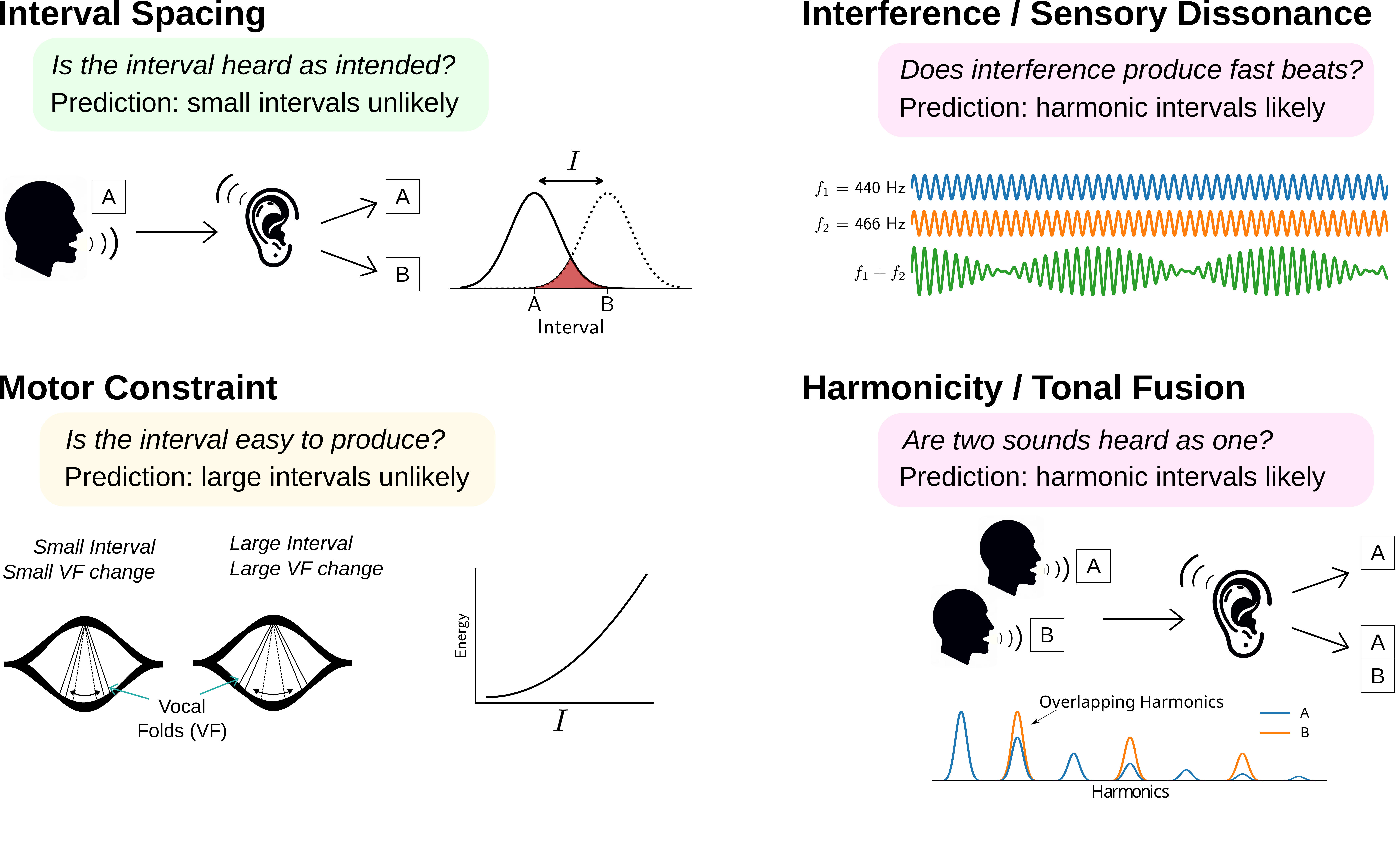}
  \caption{\textbf{Mechanisms by which the perception and production of
  musical intervals may bias scale evolution.} 
  \textit{Interval Spacing}: intervals are imprecisely produced and
  perceived, so two interval categories `A' and `B' that are too close are
  easily confused (shown as shaded overlap between perceived pitch distributions);
  very small intervals are therefore unlikely.
  \textit{Motor Constraint}: producing a larger interval requires
  a larger change in vocal-folds, and hence more energy, so large
  intervals are less likely.
  \textit{Interference / Auditory Roughness}: two simultaneous tones with
  partials close in frequency produce an audible interference pattern 
  called \textit{beats} at a rate given by $|f_1 - f_2|$; harmonic intervals minimise fast
  (roughly \SIrange{10}{100}{Hz}) beats and are favoured.
  \textit{Harmonicity / Tonal Fusion}: when the partials of two
  tones coincide (overlapping partials) the tones tend to fuse into a single
  percept, again favouring harmonic intervals. The Interference and Harmonicity
  mechanisms make the same prediction -- a bias towards harmonic intervals -- and
  we treat them together. We combine these mechanisms into the three theories
  tested here: \textit{Melody} (Interval Spacing $+$ Motor Constraint),
  \textit{Harmony} (Interval Spacing $+$ harmonicity/interference), and
  \textit{Full} (Interval Spacing $+$ Motor Constraint $+$ harmonicity/interference).}
  \label{fig:1}
  \end{figure*}

  Whatever the mechanism, the evidence that scale evolution is biased towards harmonic
  intervals is less secure than its long history suggests. Previous studies on this
  topic\cite{aucouturierHypothesis2008a,gillBiological2009,
  berezovskyStructure2019,buecheleCrystals2025} share some common problems.
  First, they did not compare competing models or provide a null model, so their
  results cannot be put in context -- a model is only informative relative to a
  baseline. Second, they examined scales from only a limited number of societies,
  many of which share related theoretical traditions, so the samples were not
  culturally independent. Third, they only looked at whether extant scales were
  predicted by models, and did not consider incorrect predictions -- a model that
  predicts everything predicts nothing.
  In previous work we partially solved these problems, but issues such as sampling biases
  remained\cite{mcbrideCrosscultural2020}. Here we compare multiple harmony models,
  including those that assume that a full series of partials is important,
  to those that assign importance only to the first few partials,
  and everything in between\cite{harrisonSimultaneous2020b}.

  Another ancient Greek, the Aristotelian scholar Aristoxenus, proposed a separate
  theory that has been largely forgotten. Referring to constraints of vocal
  production and perception he wrote,
  ``The voice cannot differentiate, nor can the ear discriminate, any interval smaller
  than the smallest diesis [41 cents], so as to determine what fraction it is of a
  diesis or of any other of the known interval.''\cite{aristoxenusAristoxenou1902}
  This hypothesis was independently conceived in modern times as the Interval Spacing
  theory\cite{brownMusical2025a,pfordresherVocal2017a,phillipsVocal2022a,
  phillipsHarmonicity2022,brownVocal2023}.
  Stated simply, intervals that are too small will get mixed up in communication.
  This theory is based on human imprecision in producing
  \cite{pfordresherVocal2017a,devaneyAutomatically2011a,
  cuestaChoral2019,rosenzweigDagstuhl2020,scherbaumTonal2020}
  and perceiving \cite{burja78,zaratePitchinterval2012,mcdermottMusical2010}
  intervals. Such imprecision can lead to the miscommunication of intervals,
  and so intervals must be sufficiently large to be distinguishable.

  The Motor Constraint theory is, to the best of our knowledge, a
  recent theory\cite{tierneyMotor2011,ammiranteLowSkip2015,savageGlobal2017,satoAutomatic2019},
  which states that music is constrained by vocal anatomy.
  If we assume that humans aim to be energy efficient when singing, then
  larger intervals will be used less often because they cost more energy
  to produce. This is consistent with the fact that melodies tend to follow
  scalar motion (melodies mainly use small steps), in every tradition studied to
  date\cite{mcbrideInformation2024}.

  These three theories each make a clear prediction about an aspect of scales
  (\textbf{\fref{fig:1}}), but each is insufficient on its own (\textbf{SI Section S2}).
  Harmonicity/interference models predict harmonic intervals,
  but make no predictions about limits on scale range or
  step-sizes. Interval Spacing predicts a lower limit to step-sizes, but not
  an upper limit. Motor Constraint predicts a soft upper limit to step-sizes,
  but no lower limit. To get a coherent theory of scales, we combine
  the three components in three ways: Interval Spacing and Motor Constraint
  theories affect melodic motion, so we combine these into a \textit{Melody} theory.
  Interval Spacing and harmonicity/interference theories affect harmonic
  intervals, so we combine these into a \textit{Harmony} theory. Interval Spacing
  appears in both composite theories because the discriminability of intervals
  matters whether notes are played in sequence or simultaneously. We refer to all
  three together as the \textit{Full} theory.

  The ancient theories were not quantitative enough to test directly, but their modern
  descendants are. Here, we systematically compare the major theories of scale evolution using
  a set of \num{1314} scales from \num{96} countries, encompassing vocal, instrumental,
  and music-theoretic scales. We model scale evolution as a process of convergent
  drift and selection, and infer the likelihood of each selection
  pressure by evaluating it against the full space of possible
  scales\cite{hudsonGene1991,kandlerGenerative2018}. We find that the near-universal
  structure of scales -- step intervals of roughly one to three semitones -- is
  explained by melodic constraints (the \textit{Melody} theory), whereas the bias
  towards harmonic intervals, although real, is comparatively weak and contextual,
  becoming dominant only in mathematically designed music-theoretic scales.
  We leave for future work the alternative hypothesis that scales evolve
  through social networks\cite{creanzaCultural2017,millerThailand2015}.

\section*{Results}

  \begin{figure*}[th!]
  \centering
  \includegraphics[width=\textwidth]{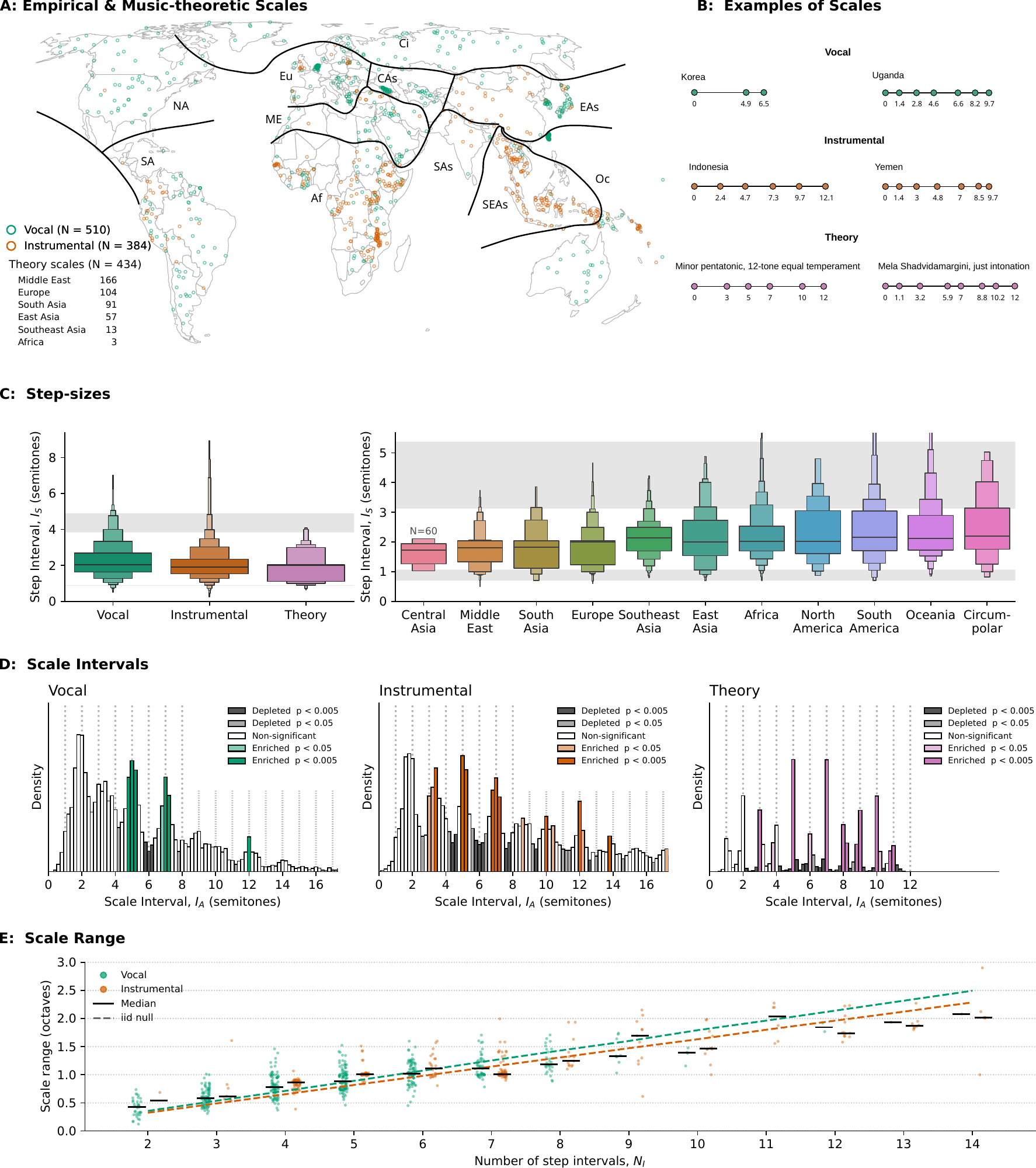}
  \caption{\textbf{Geographic and statistical structure of the scale corpus.}
  Measured scales are categorised as either Vocal or Instrumental,
  and are distinguished from music-theoretic (Theory) scales.
  \textbf{(A)} Provenance of the measured scales on an Equal~Earth projection,
  divided into \num{11} regions.
  Each scale is drawn as an open circle placed at a random point within its
  country of origin.
  Theory scales derive from broad traditions that do not map clearly to countries,
  so their per-region counts are given in the inset.
  Sample sizes $N$ are indicated for each scale type.
  \textbf{(B)} Two example scales for each type, shown as scale
  degrees above the tonic (semitones).
  \textbf{(C)} Distribution of step-sizes $I_S$, by scale type (left)
  and world region (right). Distributions are weighted by region and scale-size.
  Grey bands mark the range, taken across all categories, of
  the lower ($2.5^{\text{th}}$ percentile) and upper ($97.5^{\text{th}}$
  percentile) bounds of the step-size distribution.
  Central Asia, the most sparsely sampled region, is annotated with its number of
  step intervals $N$.
  \textbf{(D)} Probability distribution of all pairwise scale intervals $I_A$
  by scale type. Bars are coloured by whether that
  interval is significantly enriched or depleted relative to a null in which step
  intervals are resampled from the empirical pool. Dotted
  vertical lines mark integer semitones.
  \textbf{(E)} Scale range against scale-size, $N_I$,
  for Vocal and Instrumental scales (Theory scales are octave-bounded by
  construction and omitted). Points are individual scales, black ticks are the
  median range at each $N_I$, and dashed lines show the expected range under a
  null in which $N_I$ step intervals are drawn independently from the empirical
  step-size pool.}
  \label{fig:2}
  \end{figure*}

\nsb{Three types of scales.}
  We compiled a set of 1,314 scales (\textbf{\fref{fig:2}}) from two
  published datasets and one new dataset (see \textit{Scales
  Data})\cite{savageStatistical2015,mcbrideConvergent2023b,brownMusical2025a}.
  We separately analyse three types of scales
  (Vocal, Instrumental and Theory), since these are in many ways distinct, and
  have likely evolved in different ways\cite{mcbrideWhat2026}. Vocal scales are exclusively obtained
  from analysing recordings and represent the instantaneous \textit{scale of
  a performance}. Instrumental scales are primarily obtained from measuring
  how instruments are tuned, and thus can sometimes include examples that are better
  described as tuning systems -- cases where only some of the notes on the
  instrument are used for some melodies. Music-theoretic (Theory) scales are
  an entirely different object, as they are theoretical ideals which do not
  necessarily represent music as it is played. Vocal and Instrumental scales
  can be found all over the world, whereas Theory scales are mainly restricted to
  a few traditions with well-developed mathematical music theory.
  Despite this asymmetry, most of the past research and theorising on scales
  has dealt with the less common theoretical constructs, rather than measurements.
  We first describe the features of scales so that it is clear what a theory
  of scales should predict.

\nsb{Step-sizes are similar across cultures.}
  Scales can be defined by a sequence of $N_I$ scale steps
  -- where a scale step, $I_S$, is an interval between
  two adjacent scale degrees -- and the order in which the steps are arranged.
  The global step-size distribution is similar
  across scale types and geographic regions (\textbf{\fref{fig:2}C}).
  There is a clear preference for
  step-sizes of approximately 2 semitones, where the median values range from
  1.76 to 2.18 semitones across regions. There is a similar convergence in
  the lower ranges of the distributions, where the 2.5th percentile in each
  region is between 68 and 105 cents (where 1 semitone = 100 cents). The upper
  range is less consistent, with the 97.5th percentile spanning from 3 to 5
  semitones. Part of this variance in the upper range across geographic
  regions is due to sampling differences. For example, scales inferred from
  recordings typically have larger step-sizes than scales inferred from
  instrument tunings (\textbf{SI Fig.~S4}). Overall, these results suggest
  that there is a cross-cultural prevalence of step-sizes of approximately 1--3
  semitones in musical scales.

\nsb{Harmonic intervals are enriched across all scale types.}
  The same set of steps can be arranged differently to produce different \textit{scale intervals},
  $I_A$ -- the intervals between all pairs of scale degrees, not only adjacent ones.
  Theory scales are defined within a single octave, so we treat their intervals as
  octave-equivalent; Vocal and Instrumental scales often span more than an
  octave, so we measure their intervals across the full scale range (see
  \textit{Scale Data}). To test whether scales favour particular scale
  intervals, we compare the empirical distribution of scale intervals to a null
  distribution generated by resampling steps from the empirical pool.
  This approach preserves the step-size distribution but randomises how the steps are
  arranged (see \textit{Interval Significance}). Intervals that occur
  significantly more or less often than this null are enriched or depleted,
  respectively. We find that the harmonic intervals -- octaves (12 semitones),
  fifths (7 semitones) and fourths (5 semitones) -- are enriched across all
  three scale types, with the intervals immediately adjacent to them depleted
  (\textbf{\fref{fig:2}D}; region- and size-resolved counts in \textbf{SI Fig.~S5}).
  The extent of this enrichment grows from Vocal
  to Instrumental to Theory scales: in Vocal scales only the fourth, fifth and
  octave are significantly enriched; Instrumental scales additionally show
  enrichment of neutral thirds ($\sim$3.5 semitones) and minor sevenths
  (10 semitones); and in Theory scales almost every interval is significantly
  enriched or depleted. Some bias towards harmonic intervals is therefore present
  in every scale type, but is weakest in Vocal scales and strongest in the
  mathematically designed Theory scales.

\nsb{Scale range grows with size in measured scales.}
  Scales also differ in their overall range -- the interval between their
  lowest and highest degrees. Theory scales span exactly one octave by
  construction, so their step-sizes must shrink as the number of steps $N_I$
  increases. Vocal and Instrumental scales have no such constraint: their range
  instead grows with $N_I$, in such a way that step-sizes remain approximately
  constant and independent of $N_I$ (\textbf{\fref{fig:2}E}; \textbf{SI Fig.~S3}). The mean
  range of these scales is well predicted by a null in which $N_I$ steps are
  drawn independently from the empirical step-size distribution (dashed lines,
  \textbf{\fref{fig:2}E}). To a first approximation, then, the range of a
  measured scale is an emergent consequence of its step-size statistics and
  its number of steps, rather than being set by an independent limit on total
  range.

\nsb{What a theory of scale evolution must explain.}
  Together, these observations define the targets for any theory of scale
  structure. First, the near-universal concentration of step-sizes around
  1--3 semitones. Second, the enrichment of harmonic intervals above chance.
  Third, the contrasting organisation of range -- a product of step-size statistics in
  measured scales, but fixed by design in Theory scales. The consequence
  of this link between steps and range is that a theory that predicts step-sizes
  also predicts range for free. In the remainder of the paper we ask which
  combinations of Interval Spacing, Motor Constraint, and harmonicity/interference
  models can account for these features. One feature that none of the theories here
  predict directly is scale size, so we provide the size and condition predictions
  on this. See \textbf{SI Section S3} for more statistics from the dataset.

\nsb{Melody model fits step-sizes.}
  The Melody model is a composite of the Interval Spacing and Motor
  Constraint models. Both make predictions about step-sizes but not about scale
  intervals, and because they push in opposite directions (Interval Spacing
  penalises small steps, Motor Constraint penalises large steps), together they
  predict an optimal, intermediate range of step-sizes. Throughout, we describe
  each theory by the \textit{fitness} it assigns a scale: fitter scales are more
  likely to be used and passed on, and so to become common. Under the Melody
  model a scale is fit when its steps are large enough to be sung and heard
  reliably (Interval Spacing), yet small enough to produce easily (Motor
  Constraint).

  The Interval Spacing model describes the rate at which the intervals in a
  melody are miscommunicated. From signal-detection theory, the probability
  that a single interval of size $I$ is correctly distinguished from its
  nearest neighbour is $p(I) = \Phi_{0,\sigma}(I/2)$, the cumulative normal
  distribution with standard deviation $\sigma$ (in cents); this is the
  constant-variance Gaussian model\cite{greenSignal1966}, and it is the same
  psychometric function measured in interval-discrimination experiments (see
  \textbf{SI Section S4}). Larger intervals are easier to tell apart, so the
  per-interval error rate $e(I) = 1 - p(I)$ shrinks as $I$ grows. A melody is
  transmitted faithfully only if \textit{every} one of its intervals survives;
  thus, treating errors as independent, the probability that a melody of $L$ intervals
  contains no error is
  \be
  P_{\textrm{IS}}(I_S) = p(I_S)^L = \Phi_{0,\sigma}(I_S/2)^L
  \ee
  (see \textit{Interval Spacing Model} for full details).
  By fitting the model we infer two quantities: the amount of noise in
  transmission, $\sigma$, and the error rate that singers tolerate, $1/L$ --
  that is, one error per $L$ intervals.

  We lack a mechanistic account of how vocal anatomy shapes interval
  preference, so for the Motor Constraint model we posit a simple functional
  form for the likelihood that a step of size $I_S$ is used,
  \be
  P_{\textrm{MC}}(I_S) = e^{-I_S/I_0},
  \ee
  where $I_0$ is the decay constant in cents (see \textit{Motor Constraint
  Model}): larger steps are exponentially less likely, since they cost more
  energy to produce. The Melody model is the product of the two terms,
  $P_{\textrm{Melody}}(I_S) = P_{\textrm{IS}}(I_S) \times P_{\textrm{MC}}(I_S)$
  (see \textit{Melody Model}).

\begin{figure*}[th!]
  \centering
  \includegraphics[width=\textwidth]{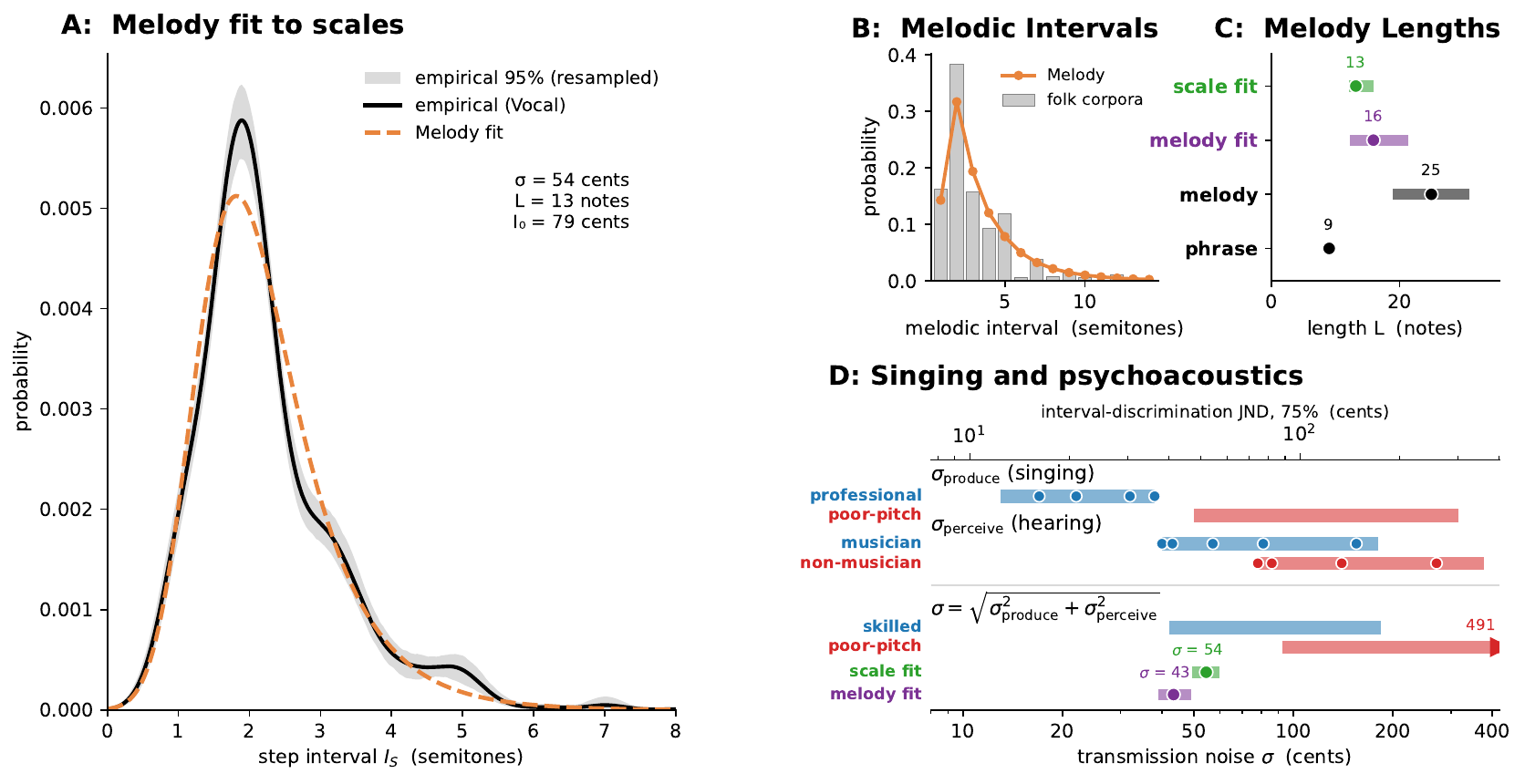}
  \caption{\textbf{Fitting the Melody model.}
  \textbf{(A)} Empirical step interval distrinution of Vocal scales
  (black, KDE with Gaussian kernel width \SI{22}{cents};
  bootstrapped 95\% CI, grey) and the distribution predicted
  by the fitted Melody model (orange); fitted parameters $\sigma$, $L$, $I_0$
  inset.
  \textbf{(B)} Empirical melodic-interval distribution (grey), averaged over \num{62}
  cross-cultural folk-music corpora. Predicted melodic-interval distribution (orange),
  a convolution over the empirical melodic-step distribution and the Melody model
  step-size distribution.
  \textbf{(C)} Melody length $L$ (notes). Each row is a confidence interval
  (bar) with its median (circle): the scale-step fit (\textit{scale-fit}, profile-L1 landscape
  interval), per-corpus melodic fits (\textit{melody-fit}, IQR across
  corpora), the empirical distribution of melody lengths (\textit{melody}, IQR of weighted
  percentiles, corpora weighted equally) and the mean Essen-corpus phrase length
  (\textit{phrase}, single value $\approx 9$)\cite{shanahanInterval2011}.
  \textbf{(D)} Transmission noise $\sigma$ (cents, log axis; upper axis: the
  equivalent interval-discrimination JND at 75\% correct). The top three rows are
  the experiment-derived ranges: per-interval production noise
  ($\sigma_{\mathrm{produce}}$, singing), perceptual discrimination noise
  ($\sigma_{\mathrm{perceive}}$, hearing), and their combination
  ($\sigma$), each split into a more-skilled blue) and a
  less-skilled (red) population. Below, the to model-fit $\sigma$
  estimates are shown as bars with median circles (\textit{scale-fit}, green,
  profile-L1 landscape interval; \textit{melody-fit}, purple, IQR
  across corpora); both sit at the lower end of the skilled transmission range.
  }
  \label{fig:3}
  \end{figure*}

  We fit the three parameters ($\sigma$, $L$, $I_0$) to the step-size
  distribution of the Vocal scales (weighted by region and scale size; see
  \textit{Weighted Sampling}). We fit to Vocal rather than Instrumental scales
  for two reasons: the two step-size distributions are nearly identical, and
  the Interval Spacing and Motor Constraint theories apply most cleanly to the
  voice, since instruments can sometimes overcome the production and perception
  limits that constrain singing. The model provides an excellent fit
  (\textbf{\fref{fig:3}A}); a few minor features are missed (\eg, the
  density at exactly 2 and 5 semitones), but the overall shape is well
  captured. The fitted values are $\sigma = 54$ cents, $L = 13$ notes, and
  $I_0 = 79$ cents.

  In plain terms, these numbers describe how carefully a melody is
  transmitted. The noise $\sigma \approx 54$ cents -- about half a semitone --
  is the typical mismatch between the interval a singer intends and the one a
  listener hears. The number of intervals $L \approx 13$ means the tolerated error rate,
  $1/L$, is roughly one mistake every thirteen notes. In other words, scales look as though they
  were shaped by musicians who accept the occasional mistake, but only about
  one per melodic phrase. This is for a single transmission event; since melodic
  phrases are typically repeated within and between performances, errors can be corrected,
  so this rate need not reflect the rate of melodic change in a tradition.
  Finally, $I_0 \approx 80$ cents sets how sharply the aversion to large
  steps increases with step-size. In the remainder of this section we
  show that each of these three values, fit here only to scale step-sizes,
  independently agrees with data of completely different kinds.

\nsb{Melody model predictions are consistent with independent data.}
  The Melody model was fit to scales, but its parameters describe how \textit{melodies}
  are produced and perceived. We test them against three independent kinds of
  data: a large collection of melodic corpora imprecision in singing, and
  psychophysical measurements of interval perception.

  We begin with melodies. We assembled $62$ cross-cultural folk melodic corpora
  (see \textit{Melody Data})\cite{mcbrideInformation2024} and, for each,
  computed the distribution of melodic intervals -- the pitch distances between
  consecutive notes. A melodic interval is not the same as a scale step: a
  melody often skips across several scale degrees at once, so a melodic interval
  spans more than one step (on average about $1.6$ steps in these corpora). The
  motor-cost parameter $I_0$ should therefore come out larger when the model is
  fit to melodic intervals than to scale step -- and it does. Fitting the
  Melody model to each corpus separately, with no reference to the scales data, gives
  an $I_0$ of \SIrange{100}{181}{cents} (interquartile range across corpora).
  The scale-based prediction falls squarely inside this range: multiplying the
  scale-step value, $I_0 = 79$ cents, by the average of about $1.6$ steps per
  melodic interval gives roughly \SI{125}{cents}. These two estimates -- one
  taken from scales, the other from melodies -- therefore agree, once the
  difference between steps and intervals is accounted for. The same agreement
  can be seen directly in the distributions: passing the scale-fitted Melody
  model through the same step-to-interval conversion reproduces the shape of the
  empirical melodic-interval distribution (\fref{fig:3}B).

  The number of intervals, $L$, has a similarly concrete interpretation as the length of a melody or phrase.
  Because $P_{\textrm{IS}}$ is the probability that $L$ consecutive intervals
  are transmitted without an error, $L$ is the span of melody over
  which singers effectively want to keep free of errors.
  The scale fit gives $L \approx 13$ notes, and fitting the model to the
  melodic corpora gives a compatible value (median $16$, interquartile range
  $12$-$21$). Both lie between the length of a single phrase -- on average $9$ notes
  in the Essen folk-song collection\cite{shanahanInterval2011} -- and that of a complete melody (median
  $25$ notes; \textbf{\fref{fig:3}C})\cite{mcbrideInformation2024}.
  The fitted $L$ value is therefore roughly a phrase: singers choose interval sizes
  in such a way that allows them to get through a phrase-length stretch of
  melody without an error.

  Finally, the noise parameter $\sigma$ can be checked against direct measurement, because
  the Interval Spacing model uses the same psychometric function as
  psychophysics. The total noise combines independent production and perception
  errors, $\sigma^2 = \sigma_{\textrm{produce}}^2 +
  \sigma_{\textrm{perceive}}^2$. We estimate $\sigma_{\textrm{perceive}}$ from
  interval-discrimination experiments, where a just-noticeable difference (JND) in
  interval size corresponds almost directly to $\sigma_{\textrm{perceive}}$ itself (\textbf{SI Section S4.2}, \textbf{Fig.~S7}).
  We estimate $\sigma_{\textrm{produce}}$ from the scatter of sung intervals in recordings,
  which is appropriate because production error feeds straight into
  transmission (\textbf{SI Section S4.1}, \textbf{Fig.~S6}). Both values span a wide range, from skilled musicians at the low-noise
  end to non-musicians and poor-pitch singers at the high-noise end. The
  scale-fit value, $\sigma = 54$ cents, and the melodic-fit value (median
  $43$ cents) both sit at the lower end of the combined range for
  musically-trained individuals (\textbf{\fref{fig:3}}).

  Taken together, the three parameters tell one simple and consistent story.
  Although it was fit only to the step-sizes of scales, the Melody model
  recovers values that match how people actually sing and hear: an interval
  error of about half a semitone ($\sigma$), a strong preference for small,
  easily-sung steps ($I_0$), and the relevant melody length consists of about
  a phrase ($L$). In everyday terms, the model describes a singer who chooses intervals large
  enough to be heard correctly, yet no larger than needed to achieve 
  an error rate of about one mistake per phrase. This of course is an average over
  many traditions, and must smooth over the variance that surely exists from
  society to society. Nevertheless, that this realistic picture of an idealised musician
  falls out of scales alone, yet agrees with independent measurements of
  singing, hearing, and the length of melodies, is strong evidence that the
  Melody model captures something real about how scales are shaped.

\nsb{Harmony model fails to predict step-sizes or scale range.}
  We have characterised only the Melody model so far, for a reason that is
  central to what follows: the Melody model predicts step-sizes directly,
  whereas the Harmony model does not. Harmonicity and interference theories
  predict \textit{which} intervals are favoured -- the harmonic ones -- but say
  nothing about how large steps should be, not about the range a scale should span.
  The Harmony model studied here is a composite of harmonicity/interference models
  and the Interval Spacing model. The addition of Interval Spacing bounds the lower end
  of likely step-sizes, but without Motor Constraint there is still no upper bound.
  This means the Harmony model requires an additional assumption about
  vocal range to limit the overall span. The harmony model on its own therefore
  does not define a complete scale: it cannot generate or score scales without
  importing constraints on step-size and range from somewhere else.

  To apply a harmonic model to Vocal or Instrumental scales, one must supply exactly the
  missing constraints that the Melody model already provides. Granting Harmony those constraints
  hands it a great deal for free: part of any apparent success would in fact be
  the arbitrary constraints doing the work. The exception is Theory scales,
  which are octave-bound by definition. There the octave fixes the range and
  hence bounds the step-sizes too, so the missing
  constraints come for free from the definition of the scale rather than from a
  borrowed melodic assumption -- which is why, as we will see, the Harmony model
  performs very well on Theory scales while needing scaffolding everywhere else.
  A fair comparison therefore requires a method for fitting and evaluating the
  Harmony model that keeps track of what it is being granted; we develop that
  method now, and only then compare the two models directly.

\nsb{Harmony models.}
  The Harmony theory rests on two psychoacoustic phenomena -- harmonicity (tonal
  fusion) and interference (roughness).
  Each Harmony model takes a scale and returns a \textit{fitness} score:
  either the harmonicity or the negative auditory roughness
  averaged over its scale intervals. The outcome is that a scale built from
  octaves, fifths and fourths is fit, while one built from less harmonic
  intervals is not. Each phenomenon each can be modelled in several ways,
  differing mainly in how many harmonics of each note they take into account.
  Rather than privilege one phenomenon or one model, we study six: three
  harmonicity and three interference models. Two serve as simplistic limiting cases.
  At one extreme, the Gill-Purves model treats every note as its full
  harmonic series, weighting all harmonics equally; at the other, the Octave-Fifth model
  keeps only the first few partials, so a scale is rewarded only for
  intervals close to an octave or a fifth. The remaining four -- one harmonicity
  model (Milne-Harrison-Pearce) and the three interference models
  (Hutchinson-Knopoff, Sethares and Berezovsky) -- lie between these extremes,
  each governed by an explicit number of partials $n$ and a roll-off $\rho$ that
  sets how quickly higher partials lose influence (the interference models also
  depend on the fundamental frequency). Varying $n$ and $\rho$ moves these models
  smoothly between the few-partial limit of the Octave-Fifth model and the many-partial limit of
  the Gill-Purves model, so the six together span a range of plausible Harmony theories rather
  than committing to one. \textbf{SI Section 5} shows how each of the six scores
  an interval, and how those scores change with $n$ and $\rho$ (\textbf{SI Fig.~S8}).

  \begin{figure*}[th!]
  \centering
  \includegraphics[width=\textwidth]{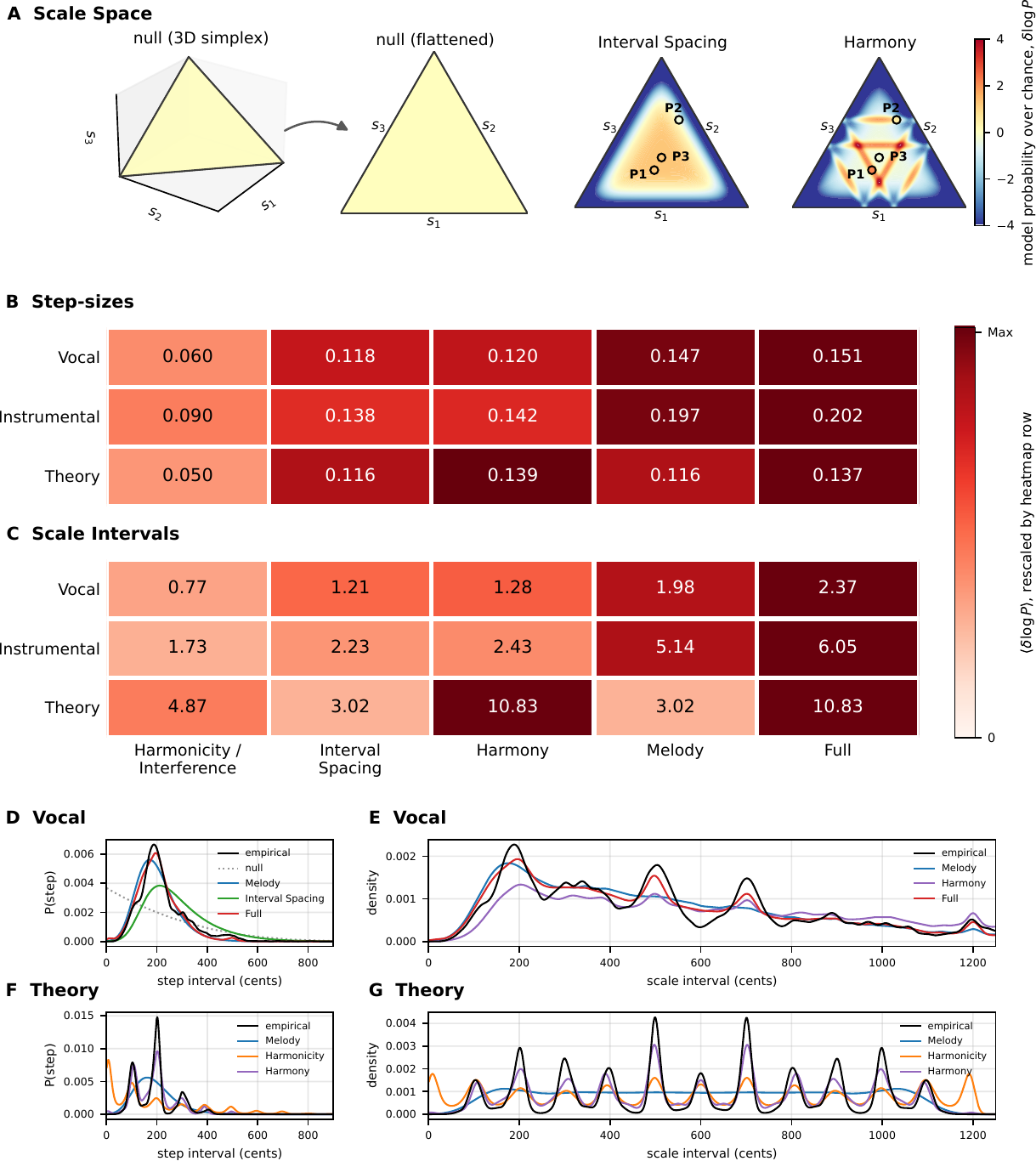}
  \caption{\textbf{Comparing models on how well they separate human scales from
  a null space of possible scales.}
  \textbf{(A)} Scale space as a simplex: for a fixed number of steps $N_I$ and
  range, every possible scale is a point on a simplex (drawn here for a
  three-step octave scale), and the null model is uniform over it. Colour shows
  the per-scale log-probability gain over chance
  $\delta \log P = \log(P_\textrm{model}/P_\textrm{chance})$
  for the Interval Spacing and Harmony models; the mean of this quantity over a
  set of real scales is the reported gain, $\langle \delta \log P \rangle$. Points P1--P3 are highlighted for
  a worked example in the text. For Theory scales, the simplex has a fixed
  octave range; for Vocal/Instrumental scales, scale space is a distribution
  of simplices, weighted by the corresponding empirical distribution of scale
  ranges.
  \textbf{(B-C)} Mean log-probability gain over chance, $\langle \delta \log P \rangle$,
  of each model against the uniform null (region-weighted), evaluated on step-sizes
  (\textbf{(B)}, per step) and scale intervals (\textbf{(C)}, per scale). Cells are
  coloured by their value relative to the Full model within each scale type; the
  printed number is the raw $\langle \delta \log P \rangle$.
  The harmonicity/interference column reports the best-performing model
  (interference, Berezovsky); differences among the harmonicity/interference models are
  mostly minor (\textbf{SI Sections S5-7}).
  \textbf{(D--G)} Illustrative model predictions: model distributions compared
  with the empirical distribution (black). Top row, Vocal scales; bottom row,
  Theory scales; left, step-sizes; right, scale intervals.
  Distributions show a weighted average over all scale-sizes $N_I$, and
  are shown as smoothed KDEs (Gaussian kernel width \SI{10}{cents}).
  Specific models are shown to illustrate points in the text: \textit{null} is
  the uniform distribution over scale space, Melody is the fitted Melody model,
  Harmony and Full models are reported for the Milne-Harrison-Pearce (harmonicity)
  model.
  }
  \label{fig:4}
  \end{figure*}

\nsb{Comparing models against chance.}
  How should we judge whether a theory explains real scales? To fit the Melody
  model we simply compared its predicted step-size distribution with the
  empirical one, which was straightforward because a step-size is a single
  number: the data lie along a line, densely enough to trace a smooth curve
  (\textbf{\fref{fig:3}A}). A whole scale is not a single number. A scale of $N_I$ steps is
  a point in an $N_I$-dimensional space -- one axis per step -- and real scales
  are scattered sparsely through it, with the space itself changing shape as
  $N_I$ changes. We cannot draw the empirical distribution of scales and fit a
  curve to it as we did for step-sizes: too many dimensions, too few scales in
  each.

  Our solution is in a way simple, yet perhaps not immediately intuitive.
  A model assigns every possible scale a
  fitness and hence a likelihood. We effectively sum the likelihoods over
  the space of all possible scales, and this sum is a normalisation constant
  that converts a likelihood to a normalised probability -- such that the probabilities
  of all scales add up to one. This allows us to quantify how probable
  any scale is given a model, and thus to compare models.
  We evaluate for each scale in a set (Vocal, Instrumental, Theory) how probable it is
  given the model, and compare it with the probability under a null model that
  treats all scales as equally likely. We report this as the model's
  \emph{log-probability gain over chance}, written $\delta \log P$:
  how much more probable a real scale is under the model than under the
  null. We will use a simple example to illustrate the approach; for
  mathematical details see \textit{Model Comparison}.

  What does this space of possible scales look like? It is easiest to picture for
  Theory scales, which span exactly one octave: such a scale is just a way of
  dividing the octave into $N_I$ steps, and the space of all such divisions is a
  \textit{simplex}, a generalisation of a triangle in $n$ dimensions.
  We can visualise this for a simple case of a 3-step scale bounded by the octave.
  The space for such a scale is a triangular plane with positive-valued step-sizes 
  whose range equals \SI{1200}{cents}, and a single
  scale $(s_1, s_2, s_3)$ is a point on that plane (\textbf{\fref{fig:4}A}).
  The null model is a uniform distribution over this triangle, under which any scale
  has a probability given by the reciprocal of the triangle's area.
  In \textbf{\fref{fig:4}A} we show this triangular plane
  first embedded in 3 dimensions, and then we show the plane in a 2d projection.
  The plane is coloured uniformly, and the colour corresponds to $\delta \log P = 0$, as
  the colour is $\delta \log P$ relative to uniform (\ie, comparing with itself).
  The remaining panels show the same triangle coloured by the per-scale $\delta \log P$
  of the Interval Spacing and Harmony models -- red where a model places more
  probability than chance, blue where less. The specific harmonicity model (Octave-Fifth)
  and its parameters ($w = 20$ cents, $\beta = 3$) are chosen purely for
  illustrative purposes. This illustration is a simplified, but accurate picture
  of how we evaluate models. In reality, there are no $N_I=3$ Theory scales in our data,
  but the methodology is the same in higher dimensions. In contrast,
  Vocal and Instrumental scales are not fixed to an octave. For them,
  we follow a similar methodology but we let the total range vary,
  drawing it from the empirically observed distribution of ranges.
  Fixing the range in this way is the necessary additional constraint
  that is required by the Harmony model; the Melody model does not need
  this, but we apply it to both for a fair comparison.

  To see how this plays out, consider three hypothetical three-step octave
  scales, marked P1--P3 on the Interval Spacing and Harmony panels of \textbf{\fref{fig:4}A}.
  Scale~P1 (steps of 500, 400 and 300 cents) contains both a perfect fourth and a
  perfect fifth and uses large steps that are easy to communicate, so both models place it above
  chance (per-scale $\delta \log P = 1.2$ under Interval Spacing, $2.1$ under Harmony).
  Scale~P2 (130, 370 and 700 cents) contains the same fourth and fifth, but its
  130-cent step is small enough that, repeated across a phrase-length melody, it
  would frequently be misheard; Interval Spacing therefore places it below chance
  ($-0.4$), while Harmony's reward for the fourth and fifth lifts it back above
  ($+0.5$). Scale~P3 (three equal 400-cent steps) is the opposite case: its steps
  are easy to place, so Interval Spacing favours it ($+1.2$), but it contains no
  harmonic intervals, and Harmony places it below chance ($-0.4$). Averaging the
  per-scale $\delta \log P$ over a set of scales gives a model's mean, $\langle \delta \log P \rangle$; across these three
  it is about $+0.7$ for each. It is the systematic averages over the full
  empirical sets that we report in \textbf{\fref{fig:4}B,C}.

  This approach solves other problems that previous approaches overlooked --
  a model should be evaluated not only on what predictions it gets right, but also
  on its incorrect predictions. Because a model's probabilities sum to
  one over the simplex, spreading predictions out over both successful and unsuccessful
  regions of scale space ultimately reduces the magnitude of the likelihood
  of successful predictions. A model scores well only by concentrating probability on
  the scales people actually use.

  Two conditions are necessary for fair comparisons: The space must be identical for
  every model in a comparison, otherwise we move the goalposts between them. The space must be finite,
  which is why the number of steps and the range are fixed. Fixing the range
  from the data is, once more, a leg-up handed to the Harmony models, which cannot
  set a scale's range themselves -- an advantage worth bearing in mind when
  reading the results. None of the models quantitatively predict $N_I$, so while
  the models are handed this for free, none are handed an advantage.

  The two models are fitted to different targets, and in different ways. The
  Melody model is fit just once, to the step-size distribution, and that single
  fitted model is then used to predict scale intervals as well. The Harmony
  models, by contrast, are fitted separately for each scale type -- choosing the
  number of partials, the roll-off and the selection strength $\beta$ that
  maximise $\langle \delta \log P \rangle$ -- because the scale-interval distributions differ far more
  between Vocal, Instrumental and Theory scales than the step-size distributions
  do. Once fitted, each Harmony model is in turn used to predict step-sizes as
  well as scale intervals, so the two theories can be compared on the same
  footing.

\nsb{Step-sizes are mainly predicted by melodic constraints.}
  We now score every model against the null (\textit{Model Comparison}). A scale
  has three features a theory might explain: its step-sizes, its scale intervals,
  and its overall range. Range is fixed when we define the scale space, so only
  the first two are scored. We begin with step-sizes (\textbf{\fref{fig:4}B}).

  Interval Spacing alone already captures most of the structure: its mean gain per
  step ($\langle \delta \log P \rangle = \numrange{0.12}{0.14}$ across scale types) is close to that of the full
  Melody and Harmony models, making it the single most important component.
  \textbf{\fref{fig:4}D} shows these step-size distributions for Vocal scales. The uniform
  null (grey, dashed) already places weight at small steps once range and $N_I$ are
  fixed -- a consequence of how we define scale space. Interval Spacing (green)
  concentrates this weight into a close match to the data, and adding the Motor
  Constraint component (to give Melody; blue) tightens the alignment further. How
  much the Motor Constraint adds depends on how free the scale range is: a little
  for Vocal ($0.118 \to 0.147$), more for Instrumental ($0.138 \to 0.197$), and
  nothing for Theory, whose single fixed range makes the Motor Constraint weight
  all scales equally. Adding harmonicity/interference to Melody (the Full model)
  increases $\langle \delta \log P \rangle$ by about \SI{3}{\%}, seen as a slight shift in the peak towards
  \SI{200}{cents} and a small bump at \SI{500}{cents} (\textbf{\fref{fig:4}D}, red).

  For Theory scales (\textbf{\fref{fig:4}F}), Interval Spacing -- and hence Melody (blue) --
  captures the overall spread of the distribution, while the Harmony models
  additionally reproduce the peaks at integer semitones ($0.116 \to 0.139$).
  Harmonicity/interference on its own is the poorest predictor of step-sizes
  (\numrange{0.05}{0.09}), and even this is achieved only by the interference
  models, which penalise small intervals; $\langle \delta \log P \rangle$ for pure harmonicity models sits
  close to the null (\textbf{SI Section 6}).

\nsb{Scale intervals tell two stories.}
  Repeating the comparison on scale intervals, where any harmonic bias should be
  most visible, tells a partly different story (\textbf{\fref{fig:4}C}). Interval Spacing
  again predicts scales well above chance for every type ($\langle \delta \log P \rangle = \numrange{1.2}{3.0}$).
  For Vocal and Instrumental scales, harmonicity/interference on its own is weaker
  than Interval Spacing (0.8 and 1.7, versus 1.2 and 2.2), and adding it to Interval
  Spacing (the Harmony model) improves prediction only marginally ($1.21 \to 1.28$
  for Vocal, $2.23 \to 2.43$ for Instrumental). Adding it instead to the Melody
  model (the Full model) raises $\langle \delta \log P \rangle$ by roughly 20\% ($1.98 \to 2.37$ for Vocal,
  $5.14 \to 6.05$ for Instrumental). The corresponding change in the Vocal
  scale-degree distribution is small: relative to Melody, the Full model only
  slightly sharpens the peaks at 200, 500, 700 and 1200 cents (\textbf{\fref{fig:4}E}, red).
  Harmony reproduces these same peaks, but, lacking the Motor Constraint component,
  it also predicts intervals that are too large (purple). The Motor Constraint
  remains the more important ingredient: it raises $\langle \delta \log P \rangle$ by about half for Vocal
  scales ($1.28 \to 1.98$) and more than doubles it for Instrumental scales
  ($2.43 \to 5.14$).

  For Theory scales the picture inverts. Harmonicity/interference on its own now
  predicts scale intervals better than Interval Spacing does ($4.9$ versus $3.0$),
  and combining the two (the Harmony model) gives a mean $\langle \delta \log P \rangle$ of 10.8 per scale --
  far above the Melody model ($3.0$). This gain is remarkably high, equivalent to
  predicting scales about \num{50000} times better than chance, and is reflected in
  the tight fit between predicted and empirical scale intervals (\textbf{\fref{fig:4}G},
  purple). For mathematically designed scales, then, harmonic bias is not a minor
  adjustment but the principal explanation of their structure -- something the
  Melody model misses almost entirely (blue). Still, the Interval Spacing
  component does a lot of work by excluding small steps, without which the
  scale interval distribution includes many unison intervals (\textbf{\fref{fig:4}G};
  at $0$ and also $1200$ cents -- the final octave is not included in the density)

  We report results from the Berezovsky interference model in \textbf{\fref{fig:4}B-C} and
  the Milne-Harrison-Pearce harmonicity model in \textbf{\fref{fig:4}D-G}; the other models
  are shown in \textbf{SI Section S6}. All six give similar results: we chose
  these two because they performed slightly better in places, but the differences
  are minor. We selected this set of six models specifically to probe the relative
  importance of different partials. The simplest model -- which scores a scale only
  by how many fifths and octaves it contains -- is very similar to
  the others (\textbf{SI Section S7}), suggesting that most of the work is done
  by the first few, most salient partials. The best-fitting parameters (the number
  of partials $n$ and the roll-off $\rho$) tell the same story (\textbf{SI Table~S3}): Vocal and
  Instrumental scales favour fewer partials and a higher roll-off, whereas Theory
  scales push to the highest $n$ and lowest $\rho$ we tested. In effect, the fitted
  models reduce to the simple Octave-Fifth model for Vocal and Instrumental
  scales, and to the all-partials Gill-Purves model for Theory scales.

\section*{Discussion}
\nsb{Aristoxenus and Pythagoras were each right, but about different scales.}
  We studied modern forms of two ancient Greek theories:
  predictions of harmonic intervals based on harmonicity and interference models (Pythagoras),
  and the Interval Spacing theory that small intervals are not reliably transmitted (Aristoxenus).
  Both theories turn out to be useful, but they fit different types
  of scales. On scales that are sung and played, Interval Spacing does most
  of the work and harmonicity/interference adds little (\textbf{Fig.~\ref{fig:4}B-E}).
  On Theory scales, harmonicity/interference beats Interval Spacing at
  predicting scale intervals, but it benefits massively from including
  Interval Spacing as it fails to adequately prevent steps from getting too
  small (\textbf{Fig.~\ref{fig:4}B,C,G}).

  When writing about scale structure in 1722, Jean-Philippe Rameau claimed that
  ``melody is only a consequence of harmony''\cite{rameauBook}.
  He meant that if you stack fifths and fold them back
  into a single octave you get the tones and semitones of the diatonic scale; the
  step-sizes then need no separate account, since they drop out of the harmonic relations.
  For music-theoretic scales, which are constructed from such harmonic relations, that is
  broadly the case.
  But scales from all parts of the globe use step-sizes of \numrange{1}{3}
  semitones whether or not they are built from fifths (\textbf{\fref{fig:2}C}),
  so harmony cannot be the reason the steps are that size. Rameau 
  noticed a coincidence and mistook it for a cause.

  Why, then, has the harmonic view been dominant for two millennia? We suspect that
  we lacked scale measurements precise enough for comparative testing.
  Aristoxenus's theory depends on quantities that only modern
  cross-cultural recordings and tuning measurements can resolve; in
  this sense he was ahead of his time.

\nsb{Discriminability is the backbone of scale structure.}
  Interval Spacing is the most important ingredient: it is needed to predict
  step-sizes, and through them scale intervals, for every scale type. The Motor
  Constraint matters too, but more narrowly -- it bounds the largest step-sizes, and
  hence the overall scale range. That bound is needed for measured scales, but not
  for the octave-bounded Theory scales, whose range is fixed in advance.

  The preference for step-sizes of \numrange{1}{3} semitones is near-universal,
  holding across regions (\textbf{Fig.~\ref{fig:2}C}), scale types
  (\textbf{Fig.~\ref{fig:2}C}), and scale sizes (\textbf{Fig.~\ref{fig:2}E}).
  Combining Interval Spacing and the Motor Constraint into the Melody theory
  reproduces this distribution (\textbf{Fig.~\ref{fig:3}A}). By the standards of the
  field, the Melody theory is unusually well supported. It is grounded in
  signal-detection theory, a well-established framework in psychology. And although
  its parameters were fit to scale step-sizes alone, each independently matches data
  of an entirely different kind -- interval-discrimination thresholds from
  psychophysics, the variability of sung intervals, melodic-interval statistics from
  \num{62} corpora, and phrase lengths (\textbf{Fig.~\ref{fig:3}B--D}).

  The theory is well-rounded but not finished. Interval Spacing makes quantitative
  predictions that can be tested directly with melody-discrimination and
  iterated-learning experiments\cite{anglada-tortLargescale2023a,verhoefMelodic2021a,popescuCore2022};
  a clean falsification would be a tradition that reliably uses, and distinguishes,
  intervals smaller than a semitone. The Motor Constraint is the weaker half: we
  posit a functional form rather than derive it from vocal mechanics, and a
  mechanistic account of how anatomy sets the cost of large intervals would
  strengthen it. These are refinements to what is, in our assessment, already the
  most complete account of scale step-structure available.

\nsb{Harmony's role is real but unresolved.}
  While discriminability is the most important component of scale structure, the
  contribution of harmony should not be overlooked. The fourth, fifth, and octave
  occur more often than the Melody theory predicts, across every scale type and
  region (\textbf{\fref{fig:2}D}, \textbf{SI Fig.~S5}), so a harmonic bias of
  some kind is present even in scales that are sung and played. Its strength,
  however, depends on context: it varies with scale type (\textbf{\fref{fig:4}C})
  and region (\textbf{SI Fig.~S14}).

  Beyond establishing that the bias exists, we cannot say how it arose. For Theory
  scales there is at least a known story -- the historical record, together with the
  old claim that intervals with small integer ratios are consonant -- but we should
  be wary of taking it at face value. Even today the origins (and even the
  definition)\cite{mcbrideMusical2025} of consonance are not clear. What we do have are
  two robust psychoacoustic phenomena, tonal fusion and auditory roughness, either of
  which can predict the harmonic content of scales. We cannot tell which is
  responsible, because fitting most of the data requires little more than a bias
  towards octaves and fifths, and every harmonicity and interference model
  predicts that much.

  Even acknowledging such a bias, how it shapes scale evolution is unresolved. At least
  three mechanisms are plausible: an innate preference for consonance, enhanced
  synchrony in group music-making\cite{savageMusic2021}, or more reliable tuning of
  instruments\cite{mcbrideConvergent2023b,rahnWas2022,hallPerception1984}. They make
  almost identical predictions for scale structure, so scale data alone cannot
  separate them. Their domains differ in principle -- consonance and synchrony act
  only on polyphonic music, sung or played, whereas reliable tuning acts mainly on
  instruments, monophonic or not -- but singing and instruments, and monophony and
  polyphony, co-occur within societies and influence one another, so context is not a
  clean discriminator either. Progress will require targeted ethnography rather than
  broad sampling, and direct experiments: do harmonic intervals measurably aid
  synchrony, or yield more reliable tuning, and can the nature of consonance finally
  be settled? Is the bias still present in cultures with no group singing, no harmony,
  or no instruments?

\nsb{A tale of two scales.}
  The gulf between measured and Theory scales points to two qualitatively different
  evolutionary processes. Measured scales evolve through stochastic drift and
  selection under the constraints of performance and instrument tuning:
  they are shaped, generation by generation, by what is easy to sing, to play,
  to hear, and to remember. Theory scales are not measured from anyone's performance;
  they are constructed, by deliberate mathematical design, to embody an ideal.
  The extraordinary performance of harmonic models on Theory scales -- predicting
  them roughly \num{50000} times better than chance (\textbf{Fig.~\ref{fig:4}C}) --
  is the signature of that design: these scales perhaps look harmonic because they were built to be.

  The strong performance of harmonic models on Theory scales therefore has to be
  read with care. These scales were originally constructed from simple integer ratios, and our
  harmonic models reward precisely such intervals -- so a high score largely recovers
  the rule the theorists built with. This circularity makes causality hard to infer:
  we cannot tell whether modern harmonic theory recovers scales that early theorists
  selected for genuine quality, or merely re-derives an integer-ratio convention that
  those theorists imposed and tradition preserved -- a distinction that a broad
  cross-cultural snapshot cannot settle, and that would need historical tracing of how
  musical knowledge spread\cite{millerThailand2015}. Design also
  helps such scales endure: once a scale is institutionalised -- written into a tuning
  theory and fixed by the technology that realises it -- its intervals can be
  reproduced exactly, as the old Greek and Chinese scales
  were\cite{barkerScience2007,chengEarly1996,bagleyProceedings2005}.

  There is also a plainer reason theorists may have settled on integer ratios: they
  are simple. Simple ratios are easy to reason about and easy to realise -- a fifth is
  a string divided 3:2 on a monochord, or two pipes cut to lengths in the same ratio
  (early pitch pipes). A scale that is easy to construct, reproduce, and teach enjoys a
  transmission advantage regardless of how it sounds. The integer-ratio
  character of Theory scales may therefore reflect the accessibility of simple
  mathematics and simple technology as much as any preference for consonance.

\nsb{Studying empirical scales, or theoretical constructs.}
  We ought to study all of the available evidence, but with an awareness
  of what type of object we are studying\cite{mcbrideWhat2026}. Most research
  on scales has concerned Theory scales, even though they are the rarest kind:
  a handful of traditions account for nearly all of them, we have likely already
  catalogued most that exist, and the great majority of cultures have
  none at all. The scales that are actually sung and played -- the very phenomenon a
  theory of scale evolution must explain -- have received far less attention.

  This has the relationship backwards. A Theory scale is an idealisation -- what
  theorists hold up as correct, or a simplified description of what is actually
  played\cite{sig77,far04}. Such scales can be generated by theorists
  using rules (\eg, Carnatic melakarta)\cite{vijayakrishnanGrammar2007}, and may
  be rarely or never played\cite{rahnStructure1981}.
  Theory scales remain valuable as cultural artefacts but they
  should not be mistaken for the thing they idealise. The more important objects
  to study are the scales as they are performed.

\nsb{What affects the number of scale degrees?}
  None of our models sets the number of scale degrees $N_I$ directly;
  it is fixed in our models, and the values are chosen based on the empirical
  scale sizes. Theory scales are an exception, as the fixed octave range
  leaves the possibility that the Melody model, which predicts step-sizes of about
  a whole tone on average, can plausibly predict $N_I$ -- a quick estimate
  would put the number of scale degrees at six on average, which is close to
  the observed 6.8 (see \textbf{SI Figs.~S1-S2} for the full scale-size
  distributions by region and type). For Vocal and Instrumental scales the range is free, so
  $N_I$ is not pinned down: the preferred step-size of roughly two semitones sets only
  an upper bound, about twelve degrees across a two-octave range. Much vocal music in
  fact spans only about ten semitones\cite{brownMusical2025a}, which would give around
  five -- consistent with the observed mean of 4.8. The number of scale degrees is
  thus consistent with the Melody theory, but prediction of the number
  of notes requires an extra constraint on (or prediction of) scale range
  that must come from elsewhere.

\nsb{Alternate hypotheses.}
  Many regularities and formal properties of scales have been proposed to
  influence scale evolution, from symmetry and structural simplicity to
  group-theoretic, information-theoretic and mathematical
  features\cite{pelofiAsymmetry2021a,harasimAxiomatic2020,balzanoPitch1982,
  veroskyHierarchizability2017a,trehubInfants1999a,mcbrideCrosscultural2020}.
  They are mainly stated as qualitative principles, or operationalised for working
  with small pitch sets rather than high-resolution measurements; they must
  be developed further for future work.

\nsb{Limitations.}
  Four limitations are worth highlighting. \textit{(i)~Data validity:} Theory and performed scales
  are different kinds of object, and we analysed them separately rather than pooling
  them; a fuller account would still need more examples of each, paired music-theoretic
  and measured versions of the same scale, and data on how performed scales drift across
  repeated performances and over time.
  \textit{(ii)~Methodological depth:} we treated scales as sequences of steps and
  neglected hierarchies. Every scale in a region is weighted equally, though some are
  far more common than others, as are the notes within a scale, despite the tonal
  hierarchies that make some pitch-classes more
  important\cite{krumhanslCognitive2001}. We did a preliminary study on one
  Vocal dataset\cite{brownMusical2025a}, where we found fifths and octaves weighted
  more heavily (\textbf{SI Fig.~S15}).
  \textit{(iii)~Data breadth:} we lack reliable tools for analysing polyphonic
  music\cite{benetosAutomatic2019}, so most of the scales that were inferred
  from recordings are from monophonic music, with some exceptions.
  This does not affect the majority of Instrumental scales, which were
  measured from instrument tunings directly and thus can be melodic or harmonic.
  This might seem to underestimate harmony, but any such concern may be overstated:
  within a society the same scales typically serve melody and harmony
  alike\cite{mcbrideWhat2026}, so a scale measured from monophonic singing is
  usually the scale that society also harmonises in;
  whether the societies also have harmonisation is not something we have examined.
  A complete picture will ultimately require measuring scales directly from
  polyphonic music, which future work should do.
  Scales inferred from recorded performances may also be incomplete, when taken from partial excerpts. 
  \textit{(iv)~Theory:} competing mechanisms often make overlapping predictions that
  scale data alone cannot separate, and we modelled only vertical transmission.
  Horizontal transmission between cultures, and the agent-based and
  geographic models needed to study it, we leave to future work.
  Theories can also take different forms: 
  the Harmony model presented here covers multiple models, yet it is possible to construct
  different models; we assumed, in line with the historical position, that maximum
  harmonicity (minimum roughness) is most fit, but this does not need to be the case.

\section*{Conclusion}

  We compared the major theories of scale evolution across 1,314 vocal,
  instrumental, and music-theoretic scales from 96 countries,
  testing each theory against the full space of scales it might have predicted.
  The most important property of a scale turns out to be the most basic:
  its intervals must be large enough to tell apart. Melody, not harmony,
  is the primary driver. The model that captures this is not some arbitrary
  curve chosen because it matches the shape of the data -- it is built on signal-detection theory,
  and its few parameters match independent data from melodies, singing,
  and psychoacoustics. Strong \textit{a priori} theoretical foundations
  and consistency across many data sources are marks of a solid theory,
  a sign that the model captures something real.

  Harmony does play a large role, but mainly for designed, music-theoretic
  scales. In fact, harmony explains these scales so well that it hints at a tautology:
  scales built from harmonic principles are explained by harmonic principles.
  The root cause may be simpler still, since elementary mathematics yields
  both the harmonic series and the most tractable music theory.
  For scales as they are performed, harmony's role is real,
  but appears to be restricted to a contextual bias towards a few harmonic intervals. 
  We also don't understand the mechanism. The historical explanation has been
  that these intervals are naturally consonant, but the nature-nurture split
  on consonance is still hotly debated, and there are other plausible explanations
  that deserve more consideration.

  Two processes thus emerge. The scales people sing and play have evolved
  under melodic constraints, converging worldwide on step-sizes of one to three
  semitones. Harmony, as operationalised here, is less a force in scale evolution than
  a signature of design, through tuning technology and music theory.

  We have missed this for so long because the field prioritised music-theoretic scales,
  which are found only in a minority of traditions, and are an abstraction from what is performed.
  Modern measurements have overturned a millennia-old narrative, and it seems likely we
  have other things about scales wrong too. It is worth a closer look.

\section*{Materials \& Methods}

\subsection*{Scale and Melody Data}

\nsb{Scale Data.}
  We define a \textit{scale}, $S = (I_1, \dots, I_{N_I})$, as a sequence of $N_I$
  \textit{steps}, $I_S$, intervals between adjacent scale degrees from low to
  high pitch. We define \textit{octave scales} as a special class of scales in
  which the pitch relations are cyclic, with a periodicity of one octave. From
  this simple representation we can derive scale degrees and all possible
  \textit{scale intervals}, $I_A$ -- the intervals between every pair of scale
  degrees. In Western music theory, scale degrees are defined as intervals
  relative to the tonic. Here we are mainly working with non-Western music
  for which we cannot unambiguously identify the tonic. As a result,
  we work instead with scale intervals, as the full set of scale degrees is
  the same no matter which note is the tonic. Scale intervals are defined
  for non-octave scales as the $N_I(N_I + 1)/2$ intervals between all pairs of
  the $N_I + 1$ scale degrees (the unison excluded). For octave scales, we take
  octave equivalence into account and include all $N_I(N_I - 1)$ sub-octave
  intervals, made by circular permutation of the $N_I$ pitch classes
  (\ie, with intervals in registers above and below the central register).

  Scales are classified here as either Vocal, Instrumental, or Theory. Theory
  scales are octave scales by definition, whereas for Vocal and Instrumental
  scales octave scales have to be inferred, and are often a subset of a larger
  scale. Instrumental scales were obtained either by measuring the pitches of
  isolated individual notes or by inferring them from a recording using
  computational methods. Vocal scales can only be obtained by inference from
  recordings, and are hence inherently ephemeral. We also group scales into 11
  geographic regions (\textbf{Fig.~\ref{fig:2}}; per-region counts by scale
  type are given in \textbf{SI Table~S2}).

  We obtained scales from three sources. The Database of Musical Scales
  (DaMuSc) contains 845 scales of all types (43 Vocal, 368 Instrumental, 434
  Theory) since it was constructed from a large range of sources spanning
  over 100 years. All of the Instrumental and Vocal scales in DaMuSc were
  obtained through physical (\eg, monochord, tuning fork), or computational
  measurements, not purely by ear. We used 409 Vocal scales that were
  previously inferred using semi-automated methods\cite{brownMusical2025a}.
  Finally, we used a subset of the Garland
  collection\cite{savageStatistical2015} (60 out of the 304 samples; 44 Vocal and 16
  Instrumental). This subset included all of the monophonic recordings as
  well as some polyphonic recordings for which a clear melody line could be
  extracted (see \textit{Scale Inference}).

\nsb{Scale Inference.}
  The scales were initially estimated by ear in Western solfège (do-re-mi)
  notation. Fundamental frequency (f0) curves were estimated using in-house
  code that allowed the use of multiple algorithms, parameter adjustment,
  and manual correction. Three algorithms (pYIN, crepe, and
  melodia)\cite{mcbrideInformation2024,mauchPYIN2014,kimCrepe2018,salamonMelody2012}
  were used to generate f0 estimates from the original audio. Algorithm
  parameters (\textit{low volume threshold}, \textit{confidence threshold},
  \textit{voicing threshold}) were adjusted to ensure that as much of the
  melody was captured as possible. The best algorithm (typically pYIN) was
  chosen based on aural evaluation -- listening to the original audio overlaid
  with a pure-tone playback of the extracted pitch. A final processing step involved deleting
  erroneous pitches (\eg, due to low-quality recording artefacts or background
  noise) and manually correcting octave errors. The python script and
  algorithm parameters used for each scale are provided in the Supplementary
  Files.

  We fitted Gaussian mixture models (GMMs) to histograms of pitch
  (log f0)\cite{scherbaumTonal2020}. Each note of a scale appears
  as a peak in the histogram, and a GMM allows us to fit overlapping
  normal distributions to each peak. We take the mean of each component
  as a scale degree. We used the manual scale estimates as
  initial guides for how many notes were in a scale, and their
  approximate pitches. In some cases, there were significant
  discrepancies between the manual scale estimate and the pitch 
  histogram, or there were other ambiguities in choosing the correct GMM fit.
  In such cases, we reassessed the number of notes, visually inspected the f0
  curves, listened to the recordings, and fitted multiple GMMs until we were
  satisfied with the result. The fits to all GMMs are available for visual
  inspection in the Supplementary Files.

\nsb{Melody Data.}
  We use a collection of 62 melodic corpora from different
  traditions\cite{mcbrideInformation2024}, which is completely separate
  from the scale data. Each corpus is a set of monophonic melodies
  notated without reference to a specific scale or tuning system. This collection
  primarily contains folk music from Europe (30) and North America (16), but also
  East Asia (6), Africa (4), and other areas (6; Türkiye, Israel, Mexico, Hawai`i,
  Kyrgyzstan). From these melodies we
  derive three quantities, defined below: the distribution of melodic interval
  sizes, the distribution of how many scale steps each melodic interval spans,
  and the distribution of melody lengths.

  A melodic interval is the pitch distance between two consecutive notes.
  The corpora do not specify an exact tuning system, so we assume equal temperament.
  For each corpus we computed the melodic interval distribution
  -- the distribution of the absolute size of these intervals, up to 14 semitones.
  A single melodic interval often skips across several scale degrees at once,
  so we also express each interval as a number of scale steps. This requires a
  scale for each melody, which the corpora do not provide. We inferred one scale
  per melody by assuming octave equivalence: we reduced the melody's notes to pitch
  classes and took its scale to be the set of pitch classes used. Each melody's
  inferred scale then lets us re-express its melodic intervals as numbers of scale
  steps, giving, for each corpus, the melodic step distribution.
  The corpora do not record how each melody was tuned, so we treat all pitches as belonging to the
  12-tone equal-tempered scale; this is an approximation, akin to rounding up or down
  to the nearest semitone, but the resulting errors should largely average out over so many melodies.

  Finally, we measured the length of each melody, counted in notes. Corpora
  differ in how they treat repetition -- some notate a passage together with its
  repeats (often slightly varied), others give the principal melody only once --
  so a raw note count is not comparable across corpora. We therefore removed
  repetition before counting, to approximate the length of the main melodic
  passage. See \citet{mcbrideInformation2024} for the full details.

\nsb{Interval Significance.}
  Some scale intervals ($I_A$, the interval between any two notes of a scale)
  are expected to be more common than others even by chance, so simply reading
  off the peaks of the empirical $I_A$ distribution is not a reliable way to
  tell which intervals occur significantly more or less often than expected. To
  test this, we compare the empirical interval counts against a null model in
  which scales carry no structure beyond their step sizes, extending the
  methodology of Ref.\cite{mcbrideConvergent2023b} to correct for the uneven
  geographic sampling of our corpus.

  To account for unbalanced geographical distribution of scales we use
  the typical region-weighting that we use throughout this work
  (see \textit{Weighted Sampling}). When building the null distribution
  we draw a subsample of 20 scales without replacement from each region
  (or all available scales, for regions with fewer than 20).
  Because any single such subsample is noisy, we repeat the entire procedure
  over 1000 independent subsamples and combine the results as described
  below; the earlier reference did not apply this correction.
  Likewise, when evaluating likelihood, we apply the same bootstrap method
  to obtain a balanced, precise average probability per interval.

  To build the null distribution, we assume the empirical distribution of step
  sizes does not depend on the number of steps $N_I$, and replace each sampled
  scale's steps with the same number of steps drawn with replacement from the
  pooled step sizes of the balanced subsample, $P_E(I_S)$. Histogramming the
  resulting scale intervals in 20-cent bins and averaging over the 1000
  subsamples gives a converged null distribution $P(I_A)$; we write $p_i$ for
  the null probability that a scale interval falls in bin $i$.

  Within each balanced subsample, the number of empirical intervals in bin $i$,
  $k_i$, is a binomial random variable with $K = \sum_i k_i$ total intervals and
  per-bin success probability $p_i$, so the probability of observing
  \emph{exactly} $k_i$ intervals there is
  \be
  q_i = \binom{K}{k_i} p_i^{k_i} (1 - p_i)^{K-k_i}.
  \ee
  We take the bin's significance within that subsample to be the one-sided tail
  probability, summing $q$ over all outcomes at least as extreme as $k_i$: the
  probability of $k_i$ or more intervals when the bin is over-represented
  ($k_i / K > p_i$), or $k_i$ or fewer when it is under-represented
  ($k_i / K < p_i$). We combine these tail probabilities across the 1000
  subsamples by their geometric mean to obtain one $p$ value per bin, and
  finally adjust for testing every bin with the Benjamini-Hochberg
  procedure\cite{benjaminiControlling1995}.

\nsb{Scale Range Statistics.}
  For Vocal and Instrumental scales we resampled steps from their
  region-weighted (see \textit{Weighted Sampling}) distributions
  to obtain a null distribution for the scale range. For each scale size $N_I$
  we drew $N_I$ steps independently, with replacement, from the empirical
  step-size pool and summed them into a range; repeating this many times, the
  mean of the resulting sums is the expected range plotted as the dashed lines
  in \textbf{\fref{fig:2}E}. Theory scales are octave-bound by construction and are
  omitted.

\subsection*{Models}

\nsb{Interval Spacing Model.}
  The Interval Spacing (IS) theory posits that intervals must be large enough to
  avoid being confused, given the limits on how precisely intervals are produced
  -- by the voice, or other non-fixed-pitch instruments -- and perceived. Two
  intervals close in size are easily mistaken for one another, whereas larger
  intervals are easier to tell apart.
  The Interval Spacing theory applies to all scale types, since limits on interval
  perception are common to all. Instrumental scales may tolerate a lower bound on
  interval size, since production variance can in principle be reduced to zero,
  but this should depend on the instrument and would need a separate study.

  We make this precise with signal detection theory\cite{greenSignal1966}. We
  treat a note as a noisy signal, which can be confused with other notes based
  on how close they are. Let's say that two notes are separated by an interval
  of size $I$, and a note is communicated with a normally-distributed error of
  standard deviation $\sigma$ (in cents). The probability that the note is
  correctly distinguished from its nearest neighbour is then
  $p(I) = \Phi_{0,\sigma}(I/2)$, the cumulative normal distribution evaluated
  at $I/2$. This is the constant-variance Gaussian model -- the same psychometric
  function measured in interval-discrimination experiments (\textbf{SI Section
  S4}). Larger intervals are easier to place, so the per-interval error rate
  $e(I) = 1 - p(I)$ shrinks as $I$ grows.

  The noise $\sigma$ combines independent perception and production errors;
  treating each as normal, their variances add,
  $\sigma^2 = \sigma_{\textrm{perceive}}^2 + \sigma_{\textrm{produce}}^2$. From
  psychophysics experiments we infer $40 \leq \sigma_{\textrm{perceive}} \leq 377$
  cents (\textbf{SI Section S4.2}); from audio recordings of singing we infer
  $13 \leq \sigma_{\textrm{produce}} \leq 315$ cents for vocal music
  (\textbf{SI Section S4.1}). In both cases the low end corresponds to trained
  musicians and the high end to non-musicians. The combined noise therefore spans
  $42 \leq \sigma \leq 491$ cents.
  These estimates come from sequential tones; for simultaneous intervals the
  relevant quantity would be the discriminability of dyads. We nonetheless
  use a single $\sigma$ for all models (see \textit{Melody Model}).

  A melody is transmitted faithfully only if \textit{every} one of its intervals
  survives. Treating errors as independent, the probability that a melody of $L$
  intervals contains no error is
  \be
  P_{\textrm{IS}}(I) = p(I)^L = \Phi_{0,\sigma}(I/2)^L ~.
  \ee
  Fitting the model infers two quantities: the transmission noise $\sigma$, and
  the error rate $1/L$ that singers tolerate -- one error per $L$ intervals.
  This error rate is not something that we can construct an \textit{a priori} model
  for. Thus, we simply infer $L$ from the scale data, and compare
  it to data on melodic lengths, to see if we arrive at a plausible answer.
  We then also infer $L$ by fitting the parameters to independent melodic data;
  the parameter ranges obtained using different data overlap quite well.

  In line with the study of scales as a cultural-evolutionary process, we can
  define the fitness as the log-probability of faithfully transmitting a single
  interval, $F = \log p(I)$. A scale whose intervals have a high log-probability
  of being correctly communicated is fit. This definition in turn gives us the probability of an
  interval being accepted in a scale according to the interval-spacing model,
  \be
  P_{\textrm{IS}}(I) = \exp LF ~,
  \ee
  where now it is clear that $L$ is equivalent to selection strength. A singer
  who is more stringent about accuracy will aim for a lower rate (larger $L$)
  of errors. We model the entire set of scales using one set of parameters,
  but in reality traditions may differ. They may differ in terms of how
  accurately they sing or perceive intervals, and they may differ in terms
  of how many errors they tolerate. We do not have such fine-grained data
  to test this with the current dataset, but this model is sufficiently specified
  that one can investigate cultural differences in this way.

\nsb{Motor Constraint Model.}
  The Motor Constraint (MC) theory posits that some musical features are
  universal due to constraints imposed by the biology of the vocal apparatus.
  For scales, the prediction is that small intervals are easier to produce than
  large ones. The $f_0$ produced by the voice depends on the subglottal pressure
  generated by the lungs and on the length and tension of the vocal folds, so
  pitch is modulated by relaxing and contracting the muscles in the diaphragm,
  abdomen, chest, and larynx\cite{sundbergPhonatory1993}. Small changes in pitch
  require little energy, larger changes require more. This theory applies
  primarily to vocal scales, though similar reasoning extends to instrumental
  production.

  We lack a mechanistic account relating energy use to interval preference, so
  rather than derive a model \textit{a priori} we assume a simple functional form
  for the likelihood that a step of size $I_S$ is used,
  \be
  P_{\textrm{MC}}(I_S) = e^{-I_S/I_0},
  \ee
  where $I_0$ is a decay constant (in cents). Given this form, larger steps are
  exponentially less likely. We choose the exponential for simplicity;
  a power law would be an alternative.

  Although the model scores one step at a time, its effect on a whole scale is
  simple. Treating a scale's steps as independent (see \textit{Melody Model}),
  the Motor Constraint probability of a scale is the product of its per-step
  terms, $\prod_{I_S \in S} e^{-I_S/I_0} = e^{-\sum_{I_S} I_S/I_0} = e^{-R/I_0}$,
  where $R = \sum_{I_S \in S} I_S$ is the scale's total range. Equivalently, the
  Motor Constraint fitness is $F_{\textrm{MC}}(S) = -\sum_{I_S \in S} I_S/I_0 =
  -R/I_0$: the cost of a scale depends only on how far it reaches from its lowest
  to its highest note, not on how that span is divided into steps. A scale that
  spans a smaller range, built from cheaper (smaller) steps, is therefore fitter.

  This has an important consequence. Because the cost depends only on the range,
  Motor Constraint cannot distinguish between scales of equal range -- it scores
  them all alike. When only used to predict step-sizes (or a 1-step scale) the Motor Constraint
  acts directly on step-size; but for scales with more steps it
  constrains the individual steps only indirectly, by limiting the range they sum
  to. This means that when the range is fixed in advance -- as it is for
  octave-bound Theory scales -- Motor Constraint contributes nothing, since every
  scale then has the same range and so the same cost.

  As with Interval Spacing, this has a natural cultural-evolutionary reading. The
  Motor Constraint fitness of a single step is $-I_S/I_0$, in which the rate
  $1/I_0$ acts as a selection strength -- how strongly large, energetically
  costly steps are penalised -- just as $L$ does for transmission errors under
  Interval Spacing. We cannot say in advance how much singers value conserving
  energy, so we infer $1/I_0$ from the scale data, fitting $I_0$ alongside the
  Interval Spacing parameters (see \textit{Melody Model}); fitting instead to
  independent melodic data gives a consistent value.

\nsb{Melody Model.}
  We combine the Interval Spacing and Motor Constraint models to form the Melody
  model, $P_{\textrm{Melody}}(I_S) = P_{\textrm{IS}}(I_S) \times
  P_{\textrm{MC}}(I_S)$: a step is favoured under the Melody model only if it is
  both reliably distinguishable (Interval Spacing) and cheap to produce (Motor
  Constraint). Because a fitness is a log-probability, and the logarithm of a
  product is a sum, multiplying the two probabilities is the same as adding the
  two fitnesses; the Melody fitness of a scale is therefore
  $F_{\textrm{Melody}}(S) = F_{\textrm{IS}}(S) + F_{\textrm{MC}}(S)$.

  We fit all three parameters ($\sigma$, $L$, $I_0$) to the
  step-size distribution of the Vocal scales, weighted by region and scale size
  (see \textit{Weighted Sampling}), obtaining $\sigma = 54$ cents, $L = 13$, and
  $I_0 = 79$ cents. 
  The fitted Melody model defines a probability distribution over step-sizes,
  $P_{\textrm{Melody}}(I_S)$. We fit the model once, and use it throughout the work. 
  This includes the Harmony and Full models, which inherit $\sigma$ and $L$ from
  this fit. We did not refit the parameters for Instrumental or Theory scales, nor to
  optimise the fit to scale intervals. Refitting $\sigma$ and $L$ within the
  Harmony model is prohibitively costly, and would largely fit them to the
  fixed scale range (see \textit{Scale space}), rather than to anything harmonic.

  We fit to the Vocal step-size distribution for the reasons given in the
  Results, weighting it by both region and scale size (see \textit{Weighted
  Sampling}): region-weighting corrects for uneven sampling across the world, and
  scale-size weighting balances the contributions of small and large scales. The
  three parameters are chosen to minimise the $L_1$ distance (the summed absolute
  difference) between the model and the (region- and size-weighted) empirical
  step-size distribution, which gives a close fit
  (\textbf{\fref{fig:3}A}). The fit changes little under other
  reasonable choices of which scales to include and how to weight them
  (\textbf{SI Fig.~S17}), and each parameter is well-constrained by the
  step-size data (\textbf{SI Section S10}, \textbf{Fig.~S16}).

  We quantify the uncertainty of this fit in two complementary ways, both shown
  in \textbf{\fref{fig:3}}. First, we show estimate \SI{95}{\%} confidence intervals on the empirical
  distribution (grey band, \textbf{\fref{fig:3}A}) by drawing \num{300} bootstrap samples
  of scales, recomputing the KDE with fixed bandwidth \SI{22}{cents}.

  Second, we estimate \SI{95}{\%} confidence intervals on each of the three
  fitted parameters ($\sigma$, $L$, $I_0$). For each parameter in turn we fix it at
  a grid of values, re-optimise the other two against the fixed full-sample
  empirical distribution, and record how far the $L_1$ loss rises above its
  minimum -- the profile-$L_1$ curve, whose width measures how flat the fit
  landscape is around the optimum. The threshold for the interval is set from
  the same cluster bootstrap: for each of \num{1000} resamples we evaluate the
  full-sample profile-$L_1$ curve at that resample's own optimum, and take the
  $95^{\text{th}}$ percentile of these loss increments as the cutoff. The
  confidence interval is the range of parameter values whose profile-$L_1$ loss
  stays below the cutoff, giving $\sigma \in [49, 60]$ cents, $L \in [12, 16]$,
  and $I_0 \in [70, 90]$ cents (\textbf{SI Section S10}).

  As a further check, we also fit the Melody model directly to the
  melodic-interval data, over intervals of \SIrange{1}{14}{semitones} and by the same
  $L_1$ procedure. Rather than pooling, we fit each of the \num{62} folk-music
  corpora separately (excluding the unison, $I=0$). This yields \num{62}
  independent estimates of each parameter; the spread across corpora, summarised
  by the interquartile range (IQR) with the median marked, gives the
  \textit{melody-fit} intervals in \textbf{\fref{fig:3}C,~D}: $\sigma$ has median
  \SI{43}{cents} (IQR \SIrange{39}{49}{cents}), $L$ median \num{16} (IQR
  \numrange{12}{21}), and $I_0$ median \SI{120}{cents} (IQR
  \SIrange{100}{181}{cents}).

\nsb{Harmony Models.}
  The Harmony theory can be realised by two families of model -- harmonicity and
  interference -- which differ mainly in how many partials of each note they take
  into account. We describe the harmonicity models here and the interference
  models below; the two families turn out to make very similar predictions, the
  best-fitting model of each differing little from the other
  (\textbf{SI Figs.~S9-S11}). We examine three harmonicity models.
  Two are limiting cases: the Gill-Purves model uses
  a note's full series of partials, whereas the Octave-Fifth model keeps only
  the first few. The third, the Milne-Harrison-Pearce model, lies between these extremes.

  The Gill-Purves model treats every note as its full harmonic series and
  weights all partials equally\cite{gillBiological2009}. It scores an interval by
  how much the harmonic series of its two notes overlap: the closer the interval
  is to a simple whole-number frequency ratio $x/y$ (for example $2/1$ for the
  octave or $3/2$ for the fifth), the more partials coincide,
  \be
  H_{\textrm{GP}}(I) = \frac{x + y - 1}{xy}.
  \ee
  Because this overlap changes sharply with tiny departures from an exact ratio,
  we allow a tolerance $w$: an interval is given the highest score of any interval
  within $w$ cents of it, written $H'_{\textrm{GP}}(I, w)$.

  The Octave-Fifth model is the opposite limiting case: it ignores the higher
  partials and rewards an interval only for lying close to an octave or a fifth,
  \be
  H_{\textrm{OF}}(I, w) = \mathcal{N}(I - 1200, w^2) + \mathcal{N}(I - 702, w^2),
  \ee
  where $\mathcal{N}$ is a Gaussian kernel of width $w$, and $1200$ and $702$
  cents are the octave and the fifth. As before, $w$ sets how far an interval may
  stray and still score well.

  The Milne-Harrison-Pearce model\cite{harrisonEnergybased2018,milneComputational2013}
  lies between these extremes by making the number of partials explicit.
  It has two parameters: the number of partials $n$ that each note carries,
  and a roll-off $\rho$ that sets how quickly higher partials lose loudness
  (larger $\rho$ down-weights the higher partials, and has little effect
  when $n$ is small). Rolloff affects the amplitudes of partials
  $a_i = 10^{-\rho \log_2 i / 20}$ (a decline of $\rho$ decibels per octave of
  partial number). We deviate from \citet{harrisonEnergybased2018}
  by allowing the harmonic template to depend on $n$ and $\rho$,
  since our purpose for using this model is to assess the degree to
  which partials are important for scale evolution. This gives a
  score $H_{\textrm{MHP}}(I, n, \rho)$.

  Each model scores a single interval; to score a whole scale we average over its
  intervals. For a scale $S$ and model
  $\textrm{m} \in \{\textrm{GP}, \textrm{OF}, \textrm{MHP}\}$,
  \be
  \langle H_\textrm{m}(S) \rangle = \frac{1}{N_S} \sum_{I \in \mathcal{I}} H_\textrm{m}(I),
  \ee
  where $\mathcal{I}$ is the set of the scale's intervals and $N_S$ its size. We
  use the scale intervals defined earlier (see \textit{Scale Data}): the
  $N_I(N_I + 1)/2$ intervals between all degrees for Vocal and Instrumental
  scales, and, for octave-bound Theory scales, the $N_I(N_I - 1)$ sub-octave
  intervals obtained by circular permutation (the octave itself is fixed and
  excluded). We discard intervals larger than $12.5$ semitones, for which these
  models were not designed -- in practice very few. Finally, we standardise each
  model's scores -- subtracting the minimum and dividing by the standard deviation
  over $0 \leq I \leq 1250$ cents -- so that the models are placed on a common
  scale. The harmonicity score is then the model's \textit{fitness} directly, a
  higher score meaning a fitter scale,
  \be
  F_\textrm{m}(S) = \langle H_\textrm{N}(S) \rangle.
  \ee

  We searched the parameter grids $w \in \{2, 4, \dots, 40\}$ cents,
  $n \in \{3, 4, \dots, 40\}$, and $\rho \in \{0, 1, 2, \dots, 10, 12, \dots, 20\}$.
  As a check, we also computed fitness using only a scale's degrees (its intervals
  from the lowest note) in place of all pairwise scale intervals. The results are
  robust to these parameter choices (\textbf{SI Figs.~S18-S19}). The scoring
  profiles of all six models, and how they vary with these parameters, are shown
  in \textbf{SI Section S5}.

\nsb{Interference Models.}
  All three interference models estimate the auditory roughness of an interval
  from the interactions between the partials (the fundamental and partials) of its two
  notes. They differ in three ingredients: the roughness contributed by a single
  pair of partials, the critical bandwidth $w_c$ -- the frequency separation
  within which two partials interact audibly -- and how strongly each partial counts,
  set by its amplitude $a$. We use the
  Hutchinson-Knopoff model\cite{hutchinsonAcoustic1978}, together with the models of
  Sethares\cite{setharesLocal1993} and Berezovsky\cite{berezovskyStructure2019}.

  In the Hutchinson-Knopoff model the critical bandwidth is
  \be
  w_c^{\textrm{HK}} = 1.72 \left( \frac{f_1 + f_2}{2} \right)^{0.65},
  \ee
  and the roughness between two partials is a function of their frequency gap
  measured in units of this bandwidth, $u_{ij} = |f_i - f_j| / w_c^{\textrm{HK}}$,
  \be
  d_{ij}^{\textrm{HK}} =
  \begin{cases}
    \left(4\,u_{ij}\, e^{\,1 - 4 u_{ij}}\right)^2, & u_{ij} \leq 1.2,\\[2pt]
    0, & u_{ij} > 1.2,
  \end{cases}
  \ee
  which rises from zero, peaks near a quarter of the critical bandwidth, and is
  set to zero once the partials are separated by more than $1.2$ bandwidths. The
  roughness of two complex tones is the amplitude-weighted mean of this quantity
  over all pairs of their $2n$ partials,
  \be
  D_{\textrm{HK}} = \frac{\sum_{i<j} a_i a_j\, d_{ij}^{\textrm{HK}}}{\sum_{i<j} a_i a_j},
  \ee
  where $i$ and $j$ run over the $2n$ partials of the two tones.

  In the Sethares model the roughness due to the interaction between two partials is
  \be
  d_{ij}^{\textrm{SE}} = e^{-3.5\, s\, |f_i - f_j|} - e^{-5.75\, s\, |f_i - f_j|},
  \ee
  where $s$ plays the role of the critical bandwidth,
  \be
  s = \frac{0.24}{0.021 \min\{f_i, f_j\} + 19}.
  \ee
  The roughness of two complex tones is the amplitude-weighted mean over all
  pairs of their $2n$ partials, each pair weighted by its quieter partial,
  \be
  D_{\textrm{SE}} = \frac{\sum_{i<j} \min\{a_i, a_j\}\, d_{ij}^{\textrm{SE}}}
                         {\sum_{i<j} \min\{a_i, a_j\}}.
  \ee

  In the Berezovsky model the critical bandwidth is
  \be
  w_c^{\textrm{BE}} = 5.5 \min\{f_i, f_j\}^{-0.68},
  \ee
  and the roughness from two interacting partials is
  \be
  d_{ij}^{\textrm{BE}} = \frac{1}{w_c^{\textrm{BE}}}
  \exp\!\left( -\left( \log \frac{|\log_2 f_i / f_j|}{w_c^{\textrm{BE}}} \right)^2 \right).
  \ee
  The roughness of two complex tones is the weighted mean over all pairs of
  their $2n$ partials, each pair weighted by the amplitude of its lower-indexed
  partial,
  \be
  D_{\textrm{BE}} = \frac{\sum_{i<j} a_i^{0.606}\, d_{ij}^{\textrm{BE}}}
                         {\sum_{i<j} a_i^{0.606}}.
  \ee

  Like we do for the harmonicity models, we score a whole scale by averaging the
  pair-of-tones roughness over its intervals. For a scale $S$ and model
  $\textrm{m} \in \{\textrm{HK}, \textrm{SE}, \textrm{BE}\}$,
  \be
  \langle D_\textrm{m}(S) \rangle = \frac{1}{m} \sum_{I \in \mathcal{I}} D_\textrm{m}(I),
  \ee
  where $\mathcal{I}$ is the set of the scale's intervals and $m$ its size,
  defined exactly as for the harmonicity models (see \textit{Harmony Models} and
  \textit{Scale Data}); we likewise discard the few intervals larger than $12.5$
  semitones and standardise each model's scores over $0 \leq I \leq 1250$ cents so
  that the models share a common scale. A more dissonant scale is considered
  less fit, so we take fitness to be the negative average roughness,
  \be
  F_\textrm{m}(S) = -\langle D_\textrm{m}(S) \rangle .
  \ee

  We searched the number of partials, roll-off, and fundamental-frequency grids
  $n \in \{3, 4, \dots, 40\}$, $\rho \in \{0, 1, 2, \dots, 10, 12, \dots, 20\}$,
  and \\
  $f_0 \in \{50, 100, 200, 500, 1000\}$~Hz.
  The roll-off sets the amplitude of the $i$-th partial as
  $a_i = 10^{-\rho \log_2 i / 20}$.

\subsection*{Model Comparison}
  Several theories propose different forces that shape musical scales 
  through cultural evolution. We cannot observe this process, but we have
  its product -- a snapshot of the scales in use across cultures
  over roughly the last century. Our job is to infer, from that snapshot,
  how strongly each proposed force has shaped scale structure,
  and to do so in a way that lets very
  different theories be compared on equal terms. The natural way to pose
  the question is to ask whether real scales are more probable under
  a given theory than they would be by chance. The catch is that a theory does
  not hand us probabilities -- it only scores each scale by how well the scale
  satisfies the theory -- so before we can compare anything, we must turn those
  scores into probabilities and say what ``by chance'' means.

  This is harder than it sounds, for several connected reasons. First, our data
  contain only scales that people actually use -- successes, never failures. To
  judge whether a theory singles out real scales in particular, we need
  something to compare them against: a defined space of \emph{possible} scales
  that includes scales that humans do and do not use. Second, this
  space must be finite, and small enough to explore by computer; yet any
  space large enough to contain all real scales is also enormous, and the scales
  are sparsely populated therein.
  Third, the size of this space depends on how many notes a scale has: a scale
  with more steps occupies a higher-dimensional space, so scales with different
  numbers of notes effectively live in different spaces and must be treated
  separately. That separation also fixes a subtler problem. The theories make no
  clear predictions about how many notes a scale should contain, and biases towards
  large or small scales can easily creep in through how scales are rated: for example,
  rating a scale by its average harmonicity rewards the smallest possible scales
  (a single octave is perfectly harmonious), while rating it by total harmonicity
  rewards the largest. By comparing scales only against others of the same size,
  $N_I$, we avoid crediting or penalising a theory for a preference it never expressed.
  Fourth, the choice of the scale space itself can bias the results, so it
  must be chosen carefully.
  Fifth, the theories are incomplete: each predicts only certain components of a
  scale, so we need a principled way to combine them and to evaluate them on
  comparable terms.

  Our approach combines four ideas. First, we adopt a standard probabilistic
  framework from evolutionary modelling: each theory is written as a fitness
  function that scores a scale, together with a selection strength that sets how
  sharply differences in fitness translate into differences in prevalence.
  Casting every theory in this single form lets us treat them alike and combine
  them -- Interval Spacing, Motor Constraint, harmonicity/interference
  and their composites -- without changing the machinery
  (see \textit{Evolutionary Model}). Second, we evaluate every
  model over one shared scale space that is finite, computationally tractable,
  and covers the full spread of real scales while adding as little bias as
  possible (see \textit{Scale space}).
  Third, we reduce each model's performance to a single number -- 
  its log-probability gain over chance, a log-likelihood ratio against a null model -- that draws on all of the
  available information, is relatively easy to interpret, and places every model
  on one common axis of goodness-of-fit. Fourth, we fit each
  model on the feature it naturally predicts: the Melody model on step-sizes, and
  the harmonicity/interference components on scale intervals, fitting only
  \numrange{3}{4} parameters at a time. Because the models are of comparable
  complexity and the dataset is large, comparing them reduces to comparing
  log-likelihoods directly -- complexity penalties such as AIC or BIC would give
  essentially the same ranking.

\nsb{Scale space.}
  How can we bound the space of possible scales? There are two natural ways:
  limit the size of the steps, or limit the total range a scale spans.
  The naive choice, to accept any scale whose steps or range fall below the
  largest values we observe (for steps, about \SI{900}{cents}, for scale intervals
  about \SI{4.5}{octaves}), would fill the space overwhelmingly with scales unlike
  anything humans use, making it both a poor reference and a prohibitively
  expensive one to sample. A more careful option is to let the Melody model
  define the space, since it already predicts a finite spread of plausible
  step-sizes; we used this approach in earlier work and retain it in the
  supplement as a consistency check (\textbf{SI Section S12}), where it gives the
  same qualitative ranking. But this approach carries a fatal flaw for our
  purposes: it builds the melodic step distribution into the space itself, so the
  harmony and interference theories can no longer be tested on their own, free of
  any melodic assumption. The same objection rules out gentler variants, such as
  drawing step-sizes from a bell-shaped distribution -- anything that shapes the
  steps in advance includes a melody-like assumption into a reference that
  is meant to be neutral.

  We therefore adopt a simple principle -- the reference should treat all possible scales
  as equally likely -- and constrain not the steps but the overall range. For
  Theory scales this is the natural choice, since they span exactly one octave by
  definition. For Vocal and Instrumental scales there is no such fixed range, so
  we take the next best option and bound them by the range of scales actually
  observed in the data. This keeps the space concentrated where human scales
  really lie, without otherwise prejudging their shape. This does, however,
  give a selective advantage to certain models -- the scale range would otherwise
  be predicted by the Motor Constraint model, so fixing the range diminishes
  the importance of this model (see \textit{Motor Constraint Model}).
  For Theory scales, fixing the range to a single point reduces the Motor
  Constraint contribution to zero by definition. For Vocal/Instrumental scales,
  since there are multiple values for scale range, the Motor Constraint model
  does still make a contribution; the Melody model by itself would otherwise predict
  the scale range without this constraint. This is the better compromise,
  as the alternative (fixing step-sizes) would be diminishing the contribution
  of \textit{both} the Interval Spacing and Motor Constraint models.

  We can now define scale space precisely. Writing a scale as its list of step intervals,
  $S = (I_1, \dots, I_{N_I})$, every step is non-negative and the steps sum to the
  range, $\sum_i I_i = R$. For a fixed range, the set of all such scales is the
  $(N_I - 1)$-dimensional simplex
  \be
  \Delta_R = \Big\{\, S \in \mathbb{R}^{N_I}_{\geq 0} : \textstyle\sum_i I_i = R \,\Big\},
  \ee
  a triangle for $N_I = 3$, a tetrahedron for $N_I = 4$, and so on in higher
  dimensions (\textbf{\fref{fig:4}A}). Our null model is the uniform distribution over this
  simplex: every way of dividing the range into $N_I$ ordered steps is equally
  likely. We draw uniform samples from $\Delta_R$ by the standard construction --
  placing $N_I - 1$ cut-points independently and uniformly along $[0, R]$, then
  taking the gaps between successive cut-points (and the two endpoints) as the
  steps. The simplex has volume $V(R, N_I) = R^{N_I - 1}/(N_I - 1)!$, but this
  never enters the model comparison itself, where it cancels
  (\textit{Evolutionary Model}).

  For Theory scales the range is fixed at one octave, $R = \SI{1200}{cents}$, so
  the scale space is a single simplex, $\Omega = \Delta_{1200}$. Vocal and
  Instrumental scales have no fixed range, so a single simplex will not do.
  Instead we use a whole family of them -- one simplex for each possible range,
  weighted by how often that range occurs among real scales of the same size. The
  null distribution is then a mixture of simplices,
  \be
  P_{\textrm{null}}(S) = \int p(R \mid N_I)\, u_{\Delta_R}(S)\, dR,
  \ee
  where $u_{\Delta_R}$ is the uniform density on $\Delta_R$ and $p(R \mid N_I)$ is
  the empirical, region-weighted distribution of ranges among real scales with
  $N_I$ steps; equivalently, a scale whose steps sum to $R$ has null density
  $p(R \mid N_I) / V(R, N_I)$. We sample from this space in two stages: first draw
  a range $R$ from $p(R \mid N_I)$, then draw a scale uniformly from $\Delta_R$.
  The mixture is less intuitive than the single octave simplex, but it is its
  natural extension -- each candidate scale is still uniform in shape once its
  range is set, while the range itself follows the empirical distribution rather
  than an arbitrary upper bound.

\nsb{Evolutionary Model.}
  We treat scale evolution within a standard probabilistic framework. Each
  proposed force is written as a \emph{fitness} function $F(S)$ that scores a
  scale $S$, paired with a \emph{selection strength} $s$ that sets how
  sharply differences in fitness translate into differences in how often a scale
  is used. A scale's likelihood under a model is
  \be
  \mathcal{L}_m(S) = \exp\!\Big[ \textstyle\sum_k s_k F_k(S) \Big],
  \ee
  where the sum runs over the fitness components $k$ that make up the model. This
  single form is what lets us treat all theories alike and combine them: a
  composite model is just the sum of its components' strength-weighted
  fitnesses, and a single-force model has only one term.

  To turn a likelihood into a probability, we normalise it over the scale space
  $\Omega$ defined above (\textit{Scale space}), which fixes both the number of
  steps $N_I$ and the range,
  \be
  P_m(S) = \frac{\mathcal{L}_m(S)}{Z}, \qquad Z = \int_\Omega \mathcal{L}_m(S)\,dS .
  \ee
  How we then score a model against this uniform null -- the same way for both
  step sizes and scale intervals -- is described in \textit{Log-probability gain
  over chance} below. The melodic
  parameters ($\sigma$, $L$, $I_0$) are fixed at the values that best fit the
  step-size distribution (\textit{Melody Model}); the harmonic strength $\beta$
  and other parameters of the harmonicity/interference fitness are fit to the
  scale intervals. The strength $\beta$ is selected over a logarithmically
  spaced grid of $68$ values from $10^{-2}$ to ${\approx}\,10^{2}$,
  and the harmonicity/interference shape
  parameters over the grids listed in \textit{Harmony Models} and
  \textit{Interference Models}.

  The fitness components, defined fully in \textit{Models}, are as follows.

\nsi{Null}
  No force acts, and every scale in $\Omega$ is equally likely. The fitness is
  constant ($F = 0$, so $\mathcal{L}_{\textrm{null}} = 1$), giving the uniform density
  $1/V$ against which every other model is measured.

\nsi{Interval Spacing}
  A force acting against small step intervals. Its
  fitness is the summed log-probability that each step is correctly
  distinguished,
  \be
  F_{\textrm{IS}}(S) = \sum_i \log p(I_i), \qquad p(I) = \Phi_{0,\sigma}(I/2),
  \ee
  where the $I_i$ are the step intervals and $\sigma$ is the transmission noise;
  its selection strength is $L$, the length of a melody aimed to be 
  transmitted without error (\textit{Interval Spacing Model}).

\nsi{Harmonicity/Interference}
  These forces favour scales enriched in harmonic intervals, by
  maximising harmonicity or minimising roughness. The fitness is
  the average over a scale's intervals of a per-interval score:
  \be
  F_{\textrm{H}}(S) =  \langle H(S) \rangle,
  \ee
  the mean harmonicity score, or, 
  \be
  F_{\textrm{I}}(S) = -\langle D(S) \rangle,
  \ee
  the (negative) mean roughness, with selection strength $\beta$; the
  per-interval scores carry each model's own shape parameters, such as the number
  of partials and roll-off (\textit{Harmony Models}, \textit{Interference
  Models}).

\nsi{Melody}
  The Melody model adds to Interval Spacing a Motor Constraint component that penalises
  large steps, which are harder to sing or play. The motor-constraint fitness is
  the negative of the scale's range,
  \be
  F_{\textrm{MC}}(S) = -\sum_i I_i = -R,
  \ee
  with selection strength $1/I_0$ (\textit{Motor Constraint Model}). The Melody
  model is the sum of the two,
  \be
  \log \mathcal{L}_{\textrm{Melody}}(S) = L\,F_{\textrm{IS}}(S) + \tfrac{1}{I_0}\,F_{\textrm{MC}}(S).
  \ee
  When the range is fixed, as for Theory scales, the motor-constraint term is
  constant and drops out of the comparison (\textit{Scale space}).

\nsi{Harmony}
  The Harmony model combines the harmonicity/interference force with interval
  spacing, so that it can account for both which intervals appear and how the
  steps are sized,
  \be
  \log \mathcal{L}_{\textrm{Harmony}}(S) = \beta\,F_{\textrm{H/I}}(S) + L\,F_{\textrm{IS}}(S).
  \ee

\nsi{Full}
  The Full model includes all three forces,
  \be
  \log \mathcal{L}_{\textrm{Full}}(S)
     = \beta\,F_{\textrm{H/I}}(S) + L\,F_{\textrm{IS}}(S) + \tfrac{1}{I_0}\,F_{\textrm{MC}}(S).
  \ee
  When $\beta \to 0$, this model reverts to the Melody model.

\nsb{Log-probability gain over chance.}
  We summarise how well a model accounts for a set of real scales with one
  number that quantifies how much more probable a real scale is under the model than under
  chance. The model assigns each scale a probability $P_m(S)$ over scale space
  (\textit{Evolutionary Model}); the null assigns a uniform probability,
  $P_{\textrm{null}}(S) = 1/V$, where $V$ is the
  volume of $\Omega$. The ratio of the two is how much more probable the model
  considers a given scale than chance does. We work with its logarithm -- so that
  independent factors add, and a typical value is not swamped by a few outliers --
  and define, per scale, the \emph{log-probability gain over chance},
  \be
  \delta \log P(S) = \log \frac{P_m(S)}{P_{\textrm{null}}(S)}
                   = \log \mathcal{L}_m(S) - \log \langle \mathcal{L}_m \rangle_\Omega ,
  \ee
  where the second equality uses $P_m = \mathcal{L}_m / Z$ and
  $\langle \mathcal{L}_m \rangle_\Omega = Z/V$, the mean likelihood of a scale drawn
  uniformly from $\Omega$: the volume cancels, so we never compute it and need only
  the mean likelihood, which we estimate by drawing scales uniformly from $\Omega$
  and averaging $\mathcal{L}_m$. We use at least 25 million scales per $N_I$ to
  get a converged estimate; we check convergence both using the weighted standard
  error, and a half-split comparison. We report,
  $\langle \delta \log P \rangle$, the region-weighted mean of $\delta \log P$
  over the real scales of each type (\textbf{\fref{fig:4}B,C}). It can be interpreted as follows:
  if $\langle \delta \log P \rangle = x$, then the model makes a typical real scale
  $e^x$ times more probable than chance ($x = 2.4$ is about $11$ times; $x = 10.8$
  about \num{50000} times).

  We compute this gain for two features of a scale. For \textbf{scale intervals},
  $P_m(S)$ is the probability of the \emph{whole} scale, which depends on all of its
  intervals through its relative fitness -- so
  $\langle \delta \log P \rangle$ is a gain \emph{per scale} (\textbf{\fref{fig:4}C}). For
  \textbf{step sizes}, we instead compare the model's predicted distribution of a
  single step against chance's: $P_m(I_S)$ is the model's step-size distribution
  (the distribution of one step among scales drawn from the model, estimated by the
  same Monte~Carlo sampling that gives the normaliser), while
  $P_{\textrm{null}}(I_S)$ is the chance step-size distribution -- the size of one
  step when the range $R$ is cut at random into $N_I$ parts, the
  $\mathrm{Beta}(1, N_I - 1)$ density $\propto (1 - I_S/R)^{N_I - 2}$ (mixed over the
  empirical range distribution for non-octave scales). The average now runs over
  individual steps, so $\langle \delta \log P \rangle$ is a gain \emph{per step}
  (\textbf{\fref{fig:4}B}), with values correspondingly smaller than the per-scale gains. The
  quantity is the same in both cases -- the log-probability gain of the model over
  chance -- and only the unit of data differs: a whole scale, or a single step.

  Two properties make this a fair score. First, choosing a model's free parameters
  (the selection strength $\beta$ and the harmonicity/interference shape parameters)
  to maximise $\langle \delta \log P \rangle$ is maximum-likelihood estimation,
  because the null term does not depend on them. Second, because $P_m$ is normalised
  over the whole of scale space, a model that makes incorrect predictions will misallocate
  some of its limited prediction budget to unproductive areas of scale space. 
  In the language of model comparison, $\delta \log P$ is the
  per-observation log-likelihood ratio of the fitted model to the null -- the quantity
  that a log Bayes factor estimates, here evaluated at the maximum-likelihood fit rather
  than by integrating over priors. Each model is itself an evolutionary (Boltzmann)
  distribution over scale space, so this is a likelihood-ratio comparison among such
  distributions.

  Fitting $\beta$ does, however, mean that a model can never score below the
  baseline it nests: at $\beta = 0$ the Harmony model reverts to Interval Spacing
  and the Full model to Melody, so $\langle \delta \log P \rangle \geq 0$ is
  guaranteed, and a positive value is not on its own evidence of an effect. We
  calibrate against this by refitting each model to random scales drawn from its
  own $\beta = 0$ baseline (\textbf{SI Section S13}, \textbf{Table~S5}). The gain attributable to
  optimisation alone is small.

\nsb{Generative Model.}
  A fitness function tells us which features a theory rewards, but because there
  are many ways to achieve high fitness it is difficult to guess the distribution
  of scales a theory actually predicts. To make these predictions visible we
  draw populations of scales from each model and compare their step-size and
  scale-interval distributions with the empirical ones (\textbf{\fref{fig:4}C--F}).

  We generate scales over the same scale space used to normalise the models.
  For a fixed number of steps $N_I$, this is the simplex $\Delta_R$ of step
  vectors summing to a range $R$, with $R = 1$ octave for octave (Theory)
  scales and $R$ drawn per scale from the region-weighted empirical
  distribution of ranges $R \mid N_I$ for non-octave (Vocal and Instrumental)
  scales. Each model defines a target distribution over this space,
  \be
  P_m(S) \propto e^{\beta F(S)}\, \pi(S),
  \ee
  which combines two ingredients: the model's fitness $F(S)$ due to
  harmonicity/interference ($F_{\textrm{H}}(S)$, $F_{\textrm{I}}(S)$),
  scaled by the bias strength $\beta$, and a \emph{reference measure}
  $\pi(S)$ that sets the distribution of scales in the absence of any
  fitness bias ($\beta = 0$). For the Full model,
  $\pi(S) = L\,F_{\textrm{IS}}(S) + \tfrac{1}{I_0}\,F_{\textrm{MC}}(S)$,
  and we recover the Melody model when $\beta = 0$. For the Harmony model,
  $\pi(S) = L\,F_{\textrm{IS}}(S)$, and we recover the Interval Spacing
  model when $\beta = 0$. For harmonicity/interference by itself the
  $\pi(S)$ term vanishes.

  The \emph{Interval-Spacing} and \emph{Melody} models are the easiest to generate:
  a population is obtained by sampling step vectors uniformly on $\Delta_R$ and
  resampling them in proportion to $\pi(S)$. The harmonicity and interference
  models instead require $\beta > 0$, where the additional fitness term
  reshapes the distribution. In the following, $P_m(S)$ is the same per-scale
  quantity used by the evolutionary model.

  Sampling directly from $P_m$ at the fitted $\beta$ is unreliable.
  When the bias is strong, the scale space becomes sparsely populated with
  a few disconnected peaks, so we need a sampling method capable of handling this.
  We therefore use annealed sequential Monte Carlo, which
  raises the bias gradually. We first draw a large base population (twenty
  times the target size) uniformly from $\Delta_R$ and weight it by $\pi(S)$.
  We then climb the $\beta$ grid: at step $k$ we
  reweight every scale in the current population by the incremental factor
  \be
  w(S) \propto e^{(\beta_k - \beta_{k-1})\,F(S)},
  \ee
  in which the $\beta$-independent reference $\pi(S)$ cancels, normalise the
  weights so that $\sum_S w(S) = 1$, and multinomially resample the population
  back to its target size.

  Because resampling repeatedly duplicates the same high-fitness scales, after
  each step we rejuvenate the population with ten Metropolis sweeps. Each sweep
  proposes, for every scale independently, transferring a random number of cents
  (uniform, up to $50$) from one randomly chosen step to another, and accepts the move
  with the standard Metropolis probability $\min\{1,\,P_m(S')/P_m(S)\}$.
  Transferring cents between two steps leaves the total range $R$ unchanged, so
  every proposal stays on $\Delta_R$; the proposals are symmetric, so these
  sweeps restore diversity without biasing the target distribution.

  Any single chain still drifts towards a few scales at high $\beta$, so we run
  many independent chains (50 batches of 1000 scales each) and pool them. For
  each model we report the pooled population at the parameters obtained from the
  evolutionary modelling: both the fitness (harmonicity/interference) parameters
  and the bias strength $\beta$ are fixed at the maximum-likelihood values from
  the \textit{Model Comparison} fit. The predicted step-size and scale-interval
  distributions are therefore those of the already-fitted model,
  so the comparison with the empirical distributions is a visual check of the fit
  rather than a second fit.

  \nsb{Weighted Sampling.}
  Due to imbalances in scale data from different regions of the world, we
  adopt a flexible approach to sampling that interpolates between biasing
  towards underrepresented vs. overrepresented regions. A set of scales is
  taken from a non-uniform distribution over $N_R$ regions, with $c_i$ scales
  per region $i$. If we weight all scales equally, then the model is biased
  towards regions with many scales. On the other hand, if we weight all
  scales so that regions are equally-weighted, then we overweight specific
  scales in regions with few scales, as too few scales have been collected to
  achieve a converged distribution within such regions. Therefore, we assign
  an equal weight to all scales within a region $i$ as,
  $\omega_i = 1/c'_i$, where $c'_i = \min\{c_i, c_0\}$, and $c_0$ is the
  maximum weight assigned to a region. As $c_0$ varies from its minimum,
  \be
  c_{\min} = \min_i^{N_R}\{c_i\},
  \ee
  to its maximum value,
  \be
  c_{\max} = \max_i^{N_R}\{c_i\},
  \ee
  the bias shifts from underrepresented to overrepresented regions. We can
  quantify the degree of the bias towards overrepresented (but not
  underrepresented) regions using the Gini index, which is a measure of
  inequality,
  \be
  G = \frac{\sum_{i=1}^{N_R} (2i - N_R - 1)\, c_{(i)}}{N_R \sum_{i=1}^{N_R} c_i},
  \ee
  where the region counts $c_{(i)}$ are ordered from low to high. A uniform
  distribution (obtained at $c_{\max}$) gives $G = 0$, while a maximally unequal
  one approaches $G = 1$ (exactly $1 - 1/N_R$ for $N_R$ regions).

  When comparing scale theories in \textbf{Fig.~\ref{fig:3}C-D}, we use
  $c_0 = 20$, which gives $G = 0.22$, $G = 0.27$, $G = 0.26$ for Vocal,
  Instrumental and Theory scales, respectively (\textbf{SI Fig.~S20}).

\section*{Data and Code Availability}

  Code for the main analyses, model comparison, and for producing figures, can
  be found at
  \href{https://github.com/jomimc/ModellingScaleEvolution}{github.com/jomimc/ModellingScaleEvolution}.
  Code for extracting and cleaning fundamental frequency from raw audio can
  be found at
  \href{https://github.com/jomimc/F0EstimationGUI}{github.com/jomimc/F0EstimationGUI}.
  Code for extracting scales from fundamental frequency data can be found at
  \href{https://github.com/jomimc/MusicalScaleExtraction}{github.com/jomimc/MusicalScaleExtraction}.
  Archived versions of code and data can be found
  at \href{https://zenodo.org/records/15627131}{zenodo.org/records/15627131}.

\section*{Author Contributions}

  Conceptualization: JMM, EP, PES, SB, TT.
  Methodology: JMM.
  Software: JMM.
  Formal Analysis: JMM.
  Investigation: JMM.
  Resources: JMM, EP, PES, SB, TT.
  Supervision: TT.
  Writing, original draft preparation: JMM.
  Writing, review and editing: JMM, EP, PES, SB, TT.

\section*{AI Declaration}
  DALL-E was used to generate the head and ear silhouettes in
  \textbf{Fig.~\ref{fig:1}}.
  Claude was used for analysis code and for copyediting the manuscript;
  the authors assume full responsibility for both.
  
\section*{Acknowledgements}

  JMM acknowledges Steve Granick for continuous support.
  TT was supported by the National Research Foundation of Korea Grants
  NRF-RS-2025-00573354.
  PES is supported by funding from the Japanese and New Zealand
  governments (Grant-in-Aid \#19KK0064 from the Japan Society for the
  Promotion of Science; Rutherford Discovery Fellowship MFP-UOA2236 and
  Marsden FastStart Grant RDF-UOA2202 from the Royal Society Te Apārangi). We
  acknowledge discussions with Joren Six during an early stage of this
  project.

\bibliography{ScaleEvo}
\bibliographystyle{unsrtnat}

\clearpage\includepdf[pages=-]{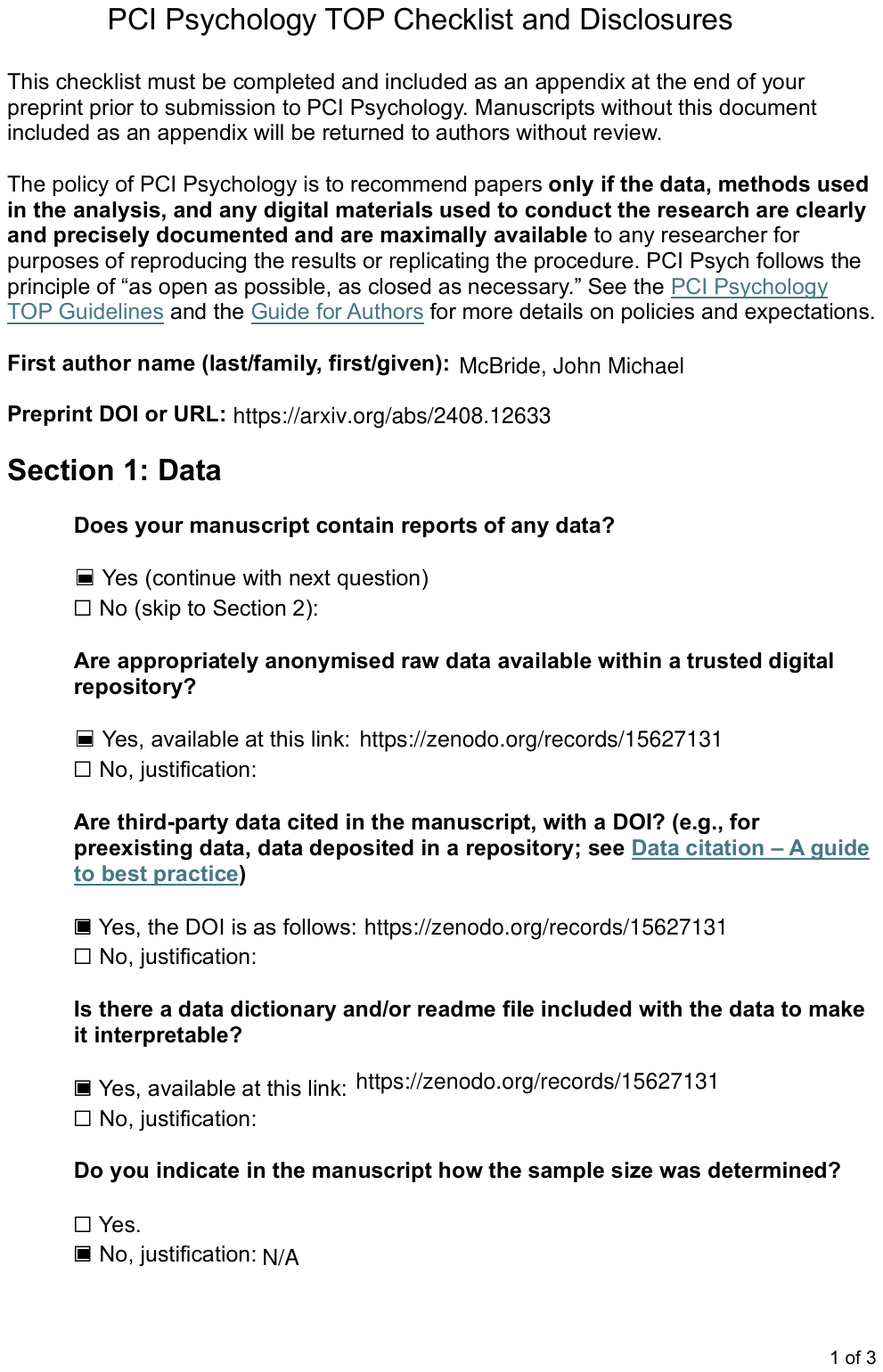}

\end{document}


\title{\LARGE\bfseries\sffamily Supporting Information\\
Evolutionary modelling reveals melodic and harmonic constraints on global scale structure.}

\author{
    John M. McBride\textsuperscript{1,*} \and
    Steven Brown\textsuperscript{2} \and
    Elizabeth Phillips\textsuperscript{2} \and
    Patrick E. Savage\textsuperscript{3,4} \and
    Tsvi Tlusty\textsuperscript{1}
}

\date{
    \small
    \textsuperscript{1}Department of Physics, Ulsan National
    Institute of Science and Technology, South Korea\\[0.5em]
    \textsuperscript{2}McMaster University, Canada\\
    \textsuperscript{3}University of Auckland, New Zealand\\
    \textsuperscript{4}Keio University, Japan\\
    \textsuperscript{*}Correspondence:
    \href{mailto:jmmcbride@protonmail.com}{jmmcbride@protonmail.com}\\
}


  \maketitle

\section{Two categories of scale evolution}

  In this section we give a fuller account of what it means for scales to
  ``evolve'', and of the mechanisms by which they can converge, elaborating on
  details omitted from the main text for brevity.

  Through evolution, scales will naturally drift randomly. The exact intervals
  between scale degrees may change due to instrument tuning naturally varying
  (due to temperature, humidity, or physical force), or from variation in sung
  performances. If selection pressures are weak, the resulting scales may
  appear random. However, scales can converge through multiple mechanisms:
  horizontal transmission processes such as conquest or contact (\ie\ we can
  learn new scales from other people); vertical transmission processes such as
  survival of the `fittest' or `flattest' (to be explained in the next
  paragraph); and conscious innovation or design. The
  most salient and consistent force appearing to act on scales is a conformity
  bias such that when humans make music together, they tend to play in tune
  with each other. This bias can explain why we see convergence within a
  population, or between populations that share borders or trade routes.
  However, it cannot explain the level of convergence observed in scales
  between distant cultures.

  In this work we investigate the role of vertical transmission, which
  explains how, within a population, scales with certain
  properties can be selected over time. Scales can be selected because
  they are ``fit'' -- because they
  have properties that aid their survival, of which we consider four: (i)
  Certain intervals may be naturally consonant, or pleasant sounding. (ii)
  Certain intervals may facilitate harmonic synchrony, which has been
  hypothesised to promote social bonding. (iii) Scales can affect the fidelity
  of melody transmission. (iv) Some scales lead to melodies that are easier to
  sing or play than others. There are more possible reasons for scales to be
  fit, which may be considered in future work.

  Scales can alternatively converge due to
  ``survival of the flattest,'' whereby, rather than the scales being selected
  for being ``good'' at something, they are effectively selected because the
  rate of change decreases. If some scales are easier to reliably tune than
  others, then they will be more stable over time and therefore last longer.
  This is facilitated by theoretical and technological development:
  A mathematical theory of scales provides an unmoving target of frequency
  ratios to aim for. Technology for measuring frequency ratios increases
  the precision with which such targets are realised.

  Scale evolution thus takes on more than one form. At one extreme there
  is a solo singer, for whom scales reside in memory. Each instantiation
  of the scale as it is sung will be unique, as it will deviate slightly
  (perhaps imperceptibly) from one song rendition to another.
  The singer's target scale -- plausibly measured as the average scale
  taken over many renditions of a song --
  may also drift over time. Certain instantiations may survive more than others,
  leading to convergence on certain scales. Here evolution resembles most closely what
  occurs in biology: the slow process of random change and subsequent selection.

  At the far extreme (\ie, modern times) we have electronic instruments,
  for which there is effectively zero change in tuning across performances.
  Here the only ``evolution'' is the process of designing new instruments,
  or consciously changing their settings.

  In between these extremes let us consider the ancient Greeks as an example.
  They had mathematical theories which essentially fixed the tuning targets,
  and they had the monochord as a simple measurement device. The measurement
  device is not perfectly precise, so scales will still drift from performance
  to performance. However, the target is fixed, so the scales will simply fluctuate
  around a stable target -- equivalent to strong stabilising selection in biology.
  As time passed, the theories of scales and their
  frequency ratios would be updated, and so the targets would eventually change.
  However, in contrast to the solo singer, the target is not changing through
  a process of drift and selection, but instead through mathematical design.
  Thus there are at least two separate evolutionary processes at play here,
  one that resembles biological evolution (drift and selection), and another
  that is better described as technological evolution.

\section{Theories of scale evolution and their predictions}

  Here we summarise the models, their mechanisms, and what they predict,
  in more detail than in Fig.~1 in the main manuscript.
  
  A model of scale evolution must, in principle, account for four features of a
  scale: the number of scale degrees, the overall scale range, the step sizes,
  and the sizes of the scale intervals measured between all pairs of degrees.
  These features are not independent.
  Fixing the step-size distribution together with the number of steps $N_I$
  determines the scale range, and vice versa. Scale intervals are
  determined by step sizes and how they are ordered. Thus, a model does not need
  to explicitly predict or explain every feature, since some are linked --
  explain one feature, and you might get one more for free.
  None of the models we consider predicts $N_I$ directly,
  so throughout we condition on the observed number of
  steps and ask what each model predicts for the remaining features.

  Two mechanisms act on step sizes alone. The \textit{Interval Spacing} theory
  holds that intervals are produced and perceived with error, so steps must be
  large enough to be reliably distinguished; it therefore penalises small steps.
  The \textit{Motor Constraint} theory holds that large intervals are harder
  to sing than small ones, and so penalises large steps. Once $N_I$ is fixed, a
  bound on step size becomes a bound on scale range; and once both $N_I$ and the
  scale range are fixed to a single value, the Motor Constraint has nothing left
  to act on and becomes inert. The two theories make opposite, individually
  unbounded predictions -- one favours large steps, the other small, and neither
  specifies a preferred scale on its own -- so they are naturally complementary.

  Two further mechanisms act on the scale intervals. \textit{Harmonicity} models
  score intervals by how well they sound like a single complex harmonic tone
  (tonal fusion), favouring harmonic intervals such as octaves and fifths.
  \textit{Interference} models score the auditory roughness of intervals; they
  likewise favour harmonic intervals, but additionally penalise intervals near a
  semitone, where beating between partials is strongest. Because harmonicity and
  interference make closely overlapping predictions for scales, we treat them
  together throughout, as in the main text.
 
  \begin{table*}[ht!]
  \centering
  \caption{\textbf{Summary of theories, mechanisms, and predictions}}
  \label{tab:theories}
  \begin{tabular}{llll}
  \toprule
  Theory & Mechanism & Predicts & Conditional predictions \\
  \midrule
  Interval Spacing & Accurate communication & Avoid small steps & \\
  Motor Constraint & Energy efficiency & Avoid large steps & \makecell{Given scale size predicts \\ a finite scale range, and vice versa} \\
  Harmonicity & \makecell{Consonance \\ Synchrony / bonding \\ Reliable tuning} & Harmonic intervals & \\
  Interference & \makecell{Consonance \\ Synchrony / bonding \\ Reliable tuning} & Harmonic intervals; avoid semitones & \\
  \bottomrule
  \end{tabular}
  \end{table*}

  \clearpage
\section{Scale dataset statistics}

\begin{figure*}[h]
  \centering
  \includegraphics[width=\textwidth]{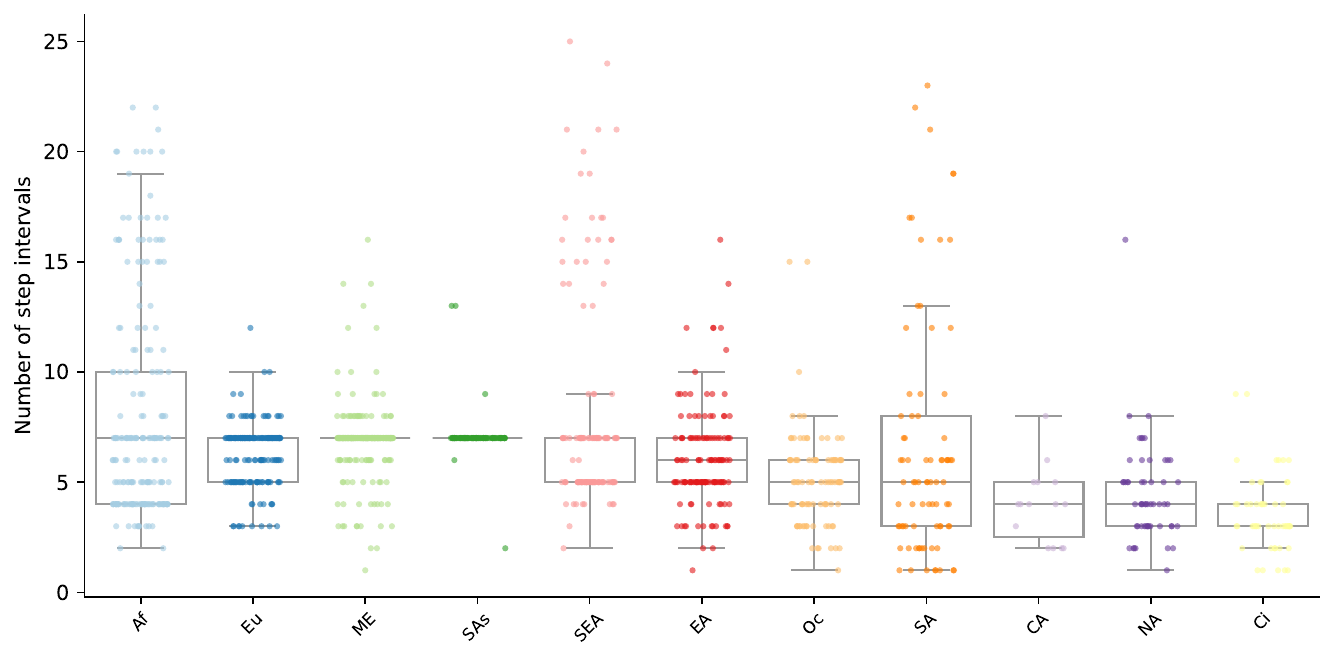}
  \caption{\textbf{Scale size by region.} Distributions of the number of step
  intervals per scale for each region, pooling across scale types (Vocal,
  Instrumental, and Theory). We count step intervals rather than scale degrees
  because we do not assume octave equivalence for Vocal and Instrumental scales:
  counting steps avoids the ambiguity of whether to include the final (octave)
  degree. For Theory scales the two counts coincide.}
  \label{fig:scale-size-region}
  \end{figure*}

\begin{figure*}[h]
  \centering
  \includegraphics[width=\textwidth]{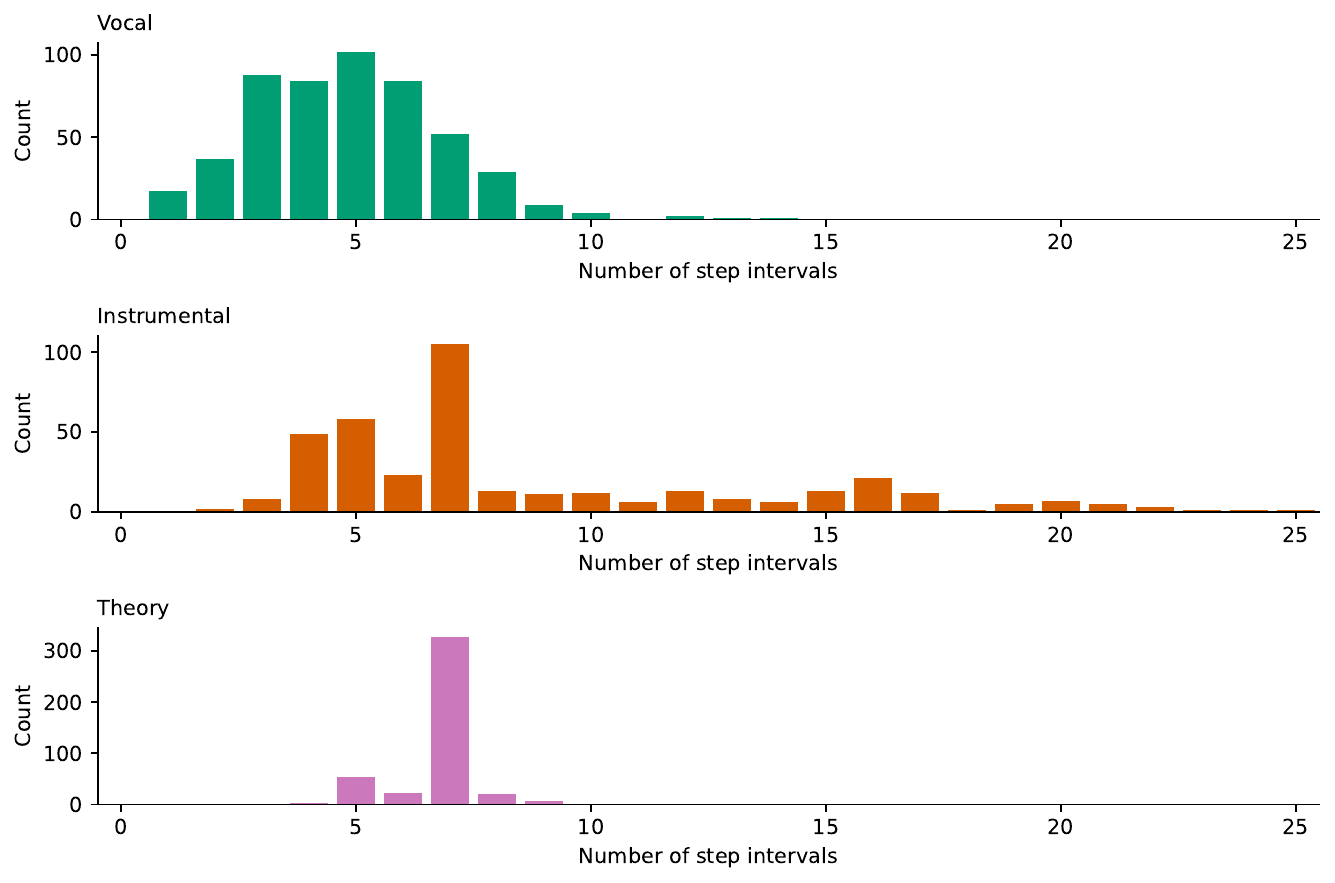}
  \caption{\textbf{Scale size by scale type.} Distributions of the number of
  step intervals per scale for each scale type, pooling across regions. As in
  \fref{fig:scale-size-region}, we count step intervals rather than scale
  degrees; for Theory scales the two counts coincide.}
  \label{fig:scale-size-type}
  \end{figure*}

\begin{figure*}[h]
  \centering
  \includegraphics[width=\textwidth]{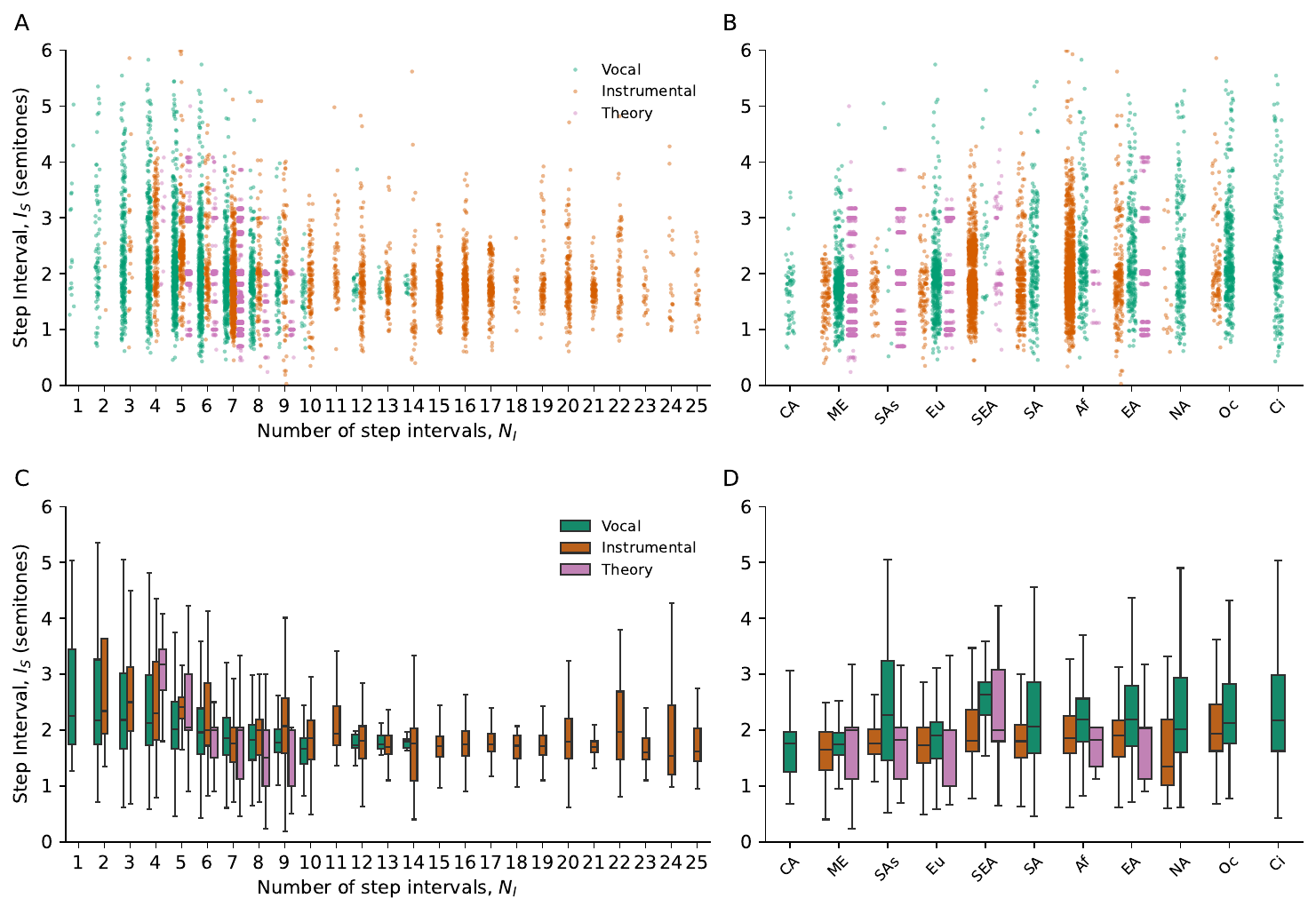}
  \caption{\textbf{Step size by scale size and region.} Distributions of the
  step interval $I_S$ (in semitones), shown both as raw points (strip plots;
  A, B) and as box-plot summaries (C, D), as a function of the number of step
  intervals $N_I$ (A, C) and of region (B, D). In all panels points are
  coloured by scale type (Vocal, Instrumental, Theory).}
  \label{fig:step-by-n}
  \end{figure*}

\begin{figure*}[h]
  \centering
  \includegraphics[width=\textwidth]{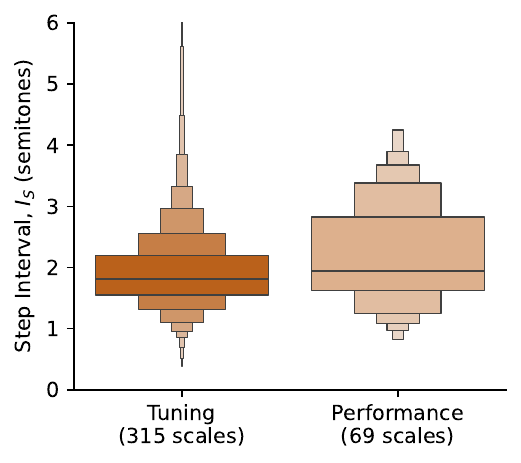}
  \caption{\textbf{Step size by measurement type.} Distributions of the
  step interval $I_S$ (in semitones) as letter-value plots for Instrumental
  scales that were either measured from their instrument tunings, or 
  from recorded performances.}
  \label{fig:step-by-inst}
  \end{figure*}

\begin{figure*}[ht!]
  \centering
  \includegraphics[width=\textwidth]{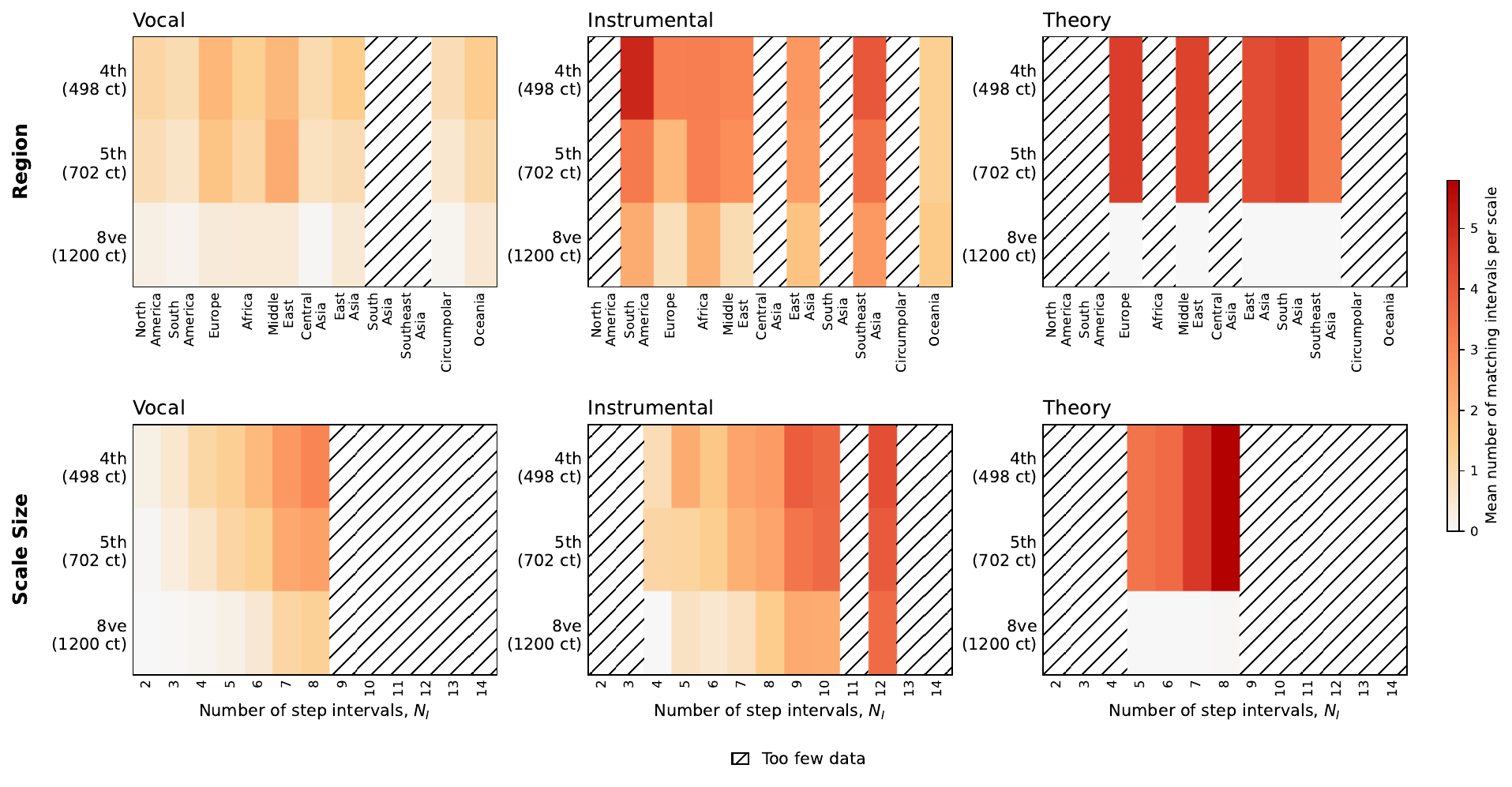}
  \caption{\textbf{Occurrence of fourths, fifths, and octaves.} Region-weighted
  mean number of intervals per scale falling within $\pm25$ cents of a perfect
  fourth, a perfect fifth, or an octave. Each block of three heatmaps corresponds to a
  scale type (Vocal, Instrumental, Theory); within a heatmap, the three rows are
  the three harmonic intervals and colour encodes the region-weighted mean
  count. The top block breaks the counts down by region, the bottom block by
  scale size $N_I$; a single colour scale, shown by the colourbar at right,
  applies to all six heatmaps. Hatched columns hold fewer than ten scales and
  are not scored.}
  \label{fig:harmonic-counts}
  \end{figure*}

  This section collects descriptive statistics of the scale dataset that were
  not reported in the main text:
  \begin{itemize}
    \item Distribution of scales across region and scale type (\tref{tab:scale-counts}).
    \item How scale size varies across regions (\fref{fig:scale-size-region})
          and scale types (\fref{fig:scale-size-type}).
    \item How step-interval sizes vary with scale size and region
          (\fref{fig:step-by-n}), and with the type of measurement made for
          Instrumental scales (\fref{fig:step-by-inst}).
    \item How often approximate fourths, fifths, and octaves
          appear across regions and scale sizes (\fref{fig:harmonic-counts}).
  \end{itemize}

  \begin{table*}[ht!]
  \centering
  \caption{\textbf{Number of scales for each scale type and region.}
  Instrumental scales are also separated by measurement type:
  measured from instrument tuning, or from a performance.}
  \label{tab:scale-counts}
  \begin{tabular}{lrrrr}
  \toprule
  Region & Vocal & Instrumental (tuning) & Instrumental (performance) & Theory \\
  \midrule
  \bfseries Overall & 510 & 315 & 69 & 434 \\
  \midrule
  North America & 54 & 0 & 1 & 0 \\
  South America & 54 & 27 & 0 & 0 \\
  Europe & 73 & 9 & 7 & 104 \\
  Africa & 45 & 118 & 36 & 3 \\
  Middle East & 79 & 2 & 12 & 166 \\
  Central Asia & 15 & 0 & 0 & 0 \\
  East Asia & 50 & 26 & 5 & 57 \\
  South Asia & 2 & 6 & 1 & 91 \\
  Southeast Asia & 5 & 112 & 5 & 13 \\
  Circumpolar & 52 & 0 & 0 & 0 \\
  Oceania & 81 & 15 & 2 & 0 \\
  \bottomrule
  \end{tabular}
  \end{table*}

  \clearpage
\section{Interval Spacing experimental parameters}
\label{si:interval-spacing-params}

  The Interval Spacing model treats scale intervals as being under
  selection for reliable transmission, with transmission error modelled as
  Gaussian noise of standard deviation $\sigma$ (in cents). Rather than leave
  $\sigma$ as a completely free parameter, we anchor it in independent measurements of
  human production and perception, so that the value fitted to scale data can
  be compared against what humans can actually do.

  An interval is \emph{produced} (\eg, sung) with error $\sprod$ and
  \emph{perceived} with error $\sper$. Modelling the two as independent
  Gaussian sources, the total transmission noise is
  \begin{equation}
  \sigma = \sqrt{\sprod^2 + \sper^2}.
  \label{eq:sigma-combine}
  \end{equation}
  We estimate ranges for $\sprod$ and $\sper$ separately below, each from
  several sources, and report both for two populations: \emph{more-skilled}
  (professional or trained singers; musically-trained listeners) and
  \emph{less-skilled} (poor-pitch singers; untrained listeners), the same split
  used in main-text Fig.~3D. Production and
  perception accuracy both depend strongly on musical training, so a single
  pooled value would be misleading. Throughout, every
  reported $\sigma$ is a per-interval standard deviation in cents; wherever a
  source reports a different quantity we explain how we converted this into
  a $\sigma$ value.

  \subsection{Variance in sung intervals}

  We compile the standard deviation of sung melodic intervals from six sources:
  recordings of the Georgian traditional singer,
  Erkomaishvilli\cite{scherbaumTonal2020}; the Anton Bruckner choir from Barcelona
  (Choral Singing Dataset)\cite{cuestaChoral2019}; an amateur choir group in
  Germany (Dagstuhl ChoirSet)\cite{rosenzweigDagstuhl2020}; a set of poor-pitch singers
  singing Happy Birthday\cite{pfordresherVocal2017a}; and a mix of graduate-level and
  professional sopranos\cite{devaneyAutomatically2011a}.

\begin{figure}[h!]
  \centering
  \includegraphics[width=\columnwidth]{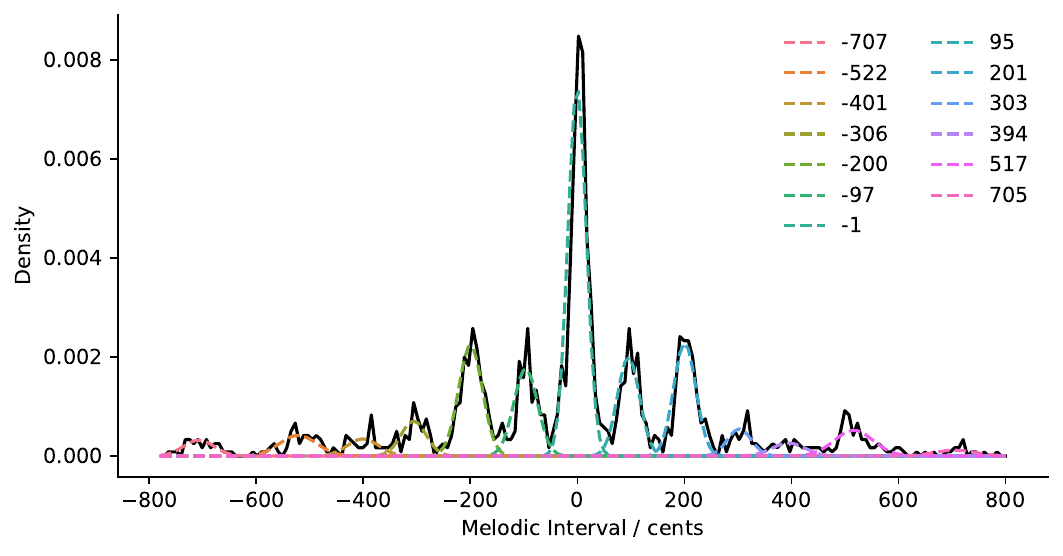}
  \caption{A log-pitch histogram of melodic intervals from the Choral Singing
  Dataset. Individual components of the best-fitting Gaussian mixture model
  are shown as dotted lines, with the means indicated in the legend in cents.}
  \label{fig:sigma-produce}
  \end{figure}

  \begin{enumerate}[label=\Roman*.]
  \item For the Erkomaishvilli data, we report the standard deviation of
  melodic step-sizes between neighbouring pitch groups taken from Figure 3
  of\cite{scherbaumTonal2020}, extracted using g3data\cite{frantzG3data}, as
  $\sprod = 32$ cents.

  \item For the Choral Singing Dataset, we have pitch annotations (in cents)
  that we use to create a melodic interval histogram. We fit a 13-component
  Gaussian Mixture Model (GMM) to this histogram, after visually estimating
  that there are 13 peaks (\textbf{SI Fig.~\ref{fig:sigma-produce}}). The arithmetic
  mean of the standard deviations of GMM components is $\sprod = 22$
  cents.

  \item For the Dagstuhl ChoirSet, we have pitch annotations aligned to a MIDI
  score, and so we do not need to fit a GMM. We group melodic intervals of
  the same magnitude and direction, and calculate the standard deviation for
  each group. The arithmetic mean of the standard deviations is
  $\sprod = 38$ cents.

  \item For poor-pitch singers, we extract mean ``semitone deviations'' per
  singer from Fig.~4a  of\cite{pfordresherVocal2017a} using g3data.
  This quantity is the mean absolute deviation (MAD) per melodic interval,
  compared to the intended interval. For a zero-mean Gaussian,
  the expected absolute deviation is $\mathbb{E}|X| = \sigma\sqrt{2/\pi}$,
  so a mean absolute deviation (MAD) maps to a standard deviation by
  \begin{equation}
  \sprod = \sqrt{\pi/2}\;\mathrm{MAD} \approx 1.25\,\mathrm{MAD}.
  \label{eq:folded-normal}
  \end{equation}
  This allows us to map ``semitone deviation'' values to get a
  range of $50 \leq \sprod \leq 315$ cents.

  \item Ref.\cite{devaneyAutomatically2011a} reports per-singer interval standard
  deviations of \SIrange{13}{22}{cents} (mean \SI{17}{cents}).
  \end{enumerate}

  Taken together, these give $\sprod \approx $ \SIrange{13}{38}{cents} for
  more-skilled (professional or trained) singers and
  $\sprod \approx$ \SIrange{50}{315}{cents} for less-skilled (poor-pitch)
  singers.

  \subsection{Variance in interval perception}

  We estimate interval-perception noise, $\sper$, from
  interval-discrimination experiments using a signal-detection-theory (SDT)
  constant-variance model. Every experiment we use has the same two-interval
  design: the listener hears two intervals and judges which is larger. Each
  interval is perceived with independent Gaussian noise of standard deviation
  $\sper$, so the difference of the two percepts has standard deviation
  $\sqrt{2}\,\sper$, and the probability of correctly ordering two intervals
  differing by $\delta I$ is
  \begin{equation}
  \mathrm{Acc}(\delta I) = \Phi\!\left(\frac{\delta I}{\sqrt{2}\,\sper}\right),
  \label{eq:sdt-acc}
  \end{equation}
  rising from chance ($0.5$) at $\delta I = 0$. This $\sqrt{2}$ is specific to the
  two-interval task, in which each trial compares two intervals; it does not
  appear in the transmission model, where $\sigma$ is the error on a single
  interval between two notes. The literature reports two kinds
  of measurement -- full accuracy curves and single-threshold JNDs -- but both
  are read off this one model, so every source is treated identically.

\begin{figure}[h!]
  \centering
  \includegraphics[width=\columnwidth]{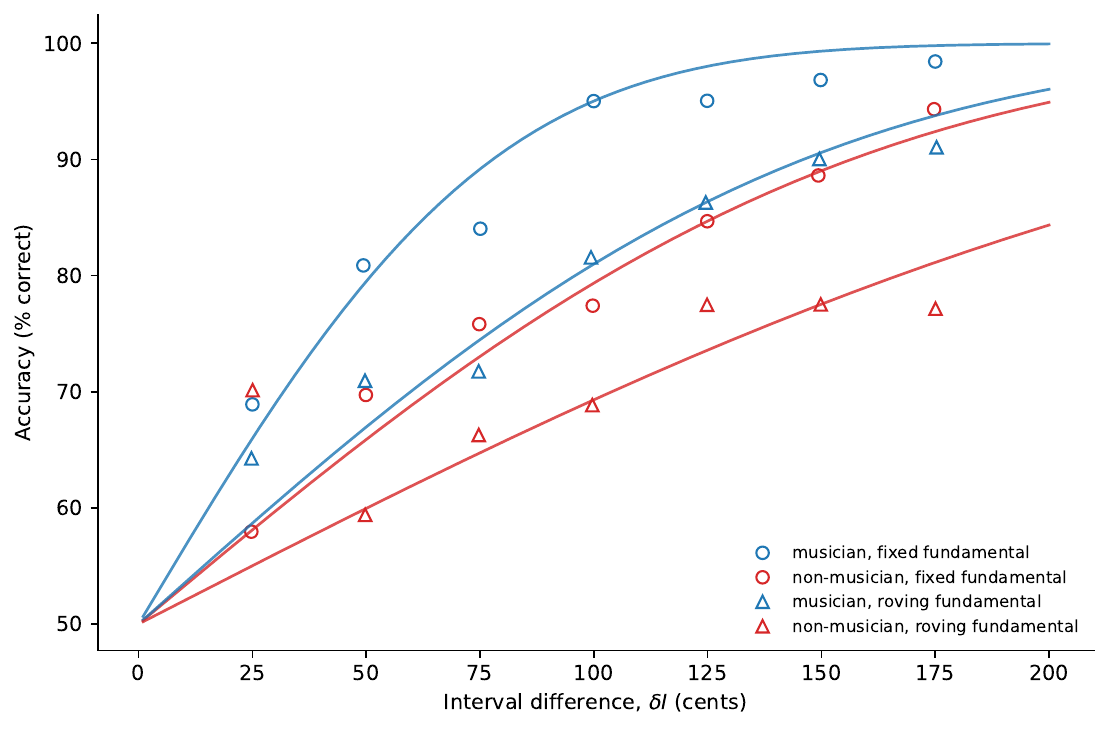}
  \caption{Interval-discrimination accuracy vs.\ interval difference $\delta I$,
  from\cite{zaratePitchinterval2012} Fig.~1 (circles), for four conditions
  (musician/non-musician $\times$ fixed/roving fundamental; colours). Lines show
  fits of the signal-detection model, Eq.~\ref{eq:sdt-acc}, giving a per-interval
  $\sper$ of $43$ and $81$ cents for musicians (fixed and roving fundamental) and
  $86$ and $140$ cents for non-musicians.}
  \label{fig:sigma-perceive}
  \end{figure}

  \nsb{Accuracy curves:} Where a study reports accuracy as a function of
  $\delta I$, we fit Eq.~\ref{eq:sdt-acc} directly for $\sper$. We do this for
  the curves of\cite{zaratePitchinterval2012} (their Fig.~1), extracted with
  g3data separately for musicians and non-musicians and for the roving and
  fixed-fundamental conditions (\textbf{SI Fig.~\ref{fig:sigma-perceive}}), giving
  $\sper \approx 43$ and $81$ cents (musicians; fixed and roving) and
  $\approx 86$ and $140$ cents (non-musicians). This study includes a screening
  procedure that removes participants who perform poorly on a basic
  pitch-discrimination task.

  \nsb{Just-noticeable differences:} Where a study reports a just-noticeable
  difference (JND) -- the $\delta I$ at a fixed criterion accuracy $p$ --
  Eq.~\ref{eq:sdt-acc} inverts to
  \begin{equation}
  \sper = \frac{\mathrm{JND}(p)}{\sqrt{2}\,\Phi^{-1}(p)}.
  \label{eq:jnd-sigma}
  \end{equation}
  The only quantity that differs between JND studies is the criterion $p$ at
  which each \emph{defines} its threshold. \citet{mcdermottMusical2010} use an
  adaptive procedure converging on $p = 0.707$, giving a factor
  $1/[\sqrt{2}\,\Phi^{-1}(0.707)] \approx 1.30$; \citet{burja78} report JNDs at
  $p = 0.75$, giving $\approx 1.05$. Applying the appropriate factor and
  averaging over interval standards per participant, we obtain
  from\cite{mcdermottMusical2010} (Fig.~2, g3data) $\sper \approx 42$--$180$
  cents for musicians (including amateurs) and $\approx 141$--$377$ cents for
  non-musicians, and from\cite{burja78} $\sper \approx 40$ cents (musicians)
  and $78$ cents (non-musicians).

  We treat musicians as one group, folding amateurs in with trained musicians,
  because the amateur/skilled boundary is too fuzzy to serve as an exclusion
  criterion. Pooling all three sources gives $\sper \approx 40$--$180$ cents for
  musicians and $\approx 78$--$377$ cents for non-musicians.

  Combining production and perception through Eq.~\ref{eq:sigma-combine} gives a
  total transmission noise of roughly $42$--$184$ cents for more-skilled
  populations and $93$--$491$ cents for less-skilled populations; the two
  together span the $42 \leq \sigma \leq 491$ cents quoted in the main text
  (\textit{Interval Spacing Model}). The $\sigma$ fitted directly to scale data
  (a point estimate of $54$ cents, with a profile interval of
  $\approx 49$--$60$ cents; \textbf{SI Section~\ref{si:melody-fit}}) falls inside
  the more-skilled band -- an independent check that the Melody model's noise
  level matches measured human ability rather than being tuned to the scales.

  \clearpage
\section{Harmonicity and Interference model scores}

\begin{figure*}[ht!]
  \centering
  \includegraphics[width=\textwidth]{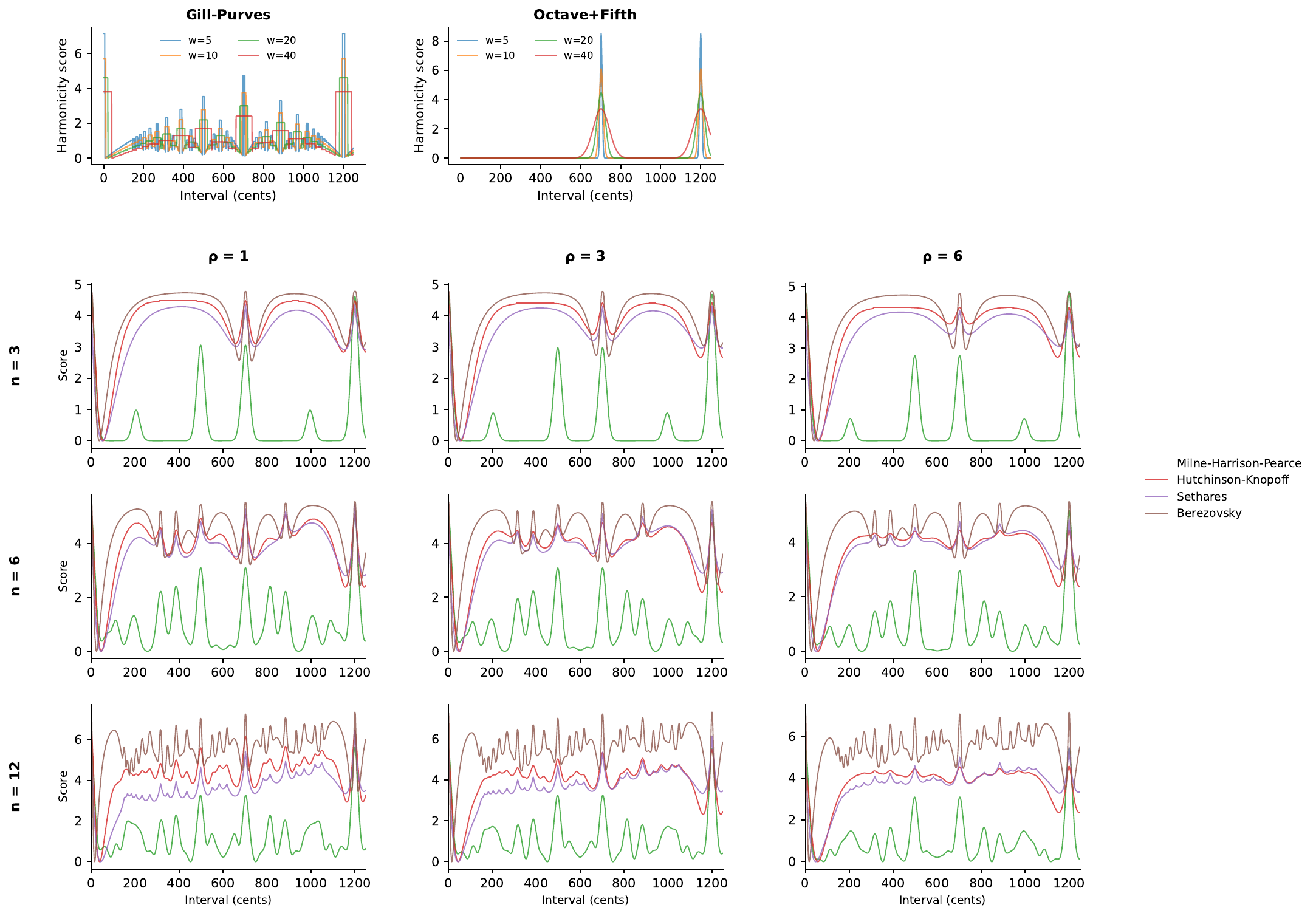}
  \caption{\textbf{How harmonicity and interference models score intervals.}
  Each curve gives the score assigned to an interval as a function of its size.
  Top row: the single-parameter harmonicity models (Gill-Purves and
  octave-fifth), each shown for several values of the tolerance $w$.
  Lower $3\times3$ grid: the four multi-partial models (Milne-Harrison-Pearce,
  Hutchinson-Knopoff, Sethares, Berezovsky), overlaid in each panel, as the
  number of partials $n \in \{3, 6, 12\}$ and the roll-off
  $\rho \in \{1, 3, 6\}$ vary; the fundamental frequency is fixed at $f_0 = 1000$~Hz.
  Scores are normalised by subtracting the minimum and dividing by the standard
  deviation: because each fitness is multiplied by the selection strength $\beta$,
  differences of scale are largely irrelevant; the reason we normalise is that
  it allows us to optimise over one range of $\beta$. What matters here
  is the relative height of peaks within a scoring function, not the absolute
  magnitudes.}
  \label{fig:hi-scoring}
  \end{figure*}

  The harmonicity and interference models all assign a score to each
  possible interval, and they differ in shape and in the parameters that control
  them. \fref{fig:hi-scoring} shows these scoring profiles across the interval
  range. The Gill-Purves and octave-fifth harmonicity models each have a
  single parameter -- the tolerance $w$, the width of the window (or kernel) used
  to account for deviations from harmonic intervals -- so we show each for a few
  values of that width (top row). The remaining four models (Milne-Harrison-Pearce
  harmonicity, and the Hutchinson-Knopoff, Sethares, and Berezovsky
  interference models) are governed by the number of partials $n$ and
  the roll-off $\rho$. The three interference models also depend
  on the fundamental frequency $f_0$. To illustrate the scoring functions we vary
  $n \in \{3, 6, 12\}$ and $\rho \in \{1, 3, 6\}$, holding the fundamental fixed at
  $f_0 = 1000$~Hz, and overlay all four in each panel of a $3\times3$ grid.

  All of these models share a key feature, that they assign the highest scores to
  unison, the octave and the fifth. The main qualitative difference is that the
  interference models additionally penalise intervals close to a semitone, where
  beating between neighbouring partials is strongest, whereas the harmonicity
  models carry no such penalty. This semitone aversion is the feature that most
  distinguishes interference from harmonicity in their predictions for scales
  when not modelled in tandem with the Interval Spacing model; otherwise,
  for the Harmony and Full models there is little extra benefit from semitone aversion.
  For the four models with explicit partial modelling,
  predictions diverge increasingly as the number of partials increases.

  \clearpage
\section{Harmonicity and Interference model full results}

\begin{figure*}[ht!]
  \centering
  \includegraphics[width=\textwidth]{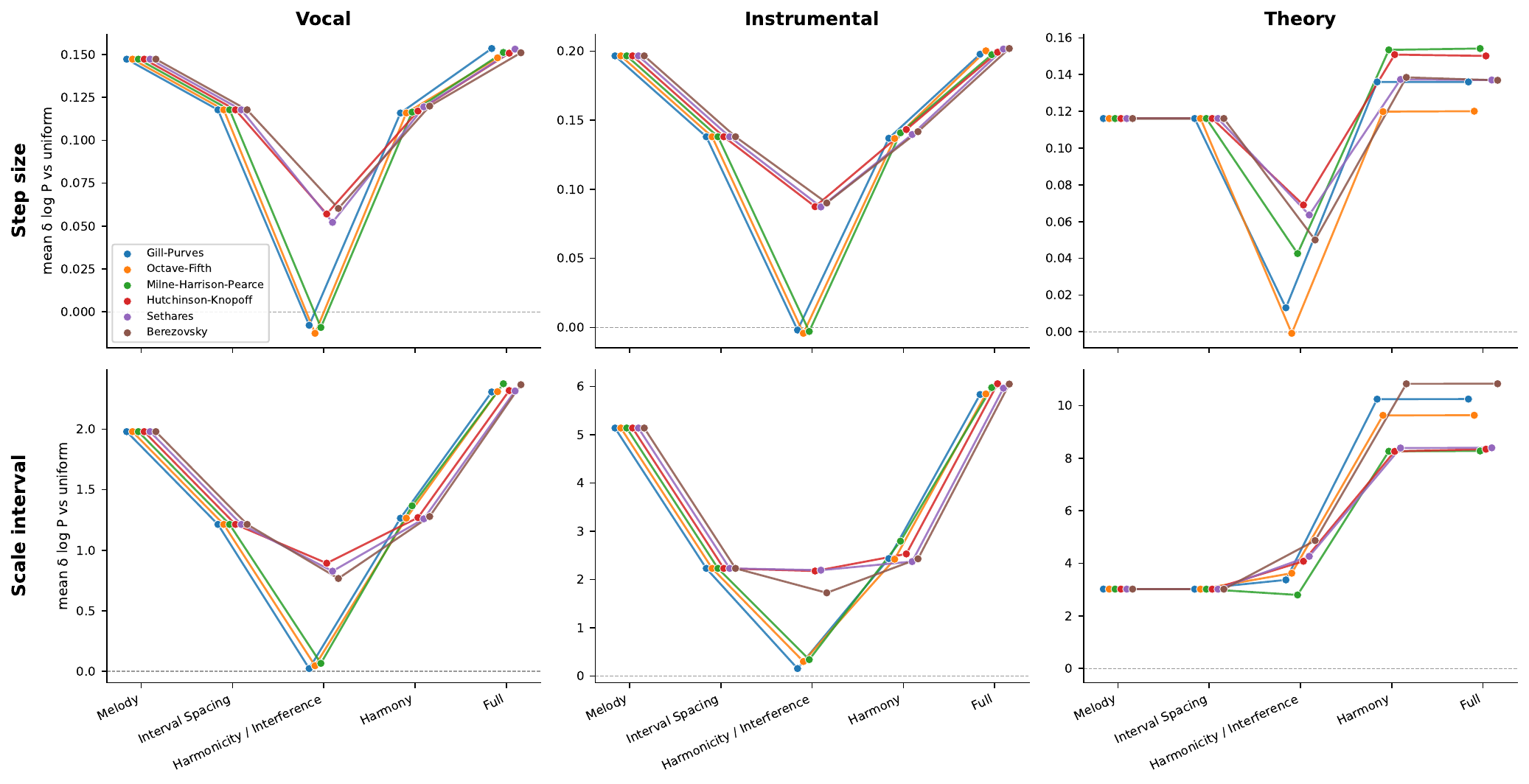}
  \caption{\textbf{Comparing every harmonicity/interference model against the null.}
  Mean log-probability gain over chance, $\langle \delta \log P \rangle$,
  against the uniform null, for three scale types (columns: Vocal,
  Instrumental, Theory) and two features (rows: step sizes, scored per step, and
  scale intervals, scored per scale). The x-axis is categorical; lines are shown
  to guide the eye.}
  \label{fig:model-grid}
  \end{figure*}

\begin{figure*}[ht!]
  \centering
  \includegraphics[width=\textwidth]{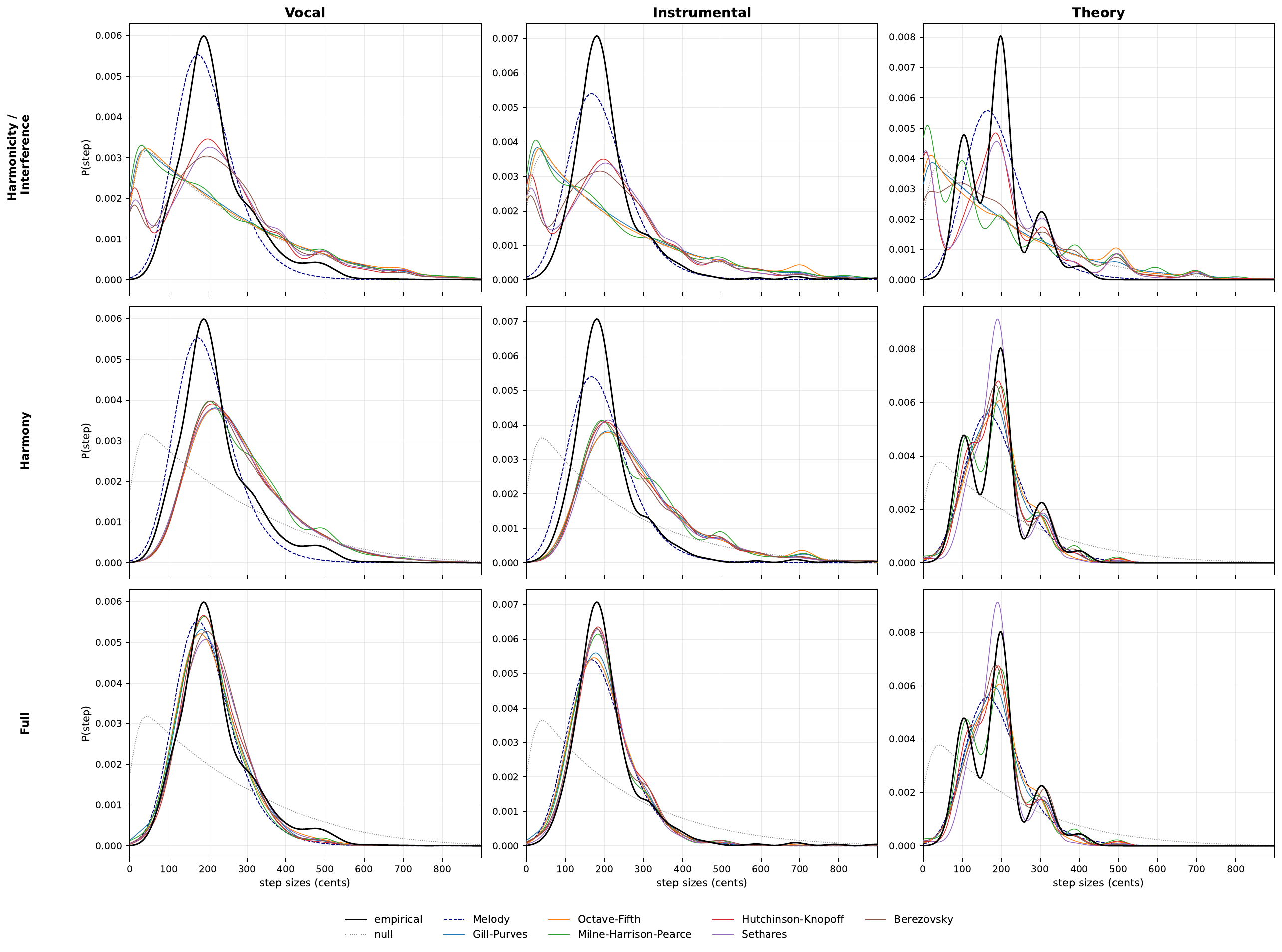}
  \caption{\textbf{Predicted step-size distributions.} Predicted step-size
  distributions at each model's best-fit parameters for three model
  groupings (rows: harmonicity/interference alone, Harmony, Full) and three
  scale types (columns: Vocal, Instrumental, Theory). Each panel overlays the
  empirical distribution (black), the uniform null (grey dotted), the Melody model
  (navy dashed), and the six harmonicity/interference models (coloured).
  Curves are smoothed for display.}
  \label{fig:jsd-steps}
  \end{figure*}

\begin{figure*}[ht!]
  \centering
  \includegraphics[width=\textwidth]{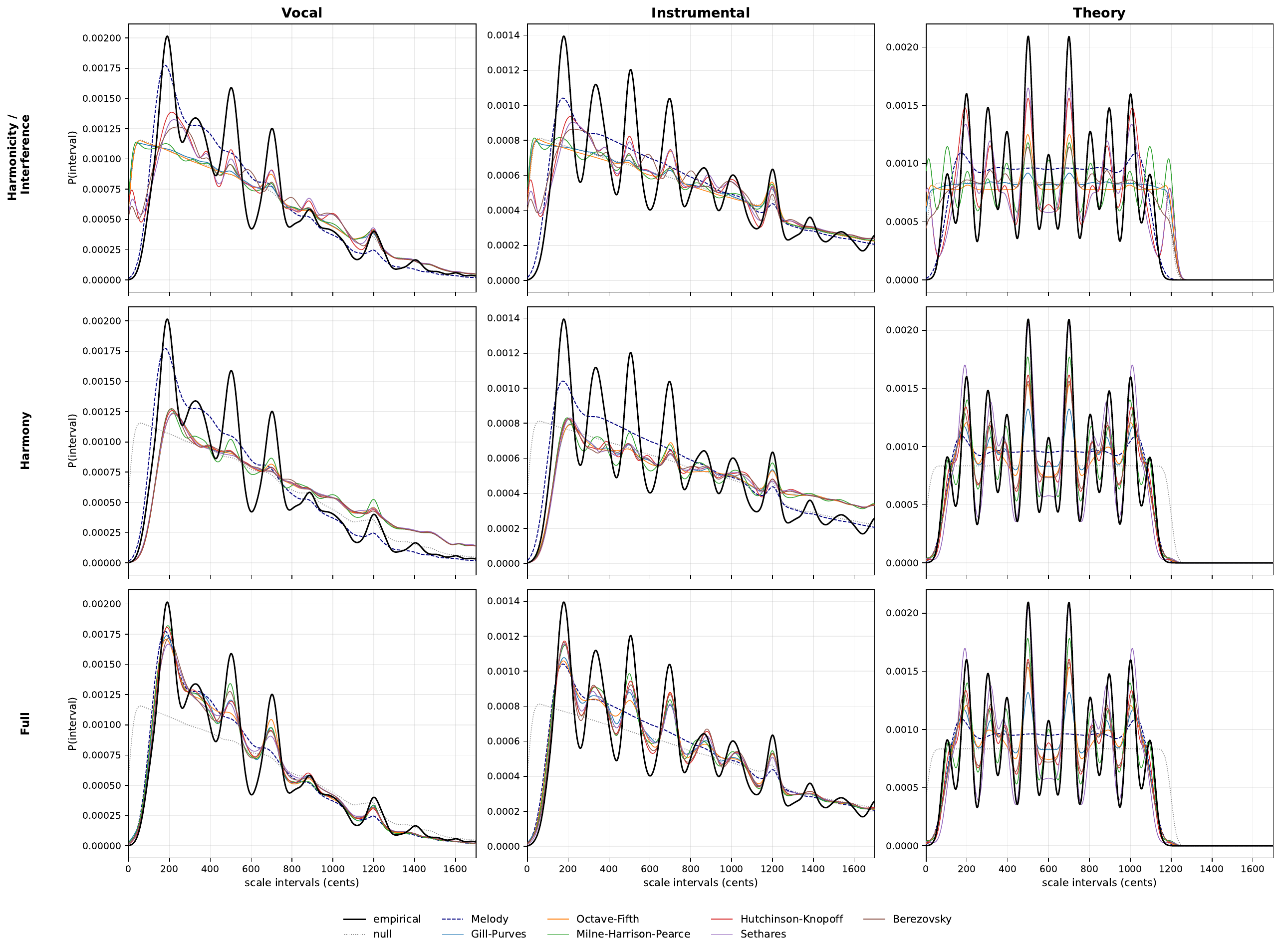}
  \caption{\textbf{Predicted scale-interval distributions.} As
  \fref{fig:jsd-steps}, but for the distribution of scale intervals (measured
  between all pairs of scale degrees) rather than step sizes.}
  \label{fig:jsd-intervals}
  \end{figure*}

  In the main text, models are compared using representative
  harmonicity/interference fitness functions, both as a log-probability
  gain over the uniform null and as predicted distributions
  overlaid on the empirical distributions. Here we reproduce both comparisons
  for all six harmonicity/interference fitness functions, confirming that the
  choice of fitness function does not change the conclusions.
  \fref{fig:model-grid} shows the log-probability gain with all six
  displayed together. \fref{fig:jsd-steps} and \fref{fig:jsd-intervals} show the
  corresponding best-fit predicted distributions of step sizes and of scale
  intervals, respectively.

  \fref{fig:model-grid} shows that the interference models outperform the
  harmonicity models when the harmonicity/interference fitness acts on its own,
  but not once it is combined into the Harmony or Full models. This is because
  the interference models' aversion to semitone-sized intervals only helps when
  there is no Interval Spacing component; where Interval Spacing is present, it
  already disfavours such intervals.

  \clearpage
\section{Octaves and Fifths are the main predictions for Harmonicity/Interference models}

  An open question is how many harmonic partials a model must account for in
  order to predict scales. This is important because earlier accounts of the evolution
  of scales referred to the importance of intervals with small-integer frequency ratios.
  Some such intervals (\eg, the octave, and the fifth) can be found within the first few
  partials, while others require a much longer harmonic series before they appear.
  This analysis directly tests this concept of small-integer ratios, as if only
  the octave and fifth are important for scale evolution, then the concept of
  small-integer ratios is certainly not general, and can be definitively ruled out
  as an important contributing factor.
  
\begin{figure}[ht!]
  \centering
  \includegraphics[width=\textwidth]{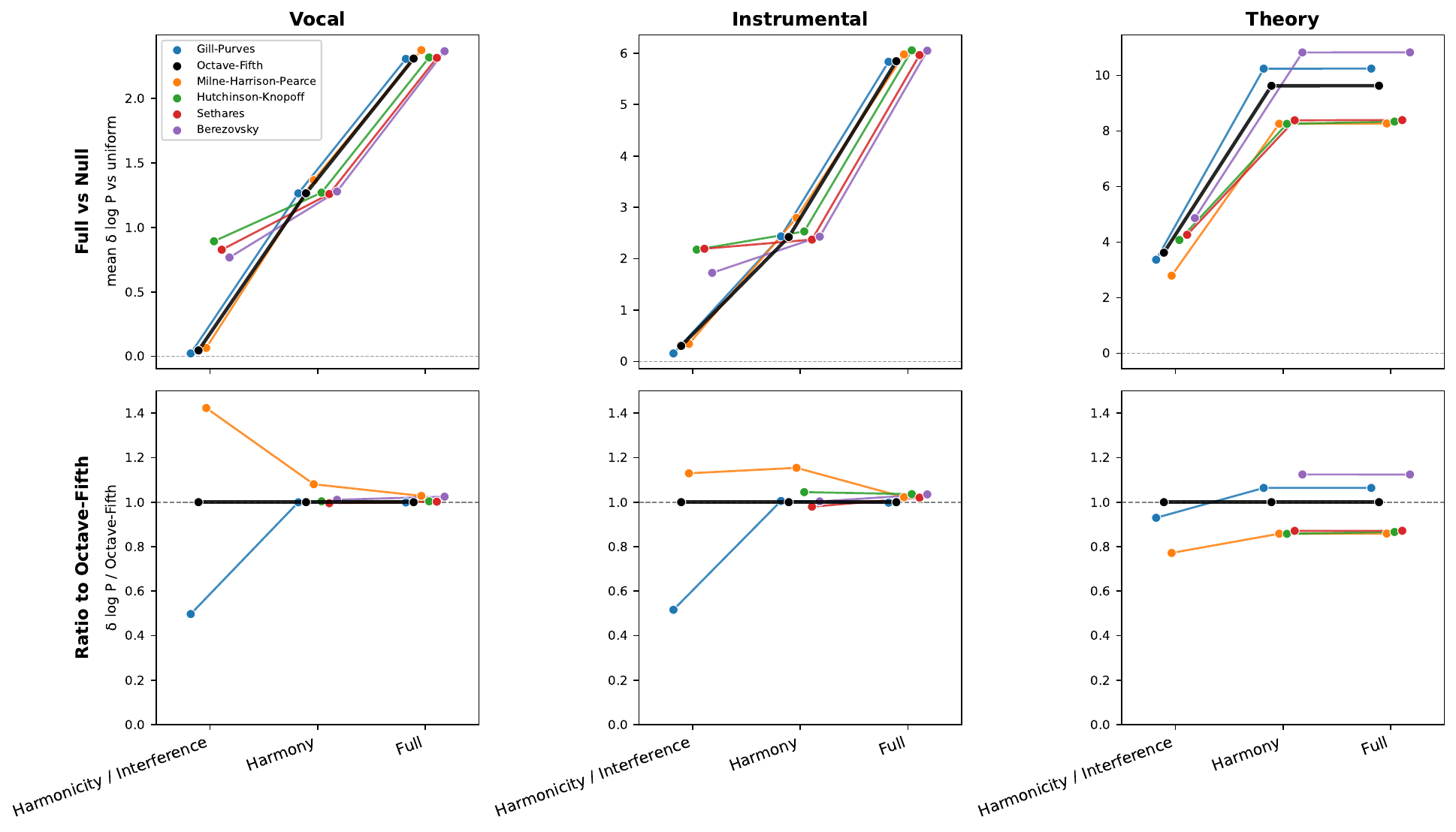}
  \caption{\textbf{Do higher-order partials improve the fit?} For
  each scale type (columns), the log-probability gain of each
  harmonicity/interference model is shown across three model groupings
  (harmonicity/interference alone, Harmony, Full). The lower row expresses each
  model's gain as a fraction of the octave-fifth model's, so that the
  octave-fifth reference sits at $1.0$ and the benefit of weighting extra
  partials can be read off directly.}
  \label{fig:overtone}
  \end{figure}

  We test this directly by comparing every
  harmonicity/interference model against the octave-fifth model, which
  credits only these two intervals. \fref{fig:overtone} shows that on the
  measured (Vocal and Instrumental) scales the more elaborate models -- which
  weight many higher partials -- perform close to the octave-fifth reference,
  and pull clearly ahead only on the mathematically designed Theory scales. This
  supports the view that a bias towards octaves and fifths captures most of what
  the harmonic models explain. Additional support for this view comes from the
  highest-scoring parameters shown in \fref{fig:param-sens-hp}:
  for measured scales, the highest-scoring models use few partials;
  one exception is the Milne-Harrison-Pearce model which performs best
  with $n \sim 12$ partials.

  The interference models differ from the harmonicity models chiefly in adding
  an aversion to intervals near a semitone. This makes them slightly better when
  the harmonicity/interference fitness is used on its own,
  but adds little once it is combined into the Harmony or Full models,
  where the Melody component already disfavours very small steps.

  To ask whether the octave-fifth model performs \emph{meaningfully} better or
  worse than the richer models, we compare region-weighted mean log-probability
  gains directly. We considered testing for statistical significance, however with
  several hundred non-independent scales a significance test is dominated by sample
  size, whereas the question of interest is the size of the difference.

  We restrict the comparison to the Harmony and Full models, on which the main
  analysis rests. We can see already that for the pure harmonicity/interference models,
  interference models perform better due to avoiding semitones; this prediction
  is not related to interactions between higher partials, so it is irrelevant
  to the question at hand. We also make the comparison within scale type, since the
  gains differ by orders of magnitude between Theory and measured scales.

  The advantage of any richer model over the Octave-Fifth model is small.
  Most lie within a factor of $1.08$ of Octave-Fifth, with a few exceptions:
  Milne-Harrison-Pearce on Instrumental scales ($1.15$); Berezovsky on
  Theory scales ($1.13$). Furthermore, for every model that beats
  octave-fifth another reports the same score or falls below it.
  Octaves and fifths thus capture the bulk of what the harmonic models explain
  for empirically attested scales.

  \clearpage
\section{Model comparison broken down by region and scale size}

\begin{figure*}[ht!]
  \centering
  \includegraphics[width=\textwidth]{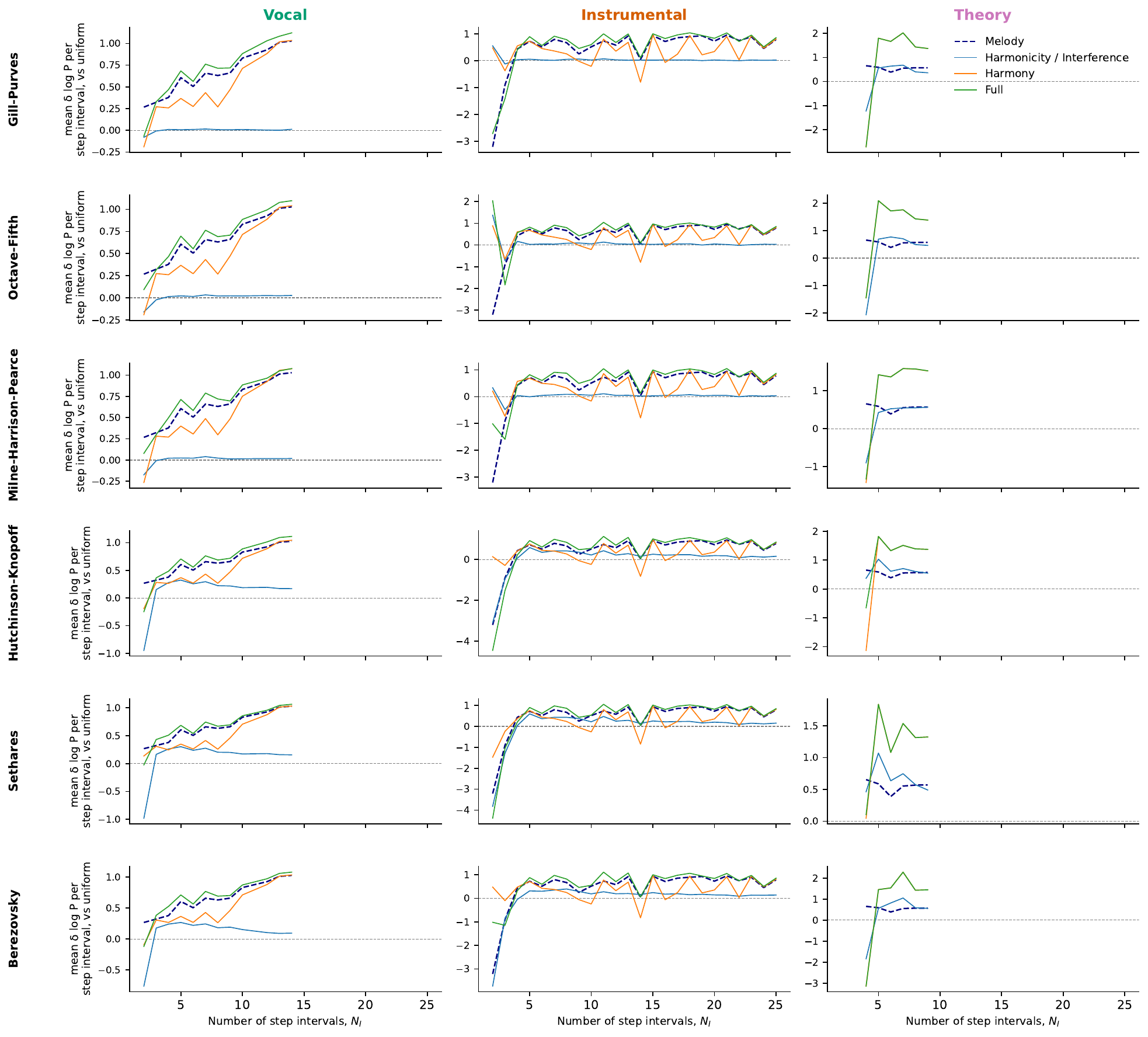}
  \caption{\textbf{Model performance by scale size.} Mean log-probability gain
  over the uniform null \emph{per step interval}, for each of the six
  harmonicity/interference models (rows) and each scale type (columns:
  Vocal, Instrumental, Theory), as a function of the number of step intervals
  $N_I$. Each panel draws the Melody model (navy dashed) as a
  cost-independent reference, together with the Harmonicity/Interference,
  Harmony and Full models evaluated at that row's cost function. The gain
  $\delta \log P$ is a per-scale quantity: a log-density ratio on the
  $(N_I - 1)$-dimensional simplex of steps summing to the scale range, so it
  grows roughly linearly with $N_I$ by construction. We therefore divide by
  $N_I - 1$, the number of free steps, which puts every scale size on a common
  footing and isolates how well the model fits, rather than how many intervals
  a scale has. The selection strength $\beta$ is optimised once per model and
  scale type, and held fixed across $N_I$; it is not refitted per scale size.
  The full range of $N_I$ is shown, and the largest sizes rest on very few
  scales. Model parameters are given in \textbf{SI Table~\ref{tab:hi-params}}.}
  \label{fig:delta-vs-nints}
  \end{figure*}

\begin{figure*}[ht!]
  \centering
  \includegraphics[width=\textwidth]{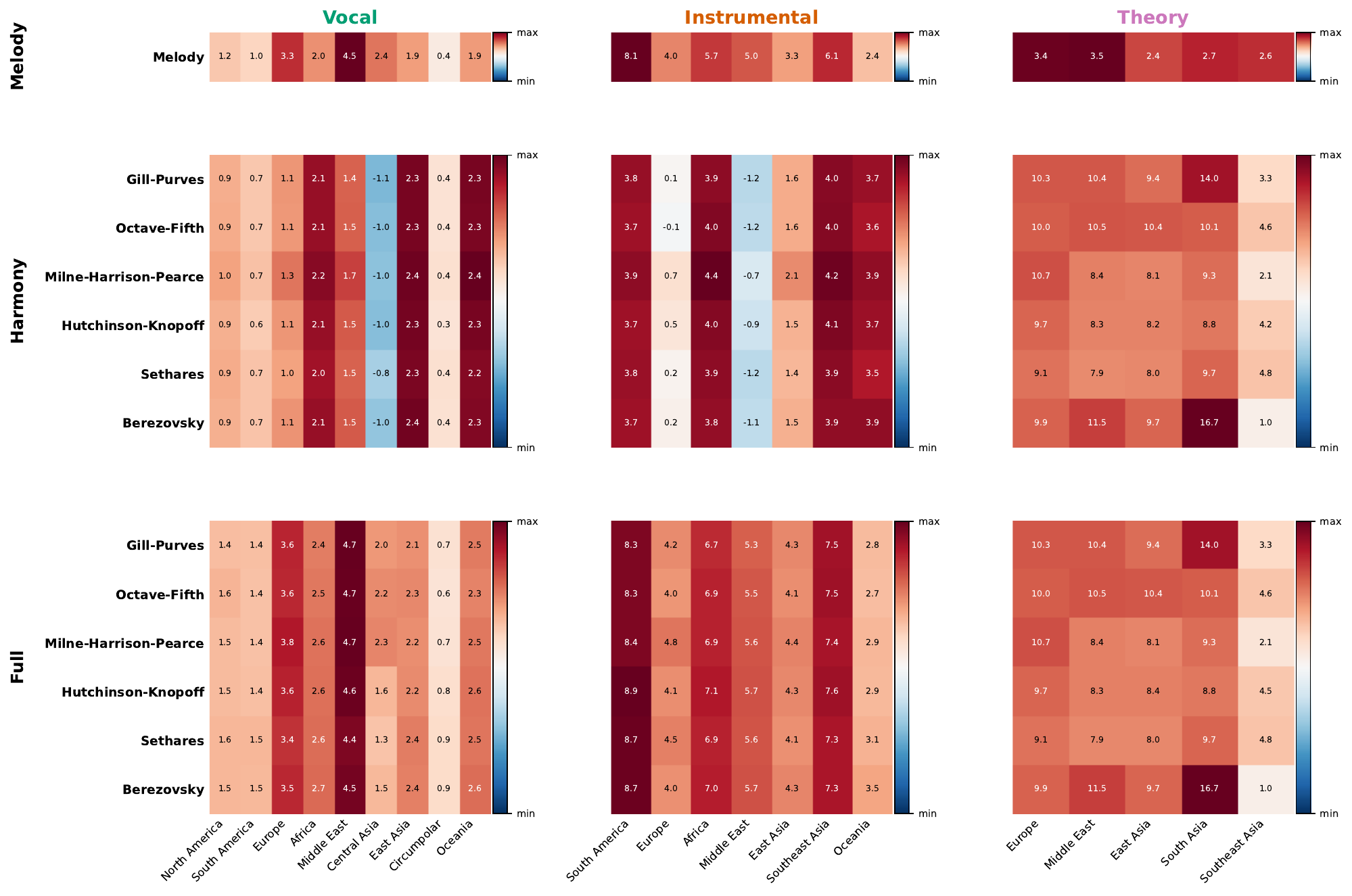}
  \caption{\textbf{Model performance by region.} Mean per-scale
  log-probability gain over the uniform null, evaluated on scale intervals and
  broken down by geographic region. The figure is a stack of three tiers -- the
  Melody model (one row), the Harmony model (six rows) and the Full model (six
  rows) -- where the six rows of the lower two tiers are the six
  harmonicity/interference models. Each tier is divided into three blocks laid
  side by side, one per scale type (Vocal, Instrumental, Theory), giving nine
  heatmap panels in all; within a panel, columns are the regions holding at
  least ten scales of that type.}
  \label{fig:delta-region}
  \end{figure*}

  We next look at how model performance varies across scale sizes
  and across geographic regions. \fref{fig:delta-vs-nints} breaks the
  comparison down by scale size, and \fref{fig:delta-region} by region.

  We found that the log-probability gain $\delta \log P$ increases a lot with
  scale size for all models, and based on this we attribute the increase primarly
  to the increased volume of the scale space. The hypothesis is that null space grows
  disproportionately to the spcae of likely scales as dimensionality increases.
  As a reminder, a scale of $N_I$ steps is a point on an $(N_I-1)$-dimensional
  simplex -- the steps are constrained to sum to the scale range --
  so $\delta \log P$ is a log-density ratio on a space whose dimension grows with $N_I$.
  Each extra step contributes its own increment to the total, and the gain therefore
  rises roughly linearly with $N_I$ for any model that is better than chance at all,
  however weakly. This is a property of the measure, not of the theory:
  it says that larger scales inhabit larger spaces, in which a model has more
  room to beat chance. It is the same fact that requires us to compare scales
  only against others of the same size (\textit{Scale Space}, main text).
  In \fref{fig:delta-vs-nints} we therefore divide by $N_I - 1$, the number of
  free steps, and plot the gain per step interval, which places every scale size
  on a common footing.

  Once normalised, the models are strikingly similar in shape, and this is the
  main thing \fref{fig:delta-vs-nints} has to say. The Melody model rises gently
  with $N_I$ (Vocal: from $\approx 0.3$ at $N_I = 3$ to $\approx 0.6$ at
  $N_I = 8$), and the Harmony and Full models rise with it; the
  harmonicity/interference models on their own are flat and close to zero
  throughout (Vocal: $-0.007$ to $0.016$ across $N_I = 3$--$8$). Because the
  Harmony and Full models contain (at least part of) the melodic factor,
  their profiles follow the Melody model's by construction,
  and what harmonicity adds on top of melody is small and does not grow with
  scale size. In contrast to an earlier version of this manuscript, we draw no
  strong conclusions about scale size from this figure. 

  Model performance varies considerably across regions, and
  \fref{fig:delta-region} shows that this is true of every theory we consider,
  not only the harmonic ones. The performance for the Meldoy model is far from uniform:
  among Vocal scales the Melody model's gain ranges from
  $0.4$ in Circumpolar scales to $4.5$ in the Middle East, 
  and among Instrumental scales from $2.4$ in Oceania to $8.1$ in
  South Asia. Regional heterogeneity is therefore a general feature of this
  comparison, and likely related to sampling differences, as differences in scale
  size clearly can produce this level of variance across regions. The strongest
  conclusion that can clearly be made is that despite the large number of scales
  studied here, they are still insufficient for a robust regional comparison.

  To a large extent, the harmonic model results simply inherit the features
  of the Melody model. For Vocal and Instrumental scales the Full model's
  regional ordering closely follows the
  Melody model's -- for Instrumental scales the ordering is identical (South
  Asia, Southeast Asia, Africa, Middle East, Europe, East Asia, Oceania). Thus,
  most of the regional variation in the Full model is melodic in origin.

  The differences are starker in for the Harmony model.
  Harmony is not merely weak in some regions but actively
  worse than chance: the Harmony model's gain is negative for Central Asian Vocal
  scales ($-0.82$ to $-1.10$ across the six cost functions) and for Middle
  Eastern Instrumental scales ($-0.68$ to $-1.20$), even though the Full model is
  positive in both cases, because the melodic component carries them. We are
  cautious about the observations about Central Asian Vocal scales, since we
  only have $15$ of them. Overall it does seem that there is greater
  variation in the results of the Harmony model across regions, but more data
  is needed to get a clearer picture.

  \clearpage
\section{Tonal hierarchy exploratory analysis}

  Throughout this work every scale is weighted equally, and every degree
  within a scale likewise, even though some scale degrees are more
  structurally important than others. Here we take a preliminary look at what
  changes when that assumption is relaxed, using the one dataset for which we
  have tonal hierarchies alongside the scales themselves.

  \begin{figure*}[ht!]
  \centering
  \includegraphics[width=\textwidth]{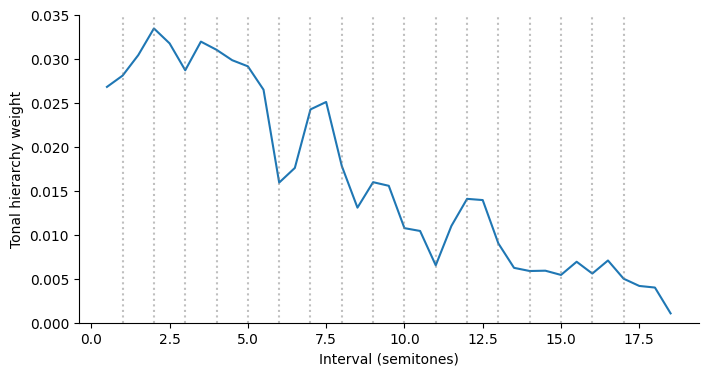}
  \caption{\textbf{Salience of intervals in tonal hierarchies.} For each
  scale in the corpus from \cite{brownMusical2024}, we have not only scale data, but
  also tonal hierarchies. For each scale, we calculate all possible scale
  intervals between all scale degrees and assign weights that are
  proportional to the joint probability of the two scale degrees (\ie,
  given a tonal hierarchy, how likely is it that a particular interval will
  be heard?). We plot the average tonal hierarchy weight for all scale
  intervals from all scales as a function of interval size, in bins of 50
  cents.}
  \label{fig:tonal-hierarchy}
  \end{figure*}

  \clearpage
\section{Fitting the Melody model}
\label{si:melody-fit}

  The Melody model has three free parameters, which we fit by minimising an
  $L_1$ objective between the model's predicted step-size distribution and the
  empirical one (see \textit{Melody Model} in the main text).

  \nsb{Which scales the fit uses.} Two choices define the fit, and both involve
  some judgement. The first is which scales to fit to. We use the Vocal scales
  only: instrument-tuning measurements are excluded because not all notes on an
  instrument are necessarily used melodically, and Theory scales are excluded
  because they are constructed rather than performed, and so are free of the
  production and perception constraints the model describes. The second is how
  those scales are weighted. We weight by region, as elsewhere in this work
  (\textbf{SI Section~\ref{si:weighted-sampling}}), and additionally by $1/N_I$,
  so that scales with many steps do not contribute more step intervals than
  scales with few. \fref{fig:canonical-steps} shows the empirical step-size
  distribution under each combination of these choices; they differ little, which
  is why the fitted model is insensitive to the choice.

\begin{figure*}[ht!]
  \centering
  \includegraphics[width=\textwidth]{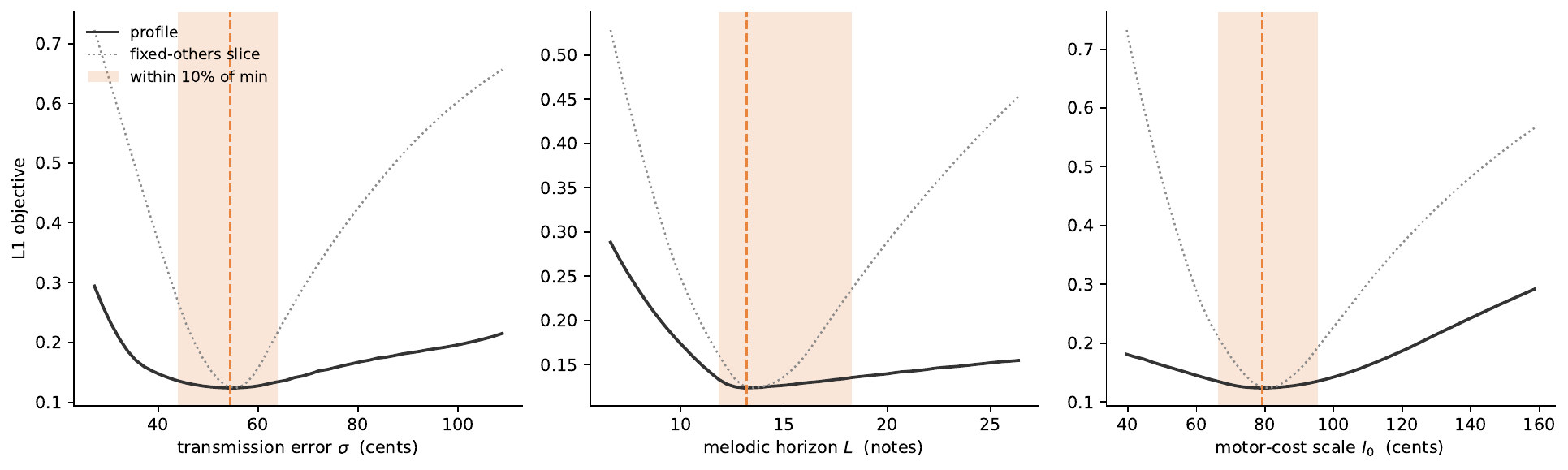}
  \caption{\textbf{The Melody-model fit is well-constrained.} For each of the
  three fitted Melody parameters ($\sigma$, $L$, $I_0$), the $L_1$ objective is
  shown as a profile (solid grey; the other two parameters re-optimised at each
  value) and as a slice (dotted; the other two held at their optimum). The orange
  band marks where the profile stays within $10\%$ of its minimum, and
  the dashed orange line marks the fitted optimum. The band indicates how sharply
  the objective is peaked; it is not a confidence interval, and it is not the
  interval reported in main-text Fig.~3C--D (see text).}
  \label{fig:melody-basin}
  \end{figure*}

\begin{figure*}[ht!]
  \centering
  \includegraphics[width=\textwidth]{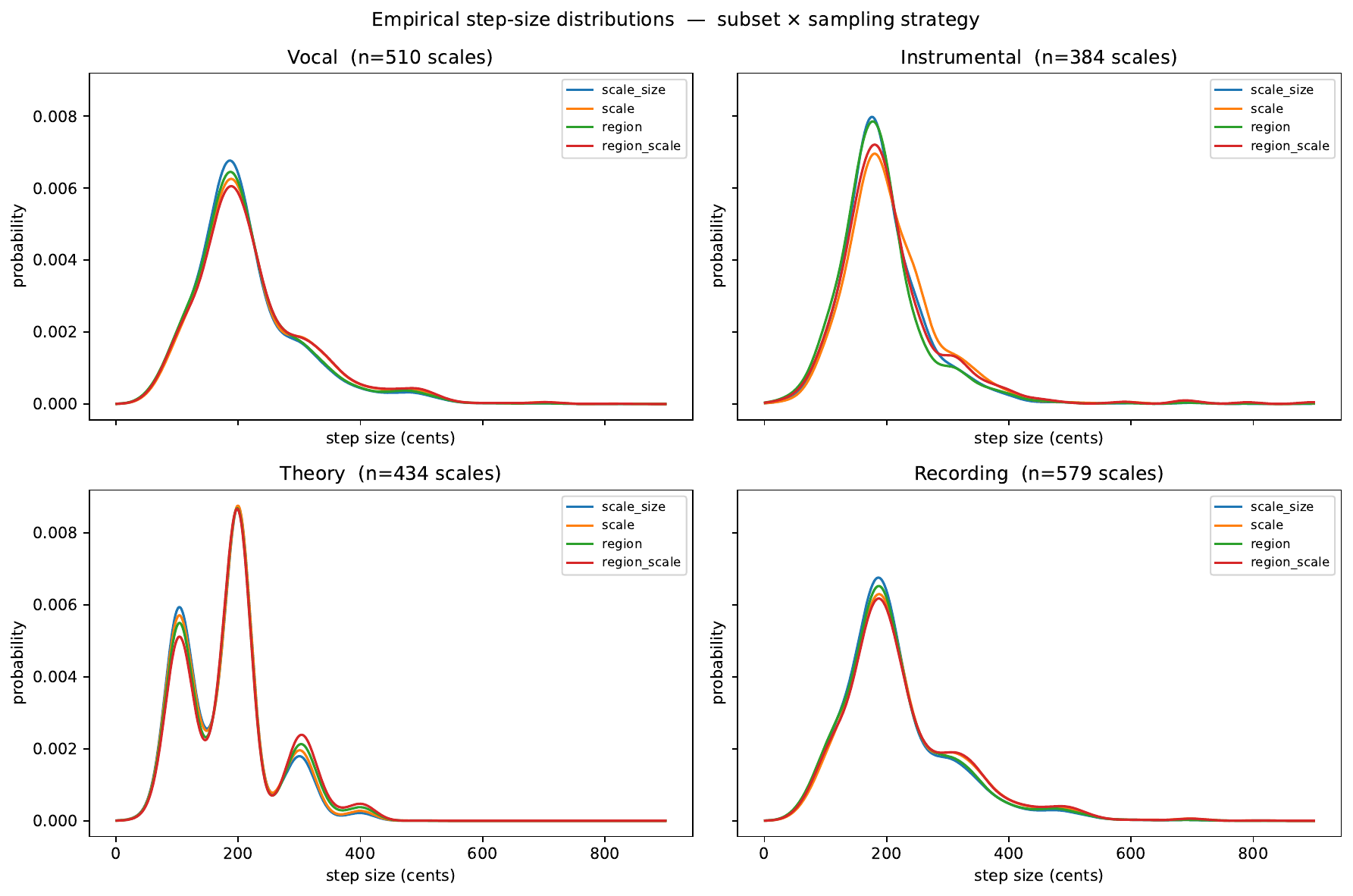}
  \caption{\textbf{Candidate step-size distributions for the Melody
  model fit.} Empirical step-size distributions computed under different choices
  of which scales to include and how to weight them. The distribution used for
  the fit reported in the main text -- Vocal scales, weighted by region and by
  $1/N_I$ -- is one of these; the others differ little from it.}
  \label{fig:canonical-steps}
  \end{figure*}

  \nsb{How well-constrained the parameters are.} \fref{fig:melody-basin} shows
  that the fit is well-behaved. For each parameter we plot the $L_1$ objective
  both as a profile (re-optimising the other two parameters at each value) and as
  a slice (holding the other two at their optimum), and shade the range over
  which the profile stays within $10\%$ of its minimum. The shaded basin is
  narrow, indicating that each parameter is reasonably well-constrained by the
  step-size data.

  This shaded basin is a visual indication of how sharply the objective is
  peaked; it is \emph{not} the interval reported in the main text, and it is not
  a confidence interval. The fit itself returns a single point estimate per
  parameter -- the optimum, marked by the dashed line -- and it is those point
  estimates ($\sigma = 54$ cents, $L = 13$, $I_0 = 79$ cents) that are used
  throughout the analysis. The intervals plotted in main-text Fig.~3C--D are a
  separate calculation: the same profile, but cut at a threshold calibrated by
  bootstrap rather than at a fixed fraction of the minimum. Because the $L_1$
  objective has no distributional theory (unlike a negative log-likelihood, where
  $2(\mathrm{NLL}_{\textrm{profile}} - \mathrm{NLL}_{\min})$ is asymptotically
  $\chi^2_1$), the threshold cannot be read off a standard table and must be
  calibrated from the data. We bootstrap the scales and refit; for each replicate
  we read off how far the full-data profile rises at that replicate's optimum,
  and take the $95^{\text{th}}$ percentile of those rises as the threshold
  $\Delta^*$ ($\alpha = 0.05$). The reported interval is then every parameter
  value whose profile lies within $\Delta^*$ of the minimum. This is what gives
  the scale-fit $\sigma$ range of approximately \SIrange{49}{60}{cents} quoted in
  \textbf{SI Section~\ref{si:interval-spacing-params}}.

  \clearpage
\section{Fitting the Harmonicity/Interference models}

  Each harmonicity/interference model has its own internal parameters,
  which we set by grid search rather than fixing them \textit{a priori}: a
  tolerance $w$ for the single-window models (Gill-Purves and octave-fifth), and
  the number of partials $n$ and the roll-off $\rho$ for the multi-partial
  models. For the interference models there is also the fundamental frequency
  $f_0$, which we fit over the grid $f_0 \in \{50, 100, 200, 500, 1000\}$~Hz
  (see \textit{Interference Models} in the main text). The best-fitting values
  per scale type are listed in \tref{tab:hi-params}. The two figures below show
  that model performance is not sensitive to the precise choice of these
  parameters.

  \begin{figure*}[ht!]
  \centering
  \includegraphics[width=\textwidth]{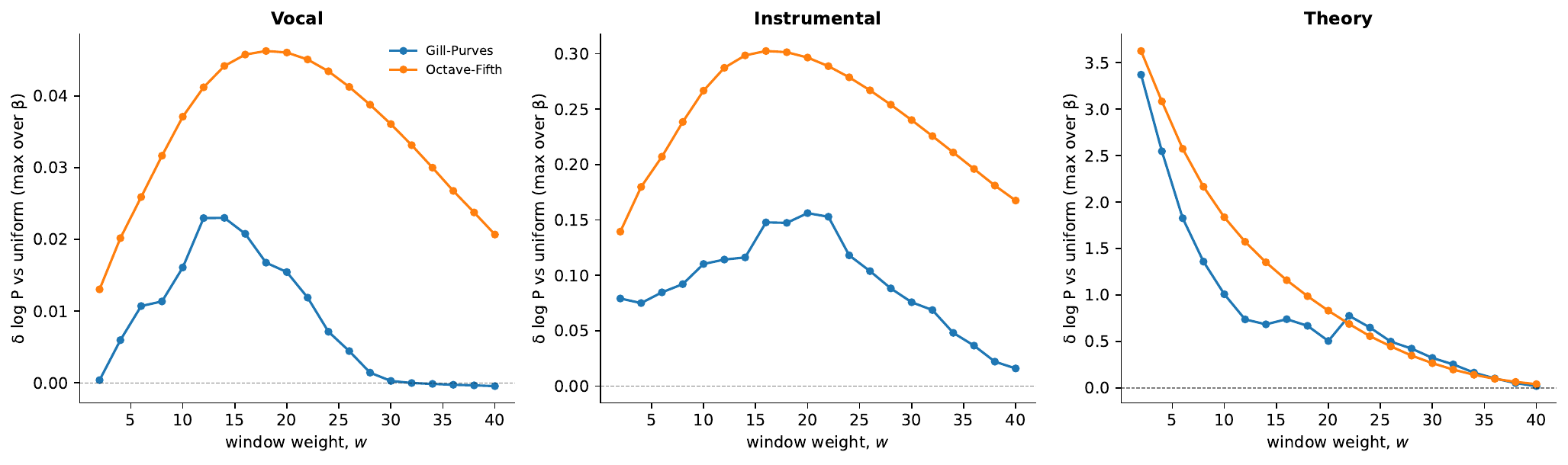}
  \caption{\textbf{Robustness to the tolerance parameter $w$.} Mean per-scale
  log-probability gain over the uniform null (maximised over the
  selection strength $\beta$) as a function of the tolerance $w$, for the two
  harmonicity models that use it (Gill-Purves and octave-fifth), shown
  separately for each scale type. The gain shown is for the
  harmonicity/interference fitness acting on its own, without an Interval
  Spacing or Motor Constraint component.}
  \label{fig:param-sens-w}
  \end{figure*}

  \begin{figure*}[ht!]
  \centering
  \includegraphics[width=\textwidth]{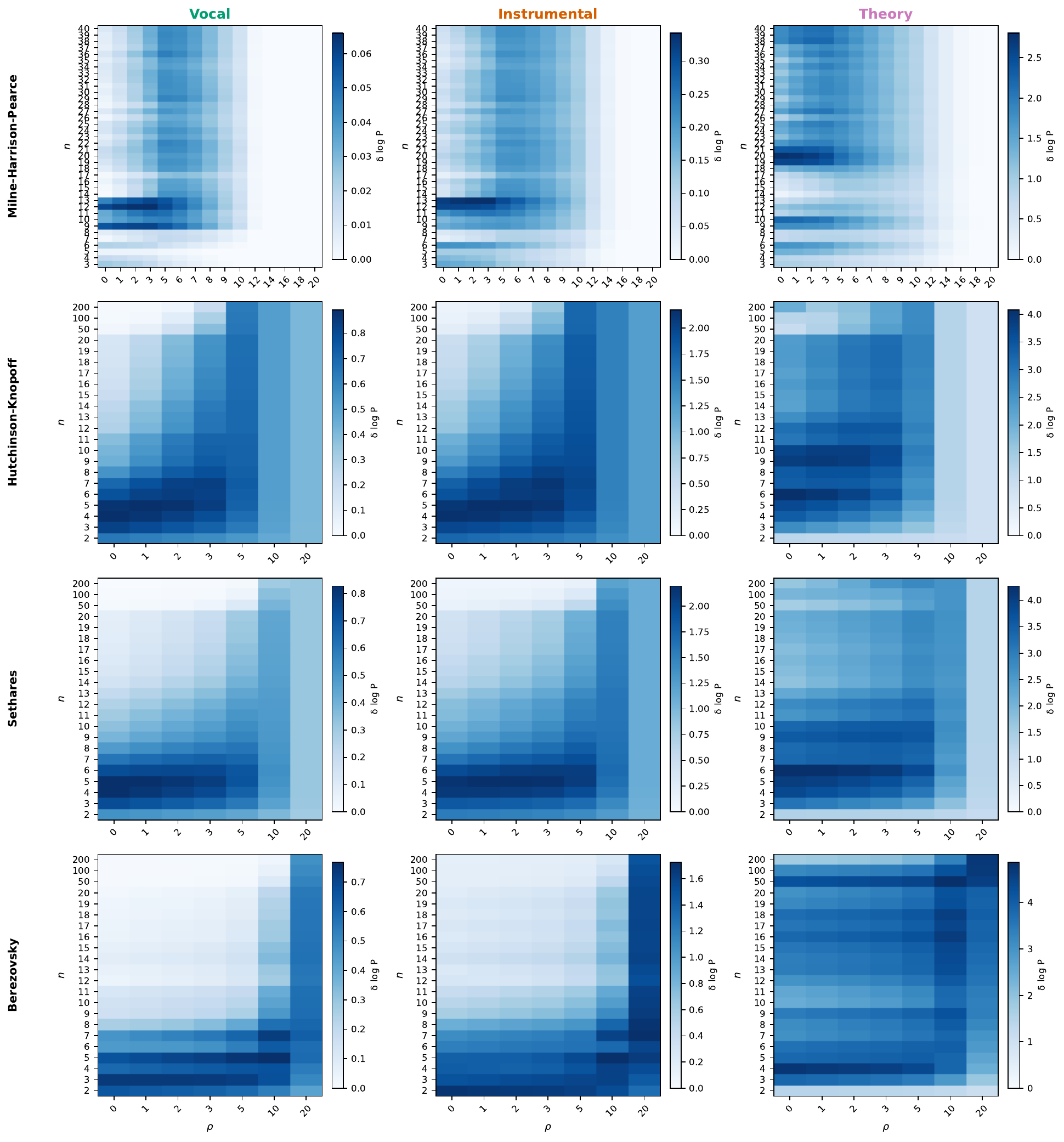}
  \caption{\textbf{Parameter sensitivity of the multi-partial models.} For each
  model (rows: the six harmonicity/interference models) and each scale
  type (columns), the mean per-scale log-probability gain over the
  uniform null (maximised over $\beta$) is shown as a heatmap over the
  number of partials $n$ (vertical) and the roll-off $\rho$
  (horizontal). For the interference models, the fundamental frequency $f_0$ is
  maximised over alongside $\beta$ at each cell. Colour is normalised separately
  within each scale type. Models that lack $n$ and $\rho$ parameters (the
  single-window models) are marked accordingly. As in
  \fref{fig:param-sens-w}, the gain shown is for the harmonicity/interference
  fitness acting on its own.}
  \label{fig:param-sens-hp}
  \end{figure*}

  \begin{longtable}{lllccccc}
  \caption{\textbf{Best-fitting parameters per scale type for each
  harmonicity/interference model.}}
  \label{tab:hi-params} \\
  Model & Cost function & Scale type & $\beta$ & $w$ (cents) & $n$ & $\rho$ & $f_0$ (Hz) \\
  \hline
  Harmonicity / Interference & Gill-Purves & Vocal & 0.63 & 14 & — & — & — \\
   & Gill-Purves & Instrumental & 1.9 & 20 & — & — & — \\
   & Gill-Purves & Theory & 3.3 & 2 & — & — & — \\
   & Octave-Fifth & Vocal & 0.83 & 18 & — & — & — \\
   & Octave-Fifth & Instrumental & 2.9 & 16 & — & — & — \\
   & Octave-Fifth & Theory & 4.4 & 2 & — & — & — \\
   & Milne-Harrison-Pearce & Vocal & 0.95 & — & 12 & 3 & — \\
   & Milne-Harrison-Pearce & Instrumental & 2.5 & — & 13 & 2 & — \\
   & Milne-Harrison-Pearce & Theory & 5.0 & — & 20 & 0 & — \\
   & Hutchinson-Knopoff & Vocal & 5.8 & — & 5 & 1 & 1000 \\
   & Hutchinson-Knopoff & Instrumental & 13.0 & — & 5 & 2 & 1000 \\
   & Hutchinson-Knopoff & Theory & 15.0 & — & 6 & 0 & 1000 \\
   & Sethares & Vocal & 5.0 & — & 5 & 0 & 1000 \\
   & Sethares & Instrumental & 13.0 & — & 5 & 2 & 1000 \\
   & Sethares & Theory & 15.0 & — & 6 & 0 & 1000 \\
   & Berezovsky & Vocal & 5.8 & — & 5 & 10 & 1000 \\
   & Berezovsky & Instrumental & 13.0 & — & 5 & 10 & 1000 \\
   & Berezovsky & Theory & 10.0 & — & 50 & 10 & 1000 \\
  \hline
  Harmony & Gill-Purves & Vocal & 0.55 & 20 & — & — & — \\
   & Gill-Purves & Instrumental & 2.9 & 22 & — & — & — \\
   & Gill-Purves & Theory & 7.6 & 2 & — & — & — \\
   & Octave-Fifth & Vocal & 0.42 & 14 & — & — & — \\
   & Octave-Fifth & Instrumental & 2.5 & 12 & — & — & — \\
   & Octave-Fifth & Theory & 7.6 & 2 & — & — & — \\
   & Milne-Harrison-Pearce & Vocal & 1.4 & — & 13 & 3 & — \\
   & Milne-Harrison-Pearce & Instrumental & 3.8 & — & 13 & 2 & — \\
   & Milne-Harrison-Pearce & Theory & 11.0 & — & 38 & 2 & — \\
   & Hutchinson-Knopoff & Vocal & 0.95 & — & 7 & 0 & 1000 \\
   & Hutchinson-Knopoff & Instrumental & 6.6 & — & 8 & 1 & 1000 \\
   & Hutchinson-Knopoff & Theory & 20.0 & — & 19 & 0 & 1000 \\
   & Sethares & Vocal & 0.72 & — & 2 & 20 & 50 \\
   & Sethares & Instrumental & 5.0 & — & 6 & 0 & 1000 \\
   & Sethares & Theory & 26.0 & — & 9 & 0 & 1000 \\
   & Berezovsky & Vocal & 1.1 & — & 7 & 0 & 1000 \\
   & Berezovsky & Instrumental & 4.4 & — & 7 & 0 & 500 \\
   & Berezovsky & Theory & 20.0 & — & 50 & 0 & 1000 \\
  \hline
  Full & Gill-Purves & Vocal & 3.8 & 38 & — & — & — \\
   & Gill-Purves & Instrumental & 7.6 & 32 & — & — & — \\
   & Gill-Purves & Theory & 7.6 & 2 & — & — & — \\
   & Octave-Fifth & Vocal & 3.3 & 34 & — & — & — \\
   & Octave-Fifth & Instrumental & 7.6 & 26 & — & — & — \\
   & Octave-Fifth & Theory & 7.6 & 2 & — & — & — \\
   & Milne-Harrison-Pearce & Vocal & 3.3 & — & 3 & 10 & — \\
   & Milne-Harrison-Pearce & Instrumental & 6.6 & — & 13 & 7 & — \\
   & Milne-Harrison-Pearce & Theory & 11.0 & — & 38 & 2 & — \\
   & Hutchinson-Knopoff & Vocal & 2.9 & — & 6 & 0 & 200 \\
   & Hutchinson-Knopoff & Instrumental & 10.0 & — & 6 & 0 & 200 \\
   & Hutchinson-Knopoff & Theory & 20.0 & — & 19 & 0 & 1000 \\
   & Sethares & Vocal & 2.9 & — & 4 & 0 & 200 \\
   & Sethares & Instrumental & 10.0 & — & 6 & 0 & 500 \\
   & Sethares & Theory & 26.0 & — & 9 & 0 & 1000 \\
   & Berezovsky & Vocal & 2.9 & — & 4 & 0 & 200 \\
   & Berezovsky & Instrumental & 8.7 & — & 6 & 0 & 200 \\
   & Berezovsky & Theory & 20.0 & — & 50 & 0 & 1000 \\
  \hline
  \end{longtable}

  \clearpage
\section{Consistency check on choice of scale space}

  The model comparison in the main text scores every model against a uniform
  distribution over the fixed-range simplex of scales (see \textit{Scale space}
  in the main text). An earlier version of the analysis instead scored the
  composite Full model against the Melody model alone.
  The main difference between the two is the scale space used: the current
  one fixes the range and draws it from a specific empirical distribution, while
  the previous one used a scale space in which the range was an emergent property
  of the combination of scale size with the step-size distribution.
  We retain that analysis here as a consistency check: the conclusions do not
  depend strongly on which scale space is used.

  In the old approach, the Harmony models cannot by themselves predict
  empirical step sizes, so each harmonicity/interference fitness is combined with
  the Melody model and the composite is compared to the Melody baseline. The
  space of possible scales is defined by the Melody model itself: a scale is a
  set of $N_I$ step intervals drawn i.i.d.\ from the empirical step-size
  distribution $P_E(I_S)$. For Theory scales, which span a single octave, the
  sampled steps are renormalised so that the scale range is exactly one octave.
  Writing the composite likelihood as $\mathcal{L}_m(S)$ and the Melody baseline
  probability of a scale as $P_{\textrm{Melody}}(S) = \prod_{I_S \in S} P_E(I_S)$,
  the normalising constant is
  \be
  Z' = \sum_{S} \mathcal{L}_m(S)\, P_{\textrm{Melody}}(S)
     \;=\; \langle \mathcal{L}_m(S) \rangle_{P_{\textrm{Melody}}},
  \ee
  which we estimate by sampling scales from $P_E$ (separately for each $N_I$).
  The per-scale score is then a log-probability gain over the Melody baseline
  rather than over chance,
  \be
  \log \frac{P_m(S)}{P_{\textrm{Melody}}(S)} \;=\; \log \frac{\mathcal{L}_m(S)}{Z'},
  \ee
  optimised over $\beta$ and the fitness parameters exactly as in the main
  analysis. Note that this is not the same quantity as the $\delta \log P$ of
  the main text, which is always measured against the uniform null; the two
  differ in their baseline, which is precisely what this check varies.

  Table~\ref{tab:legacy-baseline} reports the Full model's gain against both the
  uniform null (main text) and the old Melody baseline. The two baselines give
  the same qualitative ranking, confirming that the main conclusions are robust
  to the choice of baseline. We exclude the Berezovsky interference results
  because the previous results were wrong; the published
  equation\cite{berezovskyStructure2019} had an error that was subsequently
  reported by a separate group\cite{nasserThermodynamics2025}.

  \begin{table}[h!]
  \centering
  \caption{\textbf{Full-model log-probability gain under the uniform null
  (main text, $\delta \log P$) and under the old Melody baseline.}}
  \label{tab:legacy-baseline}
  \begin{tabular}{llrr}
  \toprule
  Scale type & Harmonicity/interference model & Uniform null & Legacy Melody baseline \\
  \midrule
  Vocal & Gill-Purves & 2.31 & 2.24 \\
  Instrumental & Gill-Purves & 5.83 & 5.68 \\
  Theory & Gill-Purves & 10.2 & 11.5 \\
  Vocal & Octave-Fifth & 2.31 & 2.25 \\
  Instrumental & Octave-Fifth & 5.85 & 5.63 \\
  Theory & Octave-Fifth & 9.63 & 10.8 \\
  Vocal & Milne-Harrison-Pearce & 2.38 & 2.32 \\
  Instrumental & Milne-Harrison-Pearce & 5.98 & 5.92 \\
  Theory & Milne-Harrison-Pearce & 8.28 & 8.58 \\
  Vocal & Hutchinson-Knopoff & 2.32 & 2.18 \\
  Instrumental & Hutchinson-Knopoff & 6.06 & 5.69 \\
  Theory & Hutchinson-Knopoff & 8.34 & 8.83 \\
  Vocal & Sethares & 2.32 & 2.14 \\
  Instrumental & Sethares & 5.96 & 5.56 \\
  Theory & Sethares & 8.4 & 9.04 \\
  \bottomrule
  \end{tabular}
  \end{table}

  \clearpage

\section{Random-scale fitness significance threshold}

  In some cases our method guarantees that $\delta \log P \geq 0$. This is a
  consequence of optimising the selection strength $\beta$ (and the fitness
  parameters) for each model: comparing the uniform null with a
  harmonicity/interference model, the two are equivalent when $\beta = 0$, so the
  fitted model can never do worse and $\delta \log P \geq 0$ follows.
  In exactly the same way, the Harmony model reverts to the Interval Spacing model
  and the Full model reverts to the Melody model when $\beta = 0$.
  A positive $\delta \log P$ is therefore not, on its own, evidence of a real
  effect. To calibrate it, we ask how large a $\delta \log P$ the same fitting
  procedure would produce from scales that carry no structure for the model to exploit.

  To estimate how much of $\delta \log P$ is attributable to fitting parameters
  alone, we bootstrap populations of random scales and re-fit the models.
  We test each of the three nested comparisons in turn: adding
  a harmonicity/interference fitness to (i)~the uniform baseline, (ii)~the
  Interval Spacing baseline (giving the Harmony model), and (iii)~the Melody
  baseline (giving the Full model). We do this for each of the six
  harmonicity/interference models and three scale types.

  For every case we draw \num{10000} replicate data sets of random scales from
  the relevant $\beta = 0$ baseline, matching the real data in the number of
  scales, their sizes $N_I$, and the regional weights. Each replicate is refitted
  exactly as the real data are, maximising the region-weighted mean
  $\delta \log P$ over $\beta$ and the fitness parameters, and the maximised value
  is recorded. The sampled values form the null distribution of $\delta \log P$
  attributable to optimisation alone; the one-sided $p$-value is the fraction of
  replicates that reach or exceed the observed $\delta \log P$. Because this
  amounts to $6 \times 3 \times 3 = 54$ tests, we control the false-discovery
  rate across the whole table with the Benjamini-Hochberg
  procedure\cite{benjaminiControlling1995} and report
  the corrected $p$-values, $p_{\mathrm{FDR}}$, in \tref{tab:random_fitness}.

  The optimisation-only null is small: its upper $97.5$th percentile stays below
  $\delta \log P \approx 0.11$ across every comparison, model, and scale type,
  and below $\approx 0.035$ for the octave-normalised Theory scales. Against
  this null, $53$ of the $54$ tests are significant at $p_{\mathrm{FDR}} < 0.05$
  and $50$ at $p_{\mathrm{FDR}} < 0.005$.

  Adding a harmonicity/interference fitness to the uniform baseline, and adding it
  to the Melody baseline to form the Full model, are significant in
  every case: for all six models and all three scale types the observed
  $\delta \log P$ exceeds every one of the \num{10000} null replicates (bar a
  single replicate in one cell), giving $p_{\mathrm{FDR}} \le 2.2 \times 10^{-4}$
  -- at or next to the smallest value resolvable at $B = \num{10000}$. The Theory
  scales, whose observed $\delta \log P$ runs from $2.8$ to $7.8$, clear the
  null by orders of magnitude.

  The one marginal case is the Harmony model for \emph{Vocal} scales -- that is,
  adding a harmonicity/interference fitness on top of the Interval Spacing
  baseline. Here the observed gains are small (all below $0.16$, a factor of less
  than $1.2$ better than chance) and sit close to the
  null. The Milne-Harrison-Pearce and Berezovsky models remain significant at
  $p_{\mathrm{FDR}} < 0.005$; octave-fifth, Gill-Purves, and Hutchinson-Knopoff
  are significant only at $p_{\mathrm{FDR}} < 0.05$; and Sethares is not
  significant ($p_{\mathrm{FDR}} = 0.16$), the sole non-significant cell in the
  table.

  \begin{longtable}{lllrr}
  \caption{\textbf{Random-scale fitness significance.} Observed $\delta \log P$
   and Benjamini-Hochberg false-discovery-rate-corrected $p$-value $p_{\mathrm{FDR}}$.
  ``$<10^{-4}$'' denotes an observed value exceeding all \num{10000} replicates.}
  \label{tab:random_fitness} \\
  \hline
  Comparison & Scale type & Model & $\delta \log P$ & $p_{\mathrm{FDR}}$ \\
  \hline
  Harmony vs Interval Spacing & Instrumental & Octave-Fifth & 0.190 & 0.000112 \\
  Harmony vs Interval Spacing & Instrumental & Gill-Purves & 0.203 & 0.000112 \\
  Harmony vs Interval Spacing & Instrumental & Milne-Harrison-Pearce & 0.564 & 0.000112 \\
  Harmony vs Interval Spacing & Instrumental & Berezovsky & 0.198 & 0.000112 \\
  Harmony vs Interval Spacing & Instrumental & Hutchinson-Knopoff & 0.300 & 0.000112 \\
  Harmony vs Interval Spacing & Instrumental & Sethares & 0.141 & 0.000112 \\
  Harmony vs Interval Spacing & Theory & Octave-Fifth & 6.611 & 0.000112 \\
  Harmony vs Interval Spacing & Theory & Gill-Purves & 7.227 & 0.000112 \\
  Harmony vs Interval Spacing & Theory & Milne-Harrison-Pearce & 5.244 & 0.000112 \\
  Harmony vs Interval Spacing & Theory & Berezovsky & 7.810 & 0.000112 \\
  Harmony vs Interval Spacing & Theory & Hutchinson-Knopoff & 5.241 & 0.000112 \\
  Harmony vs Interval Spacing & Theory & Sethares & 5.370 & 0.000112 \\
  Harmony vs Interval Spacing & Vocal & Octave-Fifth & 0.052 & 0.0108 \\
  Harmony vs Interval Spacing & Vocal & Gill-Purves & 0.051 & 0.0201 \\
  Harmony vs Interval Spacing & Vocal & Milne-Harrison-Pearce & 0.154 & 0.000112 \\
  Harmony vs Interval Spacing & Vocal & Berezovsky & 0.065 & 0.000432 \\
  Harmony vs Interval Spacing & Vocal & Hutchinson-Knopoff & 0.056 & 0.00656 \\
  Harmony vs Interval Spacing & Vocal & Sethares & 0.046 & 0.162 \\
  \hline
  Full vs Melody & Instrumental & Octave-Fifth & 0.706 & 0.000112 \\
  Full vs Melody & Instrumental & Gill-Purves & 0.694 & 0.000112 \\
  Full vs Melody & Instrumental & Milne-Harrison-Pearce & 0.838 & 0.000112 \\
  Full vs Melody & Instrumental & Berezovsky & 0.911 & 0.000112 \\
  Full vs Melody & Instrumental & Hutchinson-Knopoff & 0.917 & 0.000112 \\
  Full vs Melody & Instrumental & Sethares & 0.825 & 0.000112 \\
  Full vs Melody & Theory & Octave-Fifth & 6.612 & 0.000112 \\
  Full vs Melody & Theory & Gill-Purves & 7.227 & 0.000112 \\
  Full vs Melody & Theory & Milne-Harrison-Pearce & 5.253 & 0.000112 \\
  Full vs Melody & Theory & Berezovsky & 7.810 & 0.000112 \\
  Full vs Melody & Theory & Hutchinson-Knopoff & 5.321 & 0.000112 \\
  Full vs Melody & Theory & Sethares & 5.373 & 0.000112 \\
  Full vs Melody & Vocal & Octave-Fifth & 0.330 & 0.000112 \\
  Full vs Melody & Vocal & Gill-Purves & 0.328 & 0.000112 \\
  Full vs Melody & Vocal & Milne-Harrison-Pearce & 0.395 & 0.000112 \\
  Full vs Melody & Vocal & Berezovsky & 0.388 & 0.000112 \\
  Full vs Melody & Vocal & Hutchinson-Knopoff & 0.339 & 0.000112 \\
  Full vs Melody & Vocal & Sethares & 0.336 & 0.000112 \\
  \hline
  Harmonicity/Interference vs uniform & Instrumental & Octave-Fifth & 0.303 & 0.000112 \\
  Harmonicity/Interference vs uniform & Instrumental & Gill-Purves & 0.156 & 0.000112 \\
  Harmonicity/Interference vs uniform & Instrumental & Milne-Harrison-Pearce & 0.342 & 0.000112 \\
  Harmonicity/Interference vs uniform & Instrumental & Berezovsky & 1.725 & 0.000112 \\
  Harmonicity/Interference vs uniform & Instrumental & Hutchinson-Knopoff & 2.177 & 0.000112 \\
  Harmonicity/Interference vs uniform & Instrumental & Sethares & 2.194 & 0.000112 \\
  Harmonicity/Interference vs uniform & Theory & Octave-Fifth & 3.626 & 0.000112 \\
  Harmonicity/Interference vs uniform & Theory & Gill-Purves & 3.372 & 0.000112 \\
  Harmonicity/Interference vs uniform & Theory & Milne-Harrison-Pearce & 2.799 & 0.000112 \\
  Harmonicity/Interference vs uniform & Theory & Berezovsky & 4.869 & 0.000112 \\
  Harmonicity/Interference vs uniform & Theory & Hutchinson-Knopoff & 4.081 & 0.000112 \\
  Harmonicity/Interference vs uniform & Theory & Sethares & 4.268 & 0.000112 \\
  Harmonicity/Interference vs uniform & Vocal & Octave-Fifth & 0.046 & 0.000112 \\
  Harmonicity/Interference vs uniform & Vocal & Gill-Purves & 0.023 & 0.00022 \\
  Harmonicity/Interference vs uniform & Vocal & Milne-Harrison-Pearce & 0.066 & 0.000112 \\
  Harmonicity/Interference vs uniform & Vocal & Berezovsky & 0.767 & 0.000112 \\
  Harmonicity/Interference vs uniform & Vocal & Hutchinson-Knopoff & 0.893 & 0.000112 \\
  Harmonicity/Interference vs uniform & Vocal & Sethares & 0.827 & 0.000112 \\
  \hline
  \end{longtable}

  \clearpage
\section{Weighted sampling}
\label{si:weighted-sampling}

  Our scale dataset is unevenly distributed across geographic regions: a few
  regions contribute many scales while others contribute only a handful. Left
  uncorrected, the regions with many scales would dominate every weighted
  average. To counter this, we cap the contribution of each region at a maximum
  effective number of samples (equivalently, a maximum total weight), so that
  scales from over-represented regions are down-weighted and no region exceeds
  the cap.

\begin{figure*}[ht!]
  \centering
  \includegraphics[width=\textwidth]{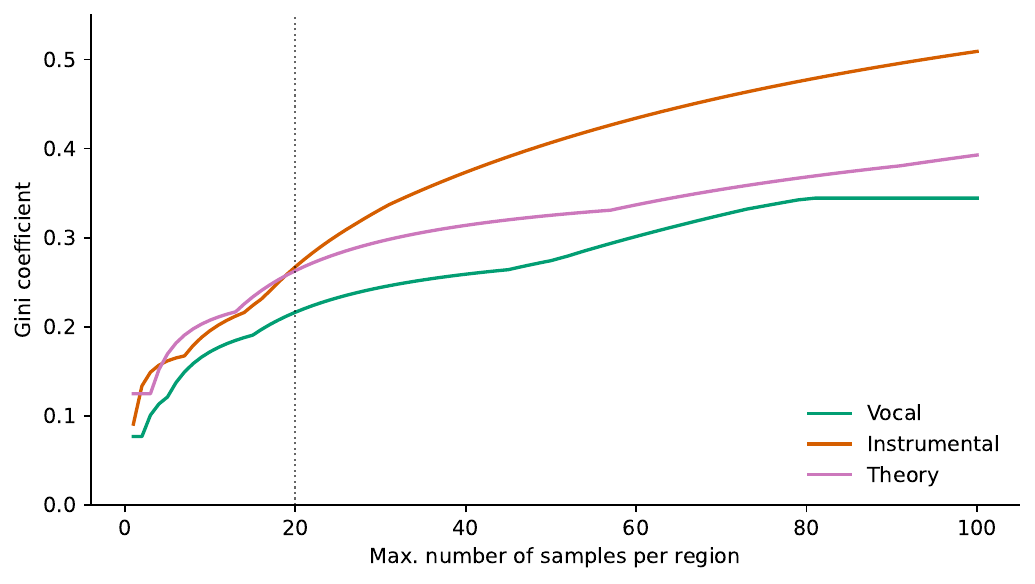}
  \caption{\textbf{Choosing the per-region sampling cap.} Gini coefficient of the
  per-region weights as a function of the maximum number of samples (maximum
  weight) allowed per region, shown separately for each scale type. The dotted
  line marks the value used in this work.}
  \label{fig:gini}
  \end{figure*}

  This cap involves a trade-off, which we quantify with the Gini coefficient of
  the per-region weights -- a standard measure of inequality ranging from $0$
  (perfect equality) to $1$ (maximal inequality). \fref{fig:gini} plots this Gini
  coefficient against the cap. A very high cap leaves the large regions dominant
  (high Gini); a very low cap discards most of the data and instead lets a few
  scales from small regions dominate. We choose a cap in the intermediate range,
  where the Gini coefficient levels off (dotted line).

  \clearpage

  \bibliography{ScaleEvo}
  \bibliographystyle{unsrtnat}